\documentclass[11pt]{article}
\pdfoutput=1
\usepackage{relsize}
\usepackage[utf8]{inputenc}
\usepackage{multirow}
\usepackage{amsmath, amsfonts, amssymb}
\usepackage{comment}
\usepackage{graphicx}
\usepackage{psfrag}
\usepackage{amsthm}
\usepackage[usenames,dvipsnames,svgnames,table]{xcolor}
\usepackage{enumerate}
\usepackage{arydshln}
\usepackage{soul}
\usepackage{slashed}
\usepackage{subfig}
\usepackage{mathrsfs}
\usepackage{a4wide}
\usepackage{tikz}
\usepackage{tikz-cd}
\usetikzlibrary{shapes.geometric}
\usepackage{tcolorbox}
\usepackage{mathtools}

\usepackage{epsf}

\usepackage{color}
\definecolor{dark-gray}{gray}{0.20}
\definecolor{gray}{gray}{0.30}
\definecolor{light-gray}{gray}{0.80}
\definecolor{dark-red}{rgb}{0.7,0,0}
\definecolor{dark-green}{rgb}{0.1,0.4,0}
\definecolor{dark-blue}{rgb}{0.3,0.3,0.7}
\definecolor{light-blue}{rgb}{0.8,0.8,1}
\definecolor{swamp}{RGB}{240, 199, 197}
\definecolor{landscape}{RGB}{180, 250, 199}
\definecolor{undecided}{RGB}{252, 252, 197}

\usepackage{pifont}

\usepackage{setspace}

\newcommand{\beq}{\begin{equation}}  \newcommand{\eeq}{\end{equation}}
\newcommand{\bal}{\begin{aligned}}   \newcommand{\eal}{\end{aligned}}
\newcommand{\be}{\begin{equation}}
\newcommand{\ee}{\end{equation}}

\def\IC{{\bf{C}}}
\def\IS{{\bf {S}}}
\def\IR{{\bf {R}}}
\def\IP{{\bf {P}}}
\def\IZ{{\bf {Z}}}
\def\IT{{\bf T}}
\def\NN{{\cal {N}}}
\def\beqa{\begin{eqnarray}}
\def\eeqa{\end{eqnarray}}

\def\tr{{\rm tr}\,}
\def\Tr{{\rm Tr}\,}
\def\id{{\bf 1}}

\newcommand{\drawsquare}[2]{\hbox{%
\rule{#2pt}{#1pt}\hskip-#2pt
\rule{#1pt}{#2pt}\hskip-#1pt
\rule[#1pt]{#1pt}{#2pt}}\rule[#1pt]{#2pt}{#2pt}\hskip-#2pt
\rule{#2pt}{#1pt}}

\newcommand{\fund}{\raisebox{-.5pt}{\drawsquare{6.5}{0.4}}}
\newcommand{\Ysymm}{\raisebox{-.5pt}{\drawsquare{6.5}{0.4}}\hskip-0.4pt%
        \raisebox{-.5pt}{\drawsquare{6.5}{0.4}}}
\newcommand{\Yasymm}{\raisebox{-3.5pt}{\drawsquare{6.5}{0.4}}\hskip-6.9pt%
        \raisebox{3pt}{\drawsquare{6.5}{0.4}}}
\newcommand{\antifund}{\overline{\fund}}
\newcommand{\bYasymm}{\overline{\Yasymm}}
\newcommand{\bYsymm}{\overline{\Ysymm}}

\def\be{\begin{equation}}
\def\ee{\end{equation}}
\def\bea{\begin{eqnarray}}
\def\eea{\end{eqnarray}}

\def\simleq{\; \raise0.3ex\hbox{$<$\kern-0.75em
      \raise-1.1ex\hbox{$\sim$}}\; }
   \def\simgeq{\; \raise0.3ex\hbox{$>$\kern-0.75em
      \raise-1.1ex\hbox{$\sim$}}\; }

\numberwithin{equation}{section}

\usepackage{jheppub}
\usepackage{hyperref}
\usepackage{cleveref}

\hypersetup{
	colorlinks=true,
	linkcolor=dark-blue,
	citecolor=dark-red,
	urlcolor=dark-green,
	linktoc=page
}

\theoremstyle{remark}

\crefname{appendix}{Appendix}{Appendices}

\title{\centering The Hitchhiker’s Guide \\to Tensionless String Limits in 4d SCFTs}

\author[1]{Jos\'e Calder\'on-Infante,}
\author[2]{Angel M. Uranga,}
\author[2,3]{and Irene Valenzuela}
\affiliation[1]{Walter Burke Institute for Theoretical Physics and Leinweber Forum for Theoretical Physics, \\  California Institute of Technology, Pasadena, CA 91125, USA}
\affiliation[2]{Instituto de F\'{i}sica Te\'{o}rica IFT-UAM/CSIC\\
C/ Nicol\'{a}s Cabrera 13-15, Campus de Cantoblanco, 28049 Madrid, Spain}
\affiliation[3]{CERN, Theoretical Physics Department, 1211 Meyrin, Switzerland}

\abstract{
We study the appearance of universality classes in overall weak-coupling infinite-distance limits in the conformal manifold of 4d ${\cal N}=4,2,1$ superconformal gauge theories at large $N$ and their relation to the type of weakly-coupled emergent string in the AdS bulk dual using brane models. We focus on the complete set of theories with simple gauge group, and provide explicit Hanany-Witten constructions for each of them, including those that have non-Einstein bulk duals. This establishes that the three universality classes found in \cite{Calderon-Infante:2024oed} are determined by the number of NS5-branes becoming coincident in a suitable double-scaling limit required to obtain the SCFT. Furthermore, all theories within a given universality class descend from the same parent theory by performing orbifold/orientifold projections (with possibly a small number of extra flavors), which are subleading effects at large $N$. This explains the origin of the universality classes from a CFT perspective. Moreover, we argue that theories in the same universality class share the same closed-string background---generated by the double-scaled NS5-brane system---in which the backreaction of D-branes is expected to generate the AdS throat holographically dual to the SCFT. We describe the corresponding worldsheet theories, and find that the three kinds of emergent tensionless strings correspond to the 10d type IIB string, a subcritical string previously studied in a related context, and a novel kind of string which we characterize in detail. Our tools are not limited to this set of SCFTs, but can be naturally applied to more general theories, and allow us to study interpolating models yielding a network of SCFTs related by partial weak-coupling limits. We explicitly discuss the generalization of our results to large classes of theories with more than one gauge factor and even to some quasi-conformal gauge theories.
}

\begin{document}
\hypersetup{pageanchor=false}
\makeatletter
\gdef\@fpheader{}
\preprint{
CALT-TH 2026-029 \\ 
\rightline{IFT-UAM/CSIC-26-82} \\ 
\rightline{CERN-TH-2026-167}
 }

\makeatother

\maketitle

\newcommand{\remove}[1]{\textcolor{red}{\sout{#1}}}
\newpage

\section{Introduction}
\label{sec:intro}

The study of infinite-distance limits in the moduli space of string compactifications has led to a number of universal results. These studies were initially motivated by the Distance Conjecture \cite{Ooguri:2006in}, which posits that an infinite tower of states should become exponentially light in Planck units in terms of the geodesic distance when approaching any infinite-distance limit of the moduli space of any theory of quantum gravity. The microscopic nature of the lightest tower and the UV-completion above its mass scale depend on the limit under consideration. This has led to a systematic top-down classification of possible infinite-distance limits in known string theory compactifications to Minkowski space, which in the context of Calabi-Yau (CY) manifolds is intimately related to the mathematical properties of the CY geometry and its moduli space \cite{Grimm:2018cpv,Corvilain:2018lgw,Joshi:2019nzi,Erkinger:2019umg,Marchesano:2019ifh,Gendler:2020dfp,Klaewer:2020lfg,Lee:2021xkn,Lee:2021vig,Alvarez-Garcia:2023qqj,Alvarez-Garcia:2023pxd,Hassfeld:2025kgo,Monnee:2025wcd,Hattab:2025gre,Monnee:2025tsl,Lee:2019wij,Castellano:2023stg,Castellano:2023jjt,Lee:2018urn,Lee:2019xtm,Baume:2019sry,Lanza:2021udy,Rudelius:2023odg,Grieco:2025slt,Kaufmann:2026fli,Kaufmann:2026mha}. In all known string-theoretic examples, the lightest tower always corresponds (in some suitable dual frame) to the excitations of a weakly-coupled critical string or to a tower of Kaluza-Klein (KK) modes signaling decompactification of extra dimensions. The expectation that this holds more generally in any theory of quantum gravity is encoded in the Emergent String Conjecture \cite{Lee:2019wij}.

More recently, these universal features at infinite-distance limits have also been investigated in a different context, namely that of AdS/CFT. Through this framework, the moduli space of a theory of quantum gravity in AdS gets mapped to the conformal manifold of a CFT. Hence, one can translate the Distance Conjecture purely into a CFT statement about the behaviour of conformal dimensions of primary operators as approaching infinite-distance limits in the conformal manifold. By considering SCFTs with conformal manifolds in four dimensions, it was found in \cite{Baume:2020dqd,Perlmutter:2020buo} that the Distance Conjecture is naturally realized at any infinite-distance limit associated to a weak-coupling limit of a (sub)sector of the field theory. Indeed, these weak-coupling limits always come with an infinite tower of higher-spin (HS) currents that become conserved exponentially fast in the Zamolodchikov geodesic distance on the conformal manifold. These results are neatly encapsulated in the CFT Distance Conjecture proposed in \cite{Perlmutter:2020buo} for the conformal manifold of any local, unitary CFT in $d>2$ dimensions. Remarkably, the first part of this conjecture, which posits that HS currents can only become conserved at infinite distance in the conformal manifold, was proven in full generality in \cite{Baume:2023msm}. Additionally, there is a similar CFT Distance Conjecture for 2d CFTs---obtained by trading the HS currents for scalar operators \cite{Kontsevich:2000yf,Acharya:2006zw,Soibelman:2011bxm,Perlmutter:2020buo}---and its first part, which asserts that the scalar gap can go to zero only at infinite distance in the conformal manifold, was proven in \cite{Ooguri:2024ofs}.

Given the results in the context of string compactifications, it is natural to ask for a similar classification of infinite-distance limits in AdS/CFT. This classification was initiated in \cite{Calderon-Infante:2024oed} by building up on results of \cite{Perlmutter:2020buo} and focusing on conformal manifolds of 4d SCFTs with any amount of supersymmetry (i.e. $\mathcal N=1$, $\mathcal N=2$ and $\mathcal N=4$). Motivated by the presence of light HS modes in the bulk, it was suggested in \cite{Calderon-Infante:2024oed} that infinite-distance limits in the conformal manifold correspond to tensionless string limits on the AdS side. 
Hence, a natural proposal is that the  classification of infinite distance limits is closely related to the classification of possible types of emergent strings becoming tensionless in the bulk picture. 

To start a classification and test this bulk interpretation, \cite{Calderon-Infante:2024oed} considered 4d superconformal gauge theories with simple gauge group and admitting a large $N$ limit. The restriction to simple gauge groups is motivated by having a complex one-dimensional conformal manifold with a cleanly isolated infinite-distance limit in which the entire theory becomes free. As pointed out in \cite{Perlmutter:2020buo}, this class of SCFTs leads only to three different types of limits characterized by the large-$N$ value of the parameter $\alpha$ that appears in the exponential law for the masses of the HS states in Planck units. This value of $\alpha$ was shown in \cite{Calderon-Infante:2024oed} to be in one-to-one correspondence with that of the ratio of the central charges $a/c$ and of the large-$N$ Hagedorn temperature of the free theory. A non-divergent large-$N$ Hagedorn temperature signals an exponential growth for the bulk density of states at high-energies, which is a common feature of the spectrum of excitations of a critical perturbative string. In fact, the Hagedorn temperature $T_H$ controls how strong this exponential growth is, which further motivates using this quantity as a way of determining the type of string theory in the bulk. 

The results of \cite{Calderon-Infante:2024oed} led to the proposal of three universality classes at large $N$ for the landscape of 4d SCFTs with simple gauge groups, that reflect the presence of three different strings in the AdS bulk becoming tensionless in the weak-coupling limit of the dual gauge theory. One of them---arising in the type 1 limit of \cite{Calderon-Infante:2024oed}---was identified with the perturbative Type IIB critical string. Fittingly, all theories in this universality class satisfy $a=c$ at large $N$ and hence can admit an Einstein gravity holographic dual. In fact, a bulk dual description in terms of the familiar 10d Type IIB string theory is known for those with $\mathcal N \geq 2$ supersymmetry \cite{Ennes:2000fu}. The other two strings are more difficult to identify, as the SCFTs with limits of type 2 and 3 in \cite{Calderon-Infante:2024oed} have $a\neq c$ at large $N$, hence their gravity duals should be non-Einstein. Interestingly, one of the SCFTs exhibiting a type 3 limit is $\mathcal N=2$ superconformal QCD (SCQCD), whose associated tensionless string limit had been studied in \cite{Gadde:2009dj,Dei:2024frl} (see \cite{Mantegazza:2026spd} for a recent detailed study from the SCFT perspective), which proposed that it corresponds to a non-critical\footnote{In this context, the name ``non-critical'' refers to a superstring theory living in less than ten spacetime dimensions. This is however a critical string in the sense of having a vanishing central charge.} string described by a specific coset worldsheet theory. This motivated the proposal in \cite{Calderon-Infante:2024oed} that all the type 3 tensionless string limits should be described by the same non-critical string, although a direct argument could not be provided. On the other hand, the nature of the string related to SCFTs with type 2 limits in \cite{Calderon-Infante:2024oed} remained elusive.

\medskip

In this work we uncover the rationale behind the existence of these universality classes of tensionless string limits, not only in the theories considered in \cite{Calderon-Infante:2024oed}, but in (at least a very large class of theories in) the general landscape of SCFTs. Focusing on the case explained above of SCFTs with simple gauge group and admitting a large $N$ limit, we realize them in terms of Type IIA Hanany-Witten brane configurations in string theory \cite{Hanany:1996ie,Witten:1997sc} (see \cite{Giveon:1998sr} for a review) and describe how they get related to precisely three underlying parent brane configurations by explicit operations. These operations typically involve orientifold and orbifold quotients, additions of flavor branes, and combinations thereof, which are well-known not to modify the behavior of correlators at large $N$. 

Moreover, our brane constructions and the above relations allow to study the infinite distance limit and to read out the physical microscopic realization of the tensionless strings. This allows us to identify the 2d worldsheet CFTs associated with the closed-string backgrounds in which the backreaction of the (large number $N$ of) D-branes would ultimately imprint an AdS$_5$ vacuum.\footnote{\label{foot:flat-space} We will use the term ``flat space'' worldsheet to refer to these closed-string backgrounds. As it will become clear later, they describe the backreaction of a certain number of NS5-branes in a double-scaling limit, and therefore contain a flat 6d spacetime piece. In the context of usual holography, this is the analogue of the worldsheet theory describing 10d flat spacetime of Type IIB string theory, namely before the backreaction of the large number of D3-branes that generate the AdS$_5\times \IS^5$ throat.} In other words, in the same way that \cite{Gadde:2009dj} provides a proposal for the dual brane configuration realizing $\mathcal N=2$ superconformal QCD, we will do the same for all the other 20 theories in this landscape of SCFTs. This includes all non-Einstein theories of type 2 and 3 in \cite{Calderon-Infante:2024oed}, as well as the type 1 $\mathcal{N}=1$ theories whose holographic dual remained unknown. 

Remarkably, we will show that what distinguishes the different classes of theories is simply the number of NS5-branes that become coincident when taking the holographic limit, so that the above ``flat space'' worldsheet theories simply correspond to the closed string background induced by the backreaction of these NS5-branes. We show that this follows because the limit is controlled by a suitable double-scaled version of the theory on coincident NS5-branes, which we describe explicitly.\footnote{Throughout this paper, we will often abuse language and refer to this as the NS5-branes `becoming coincident in the limit', leaving the required precise double-scaling limit implicit. }
The overall weak-coupling limit of the SCFT then maps to the weak-coupling limit of these closed-string backgrounds. Hence, the SCFTs associated to brane configurations arising from a given parent one (i.e. sharing the number of NS5-branes involved in the infinite distance limit) share the same kind of string becoming tensionless as one takes the overall weak-coupling limit in the dual SCFT. This ultimately stems from the fact that the operations relating the parent and descendant theories lead only to modifications subleading in the large $N$ limit, as anticipated above. 

In particular, we show that all theories with a type 3 limit in \cite{Calderon-Infante:2024oed} are controlled by a configuration of 2 NS5-branes becoming coincident in the limit, hence they all share the same tensionless string, which is given by the subcritical string proposed in \cite{Gadde:2009dj} for SCQCD. We also find that all theories with type 2 limit in \cite{Calderon-Infante:2024oed} are controlled by a configuration of 3 NS5-branes becoming coincident, and describe the worldsheet theory of the corresponding tensionless string. In view of this, we will relabel the type 2 and type 3 limits of \cite{Calderon-Infante:2024oed} as type III and type II limits, respectively, so that the notation naturally reflects the number of NS5's. Finally, our brane constructions often admit an embedding into an interpolating model as in \cite{Gadde:2009dj}, allowing for a T-dual Type IIB realization of D3-branes probing orbifold/orientifold singularities. This would allow for the identification of the string becoming tensionless as a particular subsector of an underlying critical Type IIB string theory.

\medskip

The brane toolkit we employ is by no means restricted to the case of simple gauge groups, but rather generalizes in a straightforward way to theories with multiple gauge factors.  In fact, the results of \cite{Calderon-Infante:2024oed} have been recently extended beyond simple gauge groups in \cite{Calderon-Infante:2026rkj}, to various families of 4d $\mathcal N=2$ superconformal quiver theories with unitary gauge factors, with interesting results. For instance, the Hagedorn temperature of linear quivers in the overall free limit was found to depend only on the quiver length, which is tied to the number of NS5-branes in an underlying realization in terms of NS5- and D4-brane configurations. This was argued to be related to the presence of universality classes in these limits, in the spirit of the simple gauge group case.

We hence propose that it is natural to extend our analysis of the simple gauge group case to a much larger class of theories. Following \cite{Calderon-Infante:2026rkj}, a straightforward generalization comes from considering Hanany-Witten models with $k$ NS5-branes becoming coincident in the double-scaling limit. Following the same logic as for theories with simple gauge group, we argue that they all share the same ``flat space'' worldsheet, as suggested by their coincident large-$N$ Hagedorn temperatures. Keeping the number of NS5-branes fixed, different distributions of D-branes can be regarded as providing the parent models of a large class of $\NN=2,1$ SCFTs obtained by the operations in our toolkit. Clearly, the detailed classification of all possible descendant SCFTs for a given parent theory is very involved (from the viewpoints of both the classification of SCFTs and their limits, and of their brane constructions), but the conceptual framework is very clear. For illustration, we consider the classification of large classes of $\NN=2$ SCFT descendants of some of the theories in \cite{Calderon-Infante:2026rkj}. 

In the general case with several gauge factors it is furthermore possible to consider partial limits, in which only a subsector of the theory becomes free. This will be realized by taking a subset of NS5-branes becoming coincident  (we remind the reader that these statements should be understood in terms of the corresponding double-scaling limit explained earlier). The theory emerging in the limit will contain some free subsector plus an interacting SCFT. The latter SCFT can be in a different universality class than the original one; namely, its overall weak-coupling limit can map to a different type of tensionless bulk string. This gives rise to an interesting network of SCFTs connected via partial weak-coupling limits---also referred to as interpolating models---whose values of the exponential rate $\alpha$ of the HS states get related to each other, as described in section \ref{sec:network}. To avoid confusion, we then reserve the notation of \emph{universality classes} to describe CFTs that share the same type of overall weak-coupling limit, so that they are therefore described---at weak coupling---by the same type of weakly-coupled string in the bulk. In the language used for the Distance Conjecture in flat space compactifications, these universality classes would correspond to different duality frames, i.e. different perturbative descriptions that emerge at the different asymptotic corners of the moduli space.

Our results show that the diverse landscape of 4d Lagrangian SCFTs can be enormously simplified and divided in classes according to the type of string becoming tensionless in their infinite-distance limits in the bulk. Although different examples in each class contain different open string spectrum and orientifolds, these differences become irrelevant in the weak-coupling limit of the SCFTs, as they do not contribute to leading order in string perturbation theory. This is why the perturbative spectrum of all SCFTs in each class share same properties at weak coupling, such as the value of the Hagedorn temperature or the exponential rate at which the HS towers become light. Our work confirms then the expectations of \cite{Calderon-Infante:2024oed} and paves the way for further systematic generalizations beyond simple gauge groups. The rules that we will provide in this paper (some old, some new) to construct the brane configuration can be directly used for more general SCFTs, extending this way the landscape of known AdS/CFT dual pairs beyond Einstein gravity.

\medskip

The paper is organized as follows. In section \ref{sec:overview} we overview the infinite-distance tensionless string limits of SCFTs. In section \ref{sec:review-scfts} we recall the classification of SCFTs with simple gauge group and large $N$ limits in \cite{Razamat:2020pra} and their organization in 3 universality classes in \cite{Calderon-Infante:2024oed}; in section \ref{sec:gpr} we review the subcritical string introduced \cite{Gadde:2009dj} for 4d $\NN=2$ $SU(N)$ superconformal QCD, and in section \ref{sec:wo-interpolating} we provide a variant for its derivation, useful for our more general theories. In section \ref{sec:constructions} we provide the brane constructions for the SCFTs with simple gauge group and large $N$ limit studied in \cite{Calderon-Infante:2024oed}. In section \ref{sec:classI} we study the class I theories, which correspond to the type 1 limits in \cite{Calderon-Infante:2024oed}, and are realized via Type IIA configurations with a compact circle, with no mobile NS5-branes; in section \ref{sec:classII} we study the class II theories, which correspond to the type 3 limit in \cite{Calderon-Infante:2024oed}, and are realized via Type IIA configurations with 2 NS5-branes; and in section \ref{sec:classIII} we study class III theories, which correspond to type 2 limits in \cite{Calderon-Infante:2024oed}, and are realized via Type IIA configurations with 3 NS5-branes. In section \ref{sec:strings} we describe the emergent tensionless strings in the different classes. In section \ref{sec:strings-classI} we relate class I theories to systems of D3-branes with an Einstein holographic dual, given by the 10d critical Type IIB string theory on AdS$_5\times\IS^5$ or orbifold/orientifold quotients thereof; in section \ref{sec:strings-classII}, we relate class II theories to D-branes probing the background of 2 NS5-branes in a double-scaled limit in which they become coincident (or suitable quotients thereof), which is described by the non-critical string in \cite{Gadde:2009dj} for $\NN=2$ superconformal QCD; and in section \ref{sec:strings-classIII} we relate class III theories to color D-branes probing a double-scaled limit of 3 NS5-branes becoming coincident (or quotients thereof), which is described by a new kind of string theory, whose worlsheet we describe. In section \ref{sec:generalizations} we describe several generalization of our techniques. Section \ref{sec:network} includes a general discussion of the interpolating models relating different SCFTs in our work via partial weak-coupling limits. In section \ref{sec:multiple} we explain the application of our toolkit to the study of large classes of SCFTs with more gauge factors, describe the tensionless strings in their infinite-distance limits by relating them to parent theories studied in \cite{Calderon-Infante:2026rkj}, and illustrate the discussion with several classes of $\NN=2$ SCFTs and their limits; in section \ref{sec:quasiconformal} we explain that the same limits discussed above are also realized in the weak coupling regime of quasiconformal (but not exactly conformal) theories, and illustrate it with some examples from the classification of such theories with simple gauge group and large $N$ limit in \cite{Razamat:2020pra}. Section \ref{sec:conclusions} contains some final remarks. Appendix \ref{app:brane-cooking} is a comprehensive review of Type IIA brane constructions with NS5-, D4- and D6-branes, and diverse orientifold and orbifold quotients. Appendix \ref{app:basic} describes the basic configurations with NS5- and D4-branes, giving linear or circular quiver $\NN=2$ SCFTs with unitary gauge factors; appendix \ref{app:oplanes} describes the brane configurations and gauge theories resulting from the inclusion of orientifold planes (including O4-, O6- and O8-planes); appendix \ref{app:orbifolds} introduces orbifold quotients; appendix \ref{sec:rotated-branes} describes configurations with reduced supersymmetry due to the presence of rotated branes; and appendix \ref{sec:kutasov} introduces configurations of (exactly coincident or ``glued'') NS5-branes, which lead to theories with extra adjoint chiral multiplets.

\medskip

\paragraph{Note added:} While this work was being finalized, we became aware of the work by Baume and Mantegazza \cite{Baume-Mantegazza}, which explores the presence of universality classes for general  $\mathcal N=2$ large-$N$ quiver gauge theories from the perspective of the Hagedorn temperature and group theory. We coordinated submission to appear on the same day on the arXiv. 

\section{Overview of Infinite-distance Tensionless-string Limits of SCFTs}
\label{sec:overview}

In this section we provide background material necessary for this work. We first review in section \ref{sec:review-scfts} the results in \cite{Calderon-Infante:2024oed} regarding infinite-distance limits in the classification in \cite{Razamat:2020pra} of 4d $\NN=1$ SCFTs with simple gauge group, admitting a large $N$ limit, and with a weak coupling infinite-distance limit in their conformal manifold. In section \ref{sec:gpr} we review key aspects of the study in \cite{Gadde:2009dj} about the tensionless string limit of 4d $\NN=2$ $SU(N)$ theory with $2N$ hypermultiplets in the fundamental representation. Finally, in section \ref{sec:wo-interpolating} we revisit this derivation from a perspective adapted to our more involved theories in later sections.

\subsection{The infinite-distance limits and the SCFT classification}
\label{sec:review-scfts}

In this section we review the approach and results of \cite{Calderon-Infante:2024oed}. This paper studies the weak-coupling infinite-distance limit in the conformal manifolds of 4d large $N$ superconformal gauge theories with simple gauge group (classified in \cite{Bhardwaj:2013qia,Razamat:2020pra}). It was proposed that these theories fall into three universality classes at large $N$. At leading order in large $N$, theories in an universality class share the same value of the CFT Distance Conjecture parameter $\alpha$, the same ratio of central charges $a/c$, and the same large-$N$ Hagedorn temperature in the free limit---which controls the exponential density of operators with large conformal dimensions in the large-$N$ CFT.

The CFT Distance Conjecture parameter $\alpha$ is the exponential decay rate of the mass of the bulk higher spin fields that are dual to the HS operators in the CFT. As reviewed in the Introduction, when taking a weak-coupling limit of the 4d SCFT, one always obtains a tower of HS operators whose conformal dimension $\Delta_J$ decays exponentially in terms of the Zamolodchikov distance $d(\tau,\tau')$ in the conformal manifold \cite{Baume:2020dqd,Perlmutter:2020buo} (as predicted by the CFT Distance Conjecture \cite{Perlmutter:2020buo}):
\beq
\gamma_{J}:=\Delta_J-(J+d-2)\sim f(J)\,e^{-\beta^{-1} d(\tau,\tau')}\quad \text{with } \beta=\text{dim}G_{\text{free}} \quad\text{as } d(\tau,\tau')\to \infty \, .
\label{CFTdist}
\eeq
Here $J$ is the spin, $\tau$ is the complexified gauge coupling taking to zero, and $\text{dim}G_{\text{free}}$ is the dimension of the gauge group getting free in the infinite distance/weak-coupling limit.
These operators are dual (in large $N$ theories) to higher spin fields whose mass will then decay exponentially in terms of the bulk moduli space distance (in Planck units),
\beq \label{eq:alpha-value}
m_{J}\sim e^{-\alpha \Delta\phi}\quad \text{with} \quad \alpha=\sqrt{\frac{2c}{\text{dim}G_{\text{free}}}} \quad\text{as } \Delta\phi\to \infty \, .
\eeq
This universal result for the exponential rate $\alpha$ is computed directly from the CFT using \eqref{CFTdist} and taking into account a conversion factor proportional to the CFT central charge $c$ required to write the result in bulk (AdS) Planck units.
The exponential rate $\alpha$ plays a crucial role in studies of the Distance Conjecture, as it characterizes the microscopic nature of the states becoming massless at infinite distance. From a CFT perspective, it was found to play a similar role \cite{Calderon-Infante:2024oed}, thereby the classification of weakly-coupled CFTs in universality classes. 

The one-to-one correspondence between $\alpha$ and $a/c$ follows from the simple relation
\begin{equation}
    \alpha = \frac{1}{\sqrt{2}} \frac{1}{\sqrt{2 \frac{a}{c}-1}} \, ,
\end{equation}
which holds for the overall weak-coupling limit of any superconformal gauge theory. The one-to-one correspondence with the large-$N$ Hagedorn temperature is more involved, but can be captured by the simple equation
\begin{equation}
    z_v (T_H,n) + 3 (3-4\alpha^2) z_c(T_H,n) = 1 \, ,
\end{equation}
which yields the smallest temperature $T_H$ at which the large-$N$ thermal partition function of the theory blows up. As further elaborated in \cite{Calderon-Infante:2024oed} (see \cite{Sundborg:1999ue,Aharony:2003sx} for earlier references), the functions $z_v$ and $z_c$ encode the contribution to the thermal partition function of a 4d $\mathcal N=1$ vector and chiral multiplet, respectively, and $n$ labels the multi-trace contributions. These classes were argued to correspond holographically to three types of tensionless strings: the critical type IIB string and two non-critical strings associated with non-Einstein gravity duals. In this work, we will confirm these expectations by properly characterizing these tensionless strings. The results thus provide an impressive support for the Distance Conjecture in the general context of SCFTs, even beyond those with Einstein gravity holographic duals. 

The theories in the classification in \cite{Razamat:2020pra} are shown in Table \ref{table:SCFTs}, which is adapted from that in \cite{Calderon-Infante:2024oed} with only minor notation changes, for convenience in this work. In the present table, the three different classes in which the theories are classified are called I, II, III (corresponding theories with limits of type 1, 3, 2 in \cite{Calderon-Infante:2024oed}, respectively). Also, theories within each class have been reordered, to allow for a pedagogical ordering in their construction via brane configurations in section \ref{sec:constructions} (sketched in a column in the Table). Finally, for symplectic gauge factors the rank has been renamed with respect to \cite{Calderon-Infante:2024oed}, to facilitate the comparison with the orientifolds of brane constructions in later sections. In addition, we have included a column with information about the brane construction for these theories in section \ref{sec:constructions}.

\setlength{\arrayrulewidth}{0.25mm} 
\renewcommand{\arraystretch}{1.3} 
\begin{table}[h]
\centering
    \addtolength{\leftskip} {-2cm}
    \addtolength{\rightskip}{-2cm}
    {\footnotesize
\begin{tabular}{|c| c | c c c c c c c |  c |c|c|}
    \multicolumn{12}{c}{\boldmath $\alpha = 1/\sqrt{2}$} \\
    \hline
    I&  SUSY \& Group & $n_{Ad}$ & $n_{A}$ & $n_{\bar A}$ & $n_{S}$ & $n_{\bar S}$ & $n_{F}$ & $n_{\bar F}$  & D4s on $\IS^1$ (Fig. \ref{fig:brane-cooking-classI}) & Notation \cite{Calderon-Infante:2024oed} 
    \\
    \hline
   I.1 & $\mathcal N =4 \; \; SU(N)$ & $3$ & $0$ & $0$ & $0$ & $0$ & $0$& $0$  & 
   $\IS^1$ 
   &  1.1 
   \\
    \hline
    I.2 & $\mathcal N =4 \; \; SO(N)$ & $-$ & $3$ & $-$ & $0$ & $-$ & $0$& $-$  & 
    O4$^-$ 
    &  1.8 
    \\
    \hline
    I.3 & $\mathcal N =4 \; \; USp(N)$ & $-$ & $0$ & $-$ & $3$ & $-$ & $0$& $-$  &
O4$^+$ 
& 1.5 
    \\
    \hline
    I.4 & $\mathcal N =2 \; \; USp(N)$ & $-$ & $2$ & $-$ & $1$ & $-$ & $8$& $-$  & 
    2O6$^-$ + 8D6s 
    & 1.6 
    \\
    \hline  
    I.5 & $\mathcal N =2 \; \; SU(N)$ & $1$ & $2$ & $2$ & $0$ & $0$ & $4$& $4$  & 
    (O6$^-$ + NS5)+ (O6$^-$ + NS5) + 8D6s 
    &  1.2 
    \\
    \hline
    I.6 & $\mathcal N =2 \; \; SU(N)$ & $1$ & $1$ & $1$ & $1$ & $1$ & $0$& $0$  & 
    (O6$^+$ + NS5)+ (O6$^-$ + NS5) 
    &  1.3 
    \\
    \hline
    I.7 & $\mathcal N =1 \; \; USp(N)$ & $-$ & $3$ & $-$ & $0$ & $-$ & $12$& $-$  & 
    shift orbifold of (O6$^-$+NS5) + (O6$^-$+NS5) +D6s
    & 1.7 
    \\
    \hline
    I.8 & $\mathcal N =1 \; \; SU(N)$ & $2$ & $1$ & $1$ & $0$ & $0$ & $2$& $2$  & 
    (O6$^-$+ 2 `glued' NS5s) + (O6$^-$+ NS5) + D6s & 1.4
    \\
    \hline  
\end{tabular}
\\ \medskip
\begin{tabular}{|c| c | c c c c c c c | c | c |c|}
    \multicolumn{12}{c}{\boldmath $\alpha = \sqrt{2/3}$ }  \\
    \hline
    II & SUSY \& Group & $n_{Ad}$ & $n_{A}$ & $n_{\bar A}$ & $n_{S}$ & $n_{\bar S}$ & $n_{F}$ & $n_{\bar F}$   & D4s in linear brane model (Fig. \ref{fig:brane-cooking-classII}) & Notation \cite{Calderon-Infante:2024oed}  
    \\
    \hline
    II.1 & $\mathcal N =2 \; \; SU(N)$ & $1$ & $0$ & $0$ & $0$ & $0$ & $2N$& $2N$  & 2NS5s  &  3.1 
    \\
    \hline
    II.2 & $\mathcal N =2 \; \; SO(N)$ & $-$ & $1$ & $-$ & $0$ & $-$ & $2N-4$& $-$  & 2NS5 with O4$^-$ & 3.4 
    \\
    &  &  &  &  &  &  & &  &  2NS5 with O6$^+$ & 
    \\
    \hline
    II.3 & $\mathcal N =2 \; \; USp(N)$ & $-$ & $0$ & $-$ & $1$ & $-$ & $2N+4$& $-$  & 2NS5 with O4$^+$ & 3.3 
    \\
      &  &  &  &  &  &  & &  &  2NS5 with O6$^-$ &
      \\
    \hline 
    II.4 & $\mathcal N =1 \; \; SO(N)$ & $-$ & $0$ & $-$ & $1$ & $-$ & $2N-8$& $-$  & 2 NS5s with O6'$^+$ & 3.5
    \\
    \hline 
    II.5 & $\mathcal N =1 \; \; SU(N)$ & $0$ & $0$ & $1$ & $1$ & $0$ & $2N-4$& $2N+4$  & 2NS5s, $\IC^2/\IZ_2$ + orientifold & 3.2 
    \\
    \hline
\end{tabular}
\\ \medskip
\begin{tabular}{|c| c | c c c c c c c | c| c |c|c|}
    \multicolumn{12}{c}{\boldmath $\alpha = \sqrt{7/12}$} \\
    \hline
    III & SUSY \& Group & $n_{Ad}$ & $n_{A}$ & $n_{\bar A}$ & $n_{S}$ & $n_{\bar S}$ & $n_{F}$ & $n_{\bar F}$  &  D4s in linear brane model (Fig. \ref{fig:brane-cooking-classIII}) & Notation \cite{Calderon-Infante:2024oed} 
    \\
    \hline
    III.1 & $\mathcal N =2 \; \; SU(N)$ & $1$ & $1$ & $1$ & $0$ & $0$ & $N+2$ & $N+2$  & 2 NS5s + (O6$^-$+NS5) & 2.1 
    \\
    \hline
    III.2 & $\mathcal N =2 \; \; SU(N)$ & $1$ & $0$ & $0$ & $1$ & $1$ & $N-2$ & $N-2$  & 2 NS5s + (O6$^+$+NS5) & 2.2 
    \\
    \hline
      III.3 & $\mathcal N =1 \; \; SU(N)$ & $1$ & $0$ & $1$ & $1$ & $0$ & $N-4$ & $N+4$  & 2 NS5s + (O6'$^\pm$+NS5) & 2.4 
    \\
    \hline
    III.4 & $\mathcal N =1 \; \; SU(N)$ & $2$ & $0$ & $0$ & $0$ & $0$ & $N$ & $N$  & 2 `glued' NS5 + NS5 &  2.3  
    \\
    \hline
      III.5 & $\mathcal N =1 \; \; SO(N)$ & $-$ & $1$ & $-$ & $1$ & $-$ & $N-6$ & $-$  & 2 `glued' NS5 + NS5 + O4$^-$  &  2.8 
    \\
    \hline
    III.6 & $\mathcal N =1 \; \; USp(N)$ & $-$ & $1$ & $-$ & $1$ & $-$ & $N+6$ & $-$   & 2 `glued' NS5 + NS5 + O4$^+$  & 2.5 
    \\
    \hline   
     III.7 & $\mathcal N =1 \; \; SO(N)$ & $-$ & $0$ & $-$ & $2$ & $-$ & $N-10$ & $-$  & 2 `glued' NS5 + NS5 + O8$^+$  &  2.7  
    \\
    \hline 
    III.8 & $\mathcal N =1 \; \; USp(N)$ & $-$ & $2$ & $-$ & $0$ & $-$ & $N+10$ & $-$   & 2 `glued' NS5 + NS5 + O8$^-$ & 2.6 
    \\
    \hline
\end{tabular}
}
\medskip
\caption{\small Table of SCFTs with simple gauge group and conformal manifold passing through weak coupling, classified in \cite{Razamat:2020pra}. The table is adapted from that in \cite{Calderon-Infante:2024oed}, with minor notation changes. The three different classes in which the theories are classified are called I, II, III (corresponding to limits of type 1, 3, 2 in \cite{Calderon-Infante:2024oed}, respectively). In addition, for symplectic gauge factors the rank has been renamed with respect to \cite{Calderon-Infante:2024oed}. The last column indicates the notation in \cite{Calderon-Infante:2024oed}, with the first digit indicating the type of limit, and the second indicating the position in the corresponding table in \cite{Calderon-Infante:2024oed}. Finally, we have indicated the key ingredients in the D4-brane constructions, in section \ref{sec:constructions}.}
\label{table:SCFTs}
\end{table}
\renewcommand{\arraystretch}{1} 

\subsection{The subcritical string for 4d $\NN=2$ $SU(N)$ SQCD with $N_f=2N$}
\label{sec:gpr}

We aim to understand the bulk microscopic origin of the tower of higher-spin states becoming light at the three types of infinite distances limits described in Section \ref{sec:review-scfts}. To do so, in this paper we will provide the underlying explanation of the existence of the universality classes of limits observed in \cite{Calderon-Infante:2024oed}, via a realization of the SCFTs in terms of brane constructions. We will then use these constructions to check that they indeed correspond to tensionless limits of three types of string (as proposed in \cite{Calderon-Infante:2024oed}) and, finally, we will identify which specific string realizes each limit.

This analysis is challenging for CFTs with non-Einstein gravity duals. Happily, we can adopt the strategy of \cite{Gadde:2009dj}, which was successfully used to infer the gravity dual $\NN=2$ superconformal QCD (SCQCD), i.e.  4d $\mathcal{N}=2$ $SU(N)$ SQCD with $N_f = 2N$, denoted as theory II.1 in Table~\ref{table:SCFTs}. The key idea is to construct an interpolating model that connects a holographic CFT with Einstein gravity dual to the desired CFT without it, via a weak-coupling limit. By following the fate of this limit in stringy constructions of the interpolating model, one can infer the resulting bulk string background that should be dual to the CFT of interest, which turns out to be described by a subcritical string in this case.

We review this procedure for the above example in this section, following \cite{Gadde:2009dj}. After a slight reformulation in section \ref{sec:wo-interpolating}, a similar strategy will later be applied to all other CFTs listed in Tables~\ref{table:SCFTs}. The corresponding brane constructions in section \ref{sec:constructions} explain the pattern of universality classes, and allow to identify their associated string backgrounds in section~\ref{sec:strings}.\\

Consider the SCQCD theory, i.e. 4d $\NN=2$ $SU(N)$ SYM with $2N$ hypermultiplet flavors. The relevant interpolating SCFT in \cite{Gadde:2009dj} is the $\mathcal{N}=2$ $\IZ_2$ orbifold of $\mathcal{N}=4$ SYM, with gauge group $SU(N)\times SU(N)$ and hypermultiplets in the bifundamental representation $(\mathbf{N},\overline{\mathbf{N}}) + (\overline{\mathbf{N}},\mathbf{N})$. This field theory is famously dual to Type IIB superstring theory on $\mathrm{AdS}_5\times \IS^5/\IZ_2$ \cite{Kachru:1998ys}, which arises as the near-horizon geometry of a stack of $N$ D3-branes probing the orbifold singularity ${\bf R}^2\times \IC^2/\IZ_2$.
The holographic dictionary in this setup reads
\begin{equation}
\begin{split}
&\frac{1}{g_1^2} + \frac{1}{g_2^2} = \frac{1}{2\pi g_s} , \\
&\frac{g_1^2}{g_2^2}= \frac{\beta}{1-\beta} , \qquad \beta = \int_{S^2} B_{\text{NS}} ,
\end{split}
\label{dictionarySQCD}
\end{equation}
where $g_1$ and $g_2$ are the gauge couplings of the two $SU(N)$ factors in the field theory, $g_s$ is the string coupling, and $\beta$ denotes the period of the NSNS 2-form field $B_{\text{NS}}$ over the blown-down\footnote{Notice that $\IZ_2$ acts on the $\IS^5$ by leaving an $\IS^1$ of local $\IC^2/\IZ_2$ singularities, which have a non-trivial collapsed $\IS^2$, whose $B$-field controls the relative gauge couplings \cite{Kachru:1998ys,Hanany:1998it,Gukov:1998kk}.} 2-cycle $\IS^2$ in the dual geometry \cite{Lawrence:1998ja}.

This interpolating SCFT reproduces $\NN=2$ SQCD when taking the limit in which only one of the two gauge couplings goes to zero while the other remains fixed:
\beq
SU(N)\times SU(N) \text{ with } (\fund,\antifund)+(\antifund,\fund) \xrightarrow{g_2\to 0,\ g_1\,{\rm fixed}} SU(N)  \text{ with } 2N \, \fund\, .
\label{interpolatingSQCD}
\eeq
By doing so, one of the gauge factors gets free and the remaining interacting theory is a SCFT with gauge group $SU(N)$ and $N_f=2N$ hypermultiplets in the fundamental representation, which is precisely the spectrum of $\NN=2$ SCQCD. The goal is, therefore, to follow this limit in the bulk dual and see where it bring us. Using \eqref{dictionarySQCD}, this weak-coupling limit corresponds to having simultaneously vanishing $g_s$ and vanishing $\beta$. This implies that not only the Type IIB fundamental string becomes tensionless in Planck units, but also the string arising from wrapping a D3-brane on the blow-down cycle becomes tensionless at the same rate (see e.g. \cite{Aharony:2015zea,Baume:2020dqd}). Therefore, string perturbation theory breaks down. 

\begin{figure}[htb]
\begin{center}
\includegraphics[scale=.45]{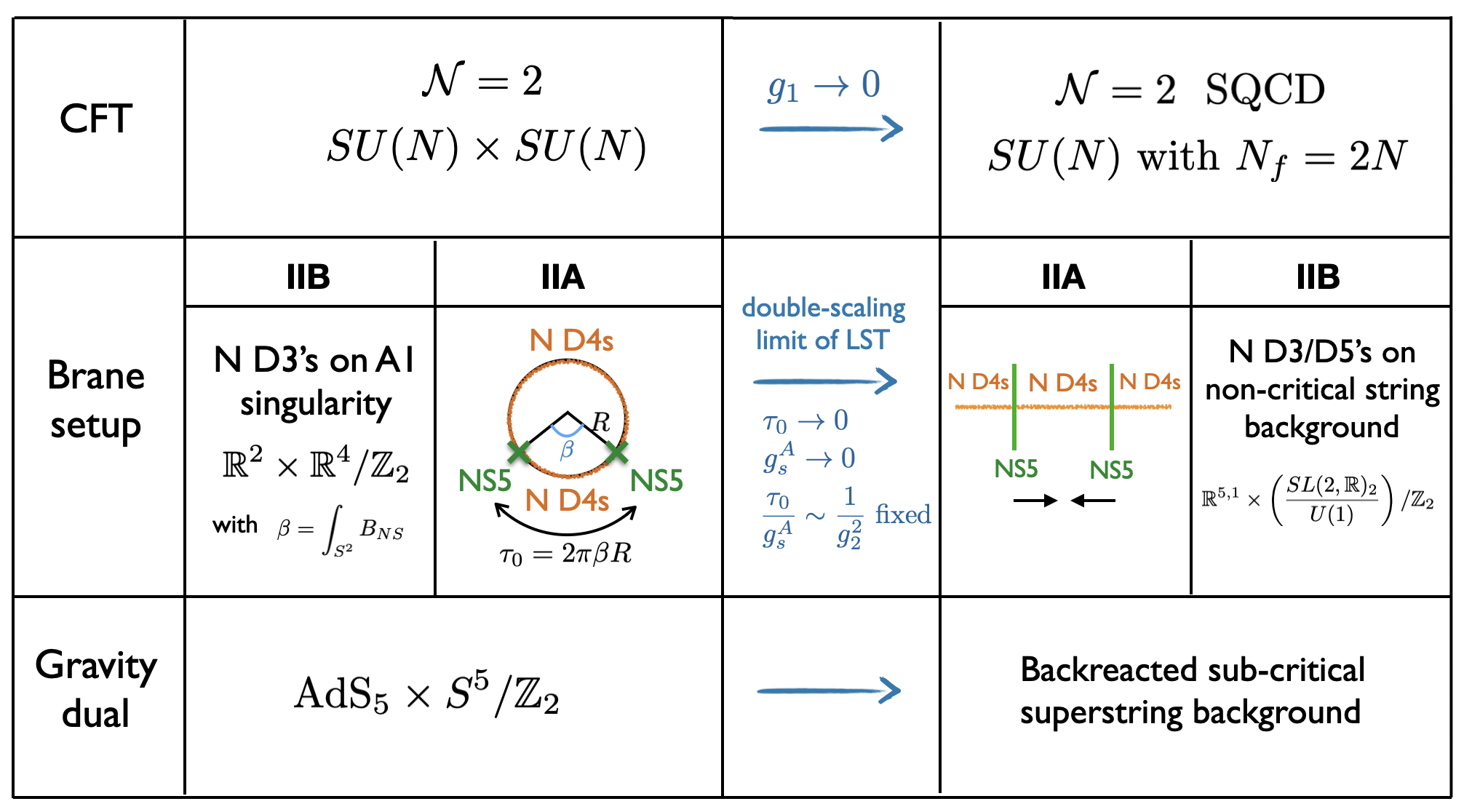}
\caption{\small Interpolating model yielding $\NN=2$ SCQCD and its double-scaling limit. The interpolating model is realized in terms of a IIA configuration of D4-branes suspended between two NS5-branes in an $\IS^1$, or a T-dual IIB stack of D3-branes at a $\IC^2/\IZ_2$ singularity, whose gravity dual is 10d IIB on AdS$_5\times \IS^5/\IZ_2$. The SCQCD theory is obtained from a IIA configuration obtained in a double-scaling limit, in which the $\IS^1$ direction decompactifies and the two NS5-branes approach each other. The IIB side is described in terms of D3/D5-branes in a non-critical string background, which should develop an AdS$_5$ region upon inclusion of the backreaction of the D-branes.}
\label{fig:interp}
\end{center}
\end{figure}

To follow the fate of this limit, it is then useful to T-dualize the D3-brane configuration to a Type IIA Hanany-Witten setup \cite{Hanany:1996ie,Witten:1997sc} with NS5- and D4-branes on the T-dual $\IS^1$ (a so-called elliptic model). To do this, we regard the $A_1$ singularity $\IC^2/\IZ_2$ as a local description of a 2-center Taub-Nut geometry with asymptotic circle radius $R_{IIB}$. T-duality along the circle direction turn the singularity into 2 NS5-branes which are coincident except in the direction of the T-dual circle of radius $R_{IIA}=\alpha'/R_{IIB}$, on which they are localized at positions separated by an angle $2\pi\beta$, where $\beta$ maps to the $B$-field in \eqref{dictionarySQCD}. The IIA string coupling is given by $g_s^A=g_s^B R/l_s$ with $l_s$ being the string scale. Moreover, the stack of $N$ D3-branes gets mapped to a stack of $N$ D4-branes extended along the circle, which actually splits into two stacks of D4-branes suspended in the two intervals between the two NS5-branes, as shown in the left panel of Figure \ref{fig:interp}. The holographic dictionary \eqref{dictionarySQCD} in terms of the IIA variables becomes
\beqa
\begin{split}
&\frac1{g_1^2}=\frac{\beta R}{2\pi g^A_s l_s} \, ,\\
&\frac{1}{g^2_2}=\frac{(1-\beta)R}{2\pi g^A_s l_s} \, .
\label{dictionarySQCD-2}
\end{split}
\eeqa
The interpolating SCFT $SU(N)\times SU(N)$ arises in the low energy limit of this Hanany-Witten setup, when the 4d field theory decouples from the higher dimensional and stringy modes in the limit $g_s^A\rightarrow 0$, $l_s\rightarrow 0$ and $R\rightarrow 0$, with $g_1$ and $g_2$ fixed.

Now we are ready to follow the weak-coupling limit in \eqref{interpolatingSQCD}. In the type IIA brane construction this limit corresponds to taking the double-scaling limit associated to bringing together the NS5s at the same rate at which the string coupling vanishes, such that
\beq
\tau_0=2\pi\beta R\rightarrow 0\ ,\quad g_s^A\to 0 \; ,\quad  \text{with } \frac{\tau_0}{l_sg_s^A}\sim \frac1{g_1^2} \text{ fixed}\, ,
\label{doublescalingLST}
\eeq
where $\tau_0$ is the distance between the two NS5s. This combined limit is also known as the double-scaling limit of little string theory (LST) \cite{Giveon:1999px,Giveon:1999tq}. Notice that it is different from the usual limit of LST in which one first takes $\tau_0\to 0$ and then $g_s^A\to 0$, such that the ratio  in \eqref{doublescalingLST} also goes to zero, and one recovers the strongly coupled theory living on the worldvolume of a stack of coincident NS5-branes. Contrarily, the double-scaling limit in \eqref{doublescalingLST} still has an effective string coupling which remains finite and is given by $g_{\text{eff}}\sim g_s^A l_s/\tau_0\sim g_1^2$. LST is formally recovered when this effective string coupling is taken to infinity. In the Hanany-Witten setup, this effective string coupling corresponds to the gauge coupling of the $SU(N)$ factor that lives in one of the stacks of D4-branes and that remains interacting in \eqref{interpolatingSQCD}.

Notice that, in this limit, we can forget about the compactness of the circle (i.e. turn the elliptic model into a linear one), and simply consider the two NS5s in $\IR^4$ at a distance $\tau_0$ from each other, which will go to zero as in \eqref{doublescalingLST}. We will have a stack of $N$ D4-branes suspended between the two NS5s and two stacks of $N$ semi-infinite D4s ending on either NS5 (see right panel of Figure \ref{fig:interp}). Before considering the fate of the D4-branes under this double-scaling limit, let us review in more detail the purely closed string background induced by the NS5s in said limit.  The region near the NS5s is described by a non-critical superstring background which admits an exact worldsheet CFT descriptions. To see this, let us first describe the string background generated by two coincident NS5-branes. This is famously given by the CHS background \cite{Callan:1991at}
\beqa
\IR^{5,1}\times \IR_\rho \times SU(2)_2\
\label{wzw-ii}
\eeqa
where $\IR^{5,1}$ is 6d spacetime, $\IR_\rho$ is the linear dilaton theory in the radial direction $\rho$ away from the NS5-branes, and $SU(2)_2$ is the  supersymmetric level 2 WZW model describing the $\IS^3$ around the NS5-branes and its NSNS 3-form flux. This background is strongly coupled in the near horizon region, as the string coupling blows up down the infinite throat when approaching the location of the NS5-branes. However, we are interested in taking the double scaling limit in \eqref{doublescalingLST}, where the string coupling goes to zero simultaneously as taking the NS5-branes coincident. In order to implement this double scaling limit, the above $\IR_\rho \times SU(2)_2$ should be replaced by the supersymmetric Liouville theory \cite{Giveon:1999px}. More details on this replacement will be given in Section \ref{ss:worldsheet}.

The inclusion of the D4-branes is hard to describe in this picture, so \cite{Gadde:2009dj} used a T-dual picture along an angular coordinate in the $\IS^3$ given by $\varphi=\arctan(x_7/x_6)$ where $x_6$ corresponds to the $\tau$-direction (along which the D4-branes are stretched). The resulting type IIB background is given by
\beq \label{eq:KS-coset}
\IR^{5,1}\times \frac{SL(2,{\bf R})_2}{U(1)}\mathlarger{\mathlarger{\mathlarger{/}}}\IZ_2
\eeq
where the $SL(2,{\bf R})_2/U(1)$ is a Kazama-Suzuki coset at level 2 describing the supersymmetric 2d Euclidean black hole (or cigar geometry) including the T-dual $\IS^1$ and the radial coordinate. The $\IZ_2$ is an orbifold quotient T-dual to the NS5-branes. In the asymptotic region far away from the NS5-branes (i.e. when the radial direction of the cigar $\rho\to \infty$), it reduces to $\IS^1$ times a linear dilaton background; while at the tip of the cigar the string coupling is bounded from above by $g_{\text{eff}}\sim g_1^2$. 

Interestingly, the above worldsheet CFT only has 8 geometric target-space dimensions. Two of the transverse dimensions to the NS5-branes end up being gapped in the near horizon limit and they do no longer contribute to the central charge. That is why this is called a non-critical (or more precisely, sub-critical) background, although the central charge of the complete worldsheet theory---including the effect of the spacelike linear dilaton background---continues to cancel.

Under the double-scaling limit and the additional T-duality, the D4-branes suspended between the two NS5s in Figure \ref{fig:interp} (right panel) become D3-branes localized at the tip of the cigar, while the semi-infinite D4 branes become flavour D5-branes extended along the cigar.\footnote{The target of this worldsheet theory is stringy, so our use of geometric notions (such as the different dimensionality of D3- and D5-branes, or of orientifold planes) in the internal CFT should be regarded as merely heuristic. A more precise description should be possible using boundary and crosscap states.}

At low energies, the worldvolume theory of the stack of D3-branes probing this non-critical string background reproduces $\NN=2$ SQCD $SU(N)$ with $N_f=2N$ hypermultiplets arising from the open strings between the D3s and the flavour D5s. To take this low energy limit and decouple the field theory from the gravitational degrees of freedom, one must still take $l_s\to 0$ in the above brane construction, which is equivalent to taking the near-horizon limit of such brane geometry. This led to \cite{Gadde:2009dj} to propose that the backreaction of the $N$ D3/D5s on the above non-critical superstring background (i.e. upon replacing the D-branes by flux) should generate an AdS$_5$ throat, hence providing the gravitational bulk dual of $\NN=2$ SCQCD. Additionally, it was argued that, apart from the AdS$_5$ factor, the bulk dual should contain a geometric $S^1$ and a stringy-size additional circle, non-trivially fibered over an interval. See \cite{Dei:2024frl} for more details on the approximate effective action of this gravitational dual.

Equipped with the above brane description, we can now provide an answer to our original question: What is the bulk interpretation of the weak-coupling limit of $\NN=2$ SCQCD? The weak-coupling limit maps to the tensionless limit of the above non-critical superstring (i.e., $g_{\text{eff}}\to 0$). Thanks to the interpolating model, we know that this closed string should be somehow `made of' of some of the degrees of freedom of the fundamental IIB string in AdS$_5\times S^5/{\bf Z}_2$ combined with the string arising from the wrapping D3-brane on the blown-down cycle. These strings give rise to two sectors of closed string states in the interpolating model, one of them becomes light in the double-scaling limit \eqref{doublescalingLST} and gives raise to higher-spin modes of AdS size (explaining why $\NN=2$ SCQCD is not an Einstein-gravity holographic theory); while the other one remains massive and only becomes light when $g_{\text{eff}}\sim g_1^2\to 0$ (see \cite{Mantegazza:2026spd} for a recent discussion on this). The latter is the tensionless string limit studied in \cite{Calderon-Infante:2024oed} and proposed to arise in all theories of class II, as we will check in section \ref{sec:strings-classII}. 

\subsection{Recovering the subcritical string without the interpolating model}
\label{sec:wo-interpolating}

For latter convenience, in this section we aim to recover the proposal reviewed in Section \ref{sec:gpr} without relying on the $SU(N)\times SU(N)$ interpolating model and its bulk dual. That is, our starting point will be the Type IIA Hanany-Witten non-elliptic model that gives rise to $\mathcal N=2$ SCQCD in the low-energy limit. Taking carefully this low-energy decoupling limit will naturally lead to the proposal of \cite{Gadde:2009dj} without reference to the interpolating elliptic model. The explanation repeats ingredients already introduced in the previous section, but with a slightly different emphasis.

\medskip

Consider two parallel NS5-branes placed in 10d flat space at a distance $\tau_0$ from each other with $N$ D4-branes stretching between them and $2N$ semi-infinite D4-branes.\footnote{Equivalently, we can replace the semi-infinite D4-branes by D6-branes placed between the two NS5s} At low energies, the worldvolume theory of the $N$ D4-branes stretched between the two NS5-branes, after dimensionally reducing on the interval, becomes $\mathcal N=2$ SCQCD. To decouple this theory from the rest of degrees of freedom in this Type IIA setting, we need to take various limits. As usual, we send $l_s \to 0$ to decouple the massive string degrees of freedom. Furthermore, we need to take $\tau_0\to 0$ to decouple the KK modes that appear in the reduction of the 5d worldvolume theory of the D4-branes to 4d. In fact, this dimensional reduction requires to treat the NS5-branes as infinitely rigid in comparison to the D4-branes, which is achieved by sending $g_s \to 0$. These three limits must be taken while keeping the $\mathcal N=2$ SCQCD coupling constant.\footnote{Even though we are ignoring it for simplicity, the same applies to the gauge theory theta angle, which is related to the difference in the VEV of the worldvolume axions carried by the NS5-branes.} All in all, we get
\begin{equation} \label{eq:decoupling-limit}
\begin{split}
	g_s \to 0 \, , \quad \tau_0\to 0\, , \quad l_s \to 0 \, , \\
	\text{with} \quad \frac{1}{g_{\rm YM}^2} \sim \frac{\tau_0}{g_s l_s} \quad \text{fixed} \, ,
\end{split}
\end{equation}
which coincides with the double-scaling limit considered in \cite{Gadde:2009dj} from the perspective of the interpolating elliptic model. From the viewpoint of the non-elliptic model, bringing the NS5-branes together is part of the usual decoupling limit that isolates the SCFT from the rest of the string theory. Going to the closed string picture---in which we replace the branes by their backreaction on the closed string sector---we can reproduce the rest of their discussion. 

The $g_s \to 0$ requirement also fits with an expected property of the closed string background in which both NS5- and D4-branes are replaced by their backreaction on the geometry in order to recover the proposal of \cite{Gadde:2009dj}. Asymptotically, i.e. far away from the brane system, this background should reduce to 10d flat space. As we move towards the interior, we expect to first feel the backreaction of the NS5-branes due their tension being parametrically larger than that of the D4-branes in the $g_s\to 0$ limit. This parametrically large region of the background then reproduces the non-critical string background discussed in Section \ref{sec:gpr}. As we keep getting close to the location of the D4-branes, we inevitable start feeling their backreaction on the geometry. As proposed in \cite{Gadde:2009dj}, this should generate an AdS throat and provide the string theory background dual to $\mathcal N=2$ SCQCD. Notice that this AdS throat is embedded into the parametrically large region described by the non-critical string background in \eqref{eq:KS-coset}. In this sense, the latter plays the same role as 10d Type IIB for the bulk dual of $\mathcal N=4$ SYM \cite{Maldacena:1997re}.

\medskip

Before concluding this section, let us comment on a crucial difference between taking the interpolating model as starting point or not. What becomes the flavor group of $\mathcal N=2$ SCQCD corresponds to a gauge group in the interpolating model. For this reason, as stressed in \cite{Gadde:2009dj}, the decoupling limit of the interpolating model describes the flavor singlet sector of this theory. On the other hand, starting directly with the Type IIA Hanany-Witten non-elliptic model and taking the decoupling limit as described above does not lead to the flavor singlet constraint. From this viewpoint, it is not clear if the restriction to the flavor singlet sector of the SCFT is necessary for the proposed holographic duality to work. Even though we are focusing on $\mathcal N=2$ SCQCD here, let us stress that the same comment applies to all SCFTs discussed here and their interpolating models (whenever one is available).

\section{Hitchhiking SCFTs with Simple Gauge Group Using Brane Constructions}
\label{sec:constructions}

As we have explained in section \ref{sec:intro}, we aim to exploit brane constructions to explain that the infinite distance tensionless string limits of large $N$ SCFTs fall in universality classes. In this section we lay the ground for this exploration by focusing on the three universality classes uncovered in \cite{Calderon-Infante:2024oed}, and provide the construction of all the gauge theories in Table \ref{table:SCFTs} using type IIA Hanany-Witten brane configurations. This will pave the way to describe in section \ref{sec:strings} the emergent tensionless strings in the three kinds of limits, and for generalizations beyond simple gauge group SCFTs in section \ref{sec:generalizations}. 

The key brane cooking techniques we employ, and further references, are collected in Appendix \ref{app:brane-cooking}. Let us emphasize that, even though they are well-established techniques in Hanany-Witten brane constructions, they had not been systematically put to work in the construction of general (i.e. including $\NN=1$)  4d  SCFTs. In this regard, the brane constructions for several theories in Table \ref{table:SCFTs} (and some brane construction rules) are, to the best of our knowledge, completely new. 

We will show that Class I theories admit an Einstein gravity dual. Accordingly, their brane configurations take the form of elliptic models built from D4-branes probing backgrounds containing various combinations of orientifold planes and additional branes. In contrast, as anticipated in \cite{Calderon-Infante:2024oed}, the limits of Class II and Class III theories correspond to non-Einstein gravitational duals. For each of them, we construct a non-elliptic Hanany–Witten configuration---analogous to the setup in the right panel of Figure~\ref{fig:interp} for 
$\NN=2$ SCQCD---that reproduces the correct spectrum of the corresponding SCFT.
A unifying feature of all theories in Class II and Class III is that the stack of D4-branes probe string backgrounds generated by either two (Class II) or three (Class III) NS5-branes (supplemented by appropriate orientifolds and flavor branes), which are mobile and become coincident in the decoupling limit in which the SCFT is obtained. This constitutes one of the main results of our work. It will play a crucial role in Section~\ref{sec:strings}, where we discuss the string backgrounds whose tensionless limits are relevant for the holographically dual to the weak-coupling limits of these SCFTs. 

\subsection{Class I: Einstein-gravity holographic SCFTs}
\label{sec:classI}

\begin{figure}[htb]
\begin{center}
\includegraphics[scale=.34]{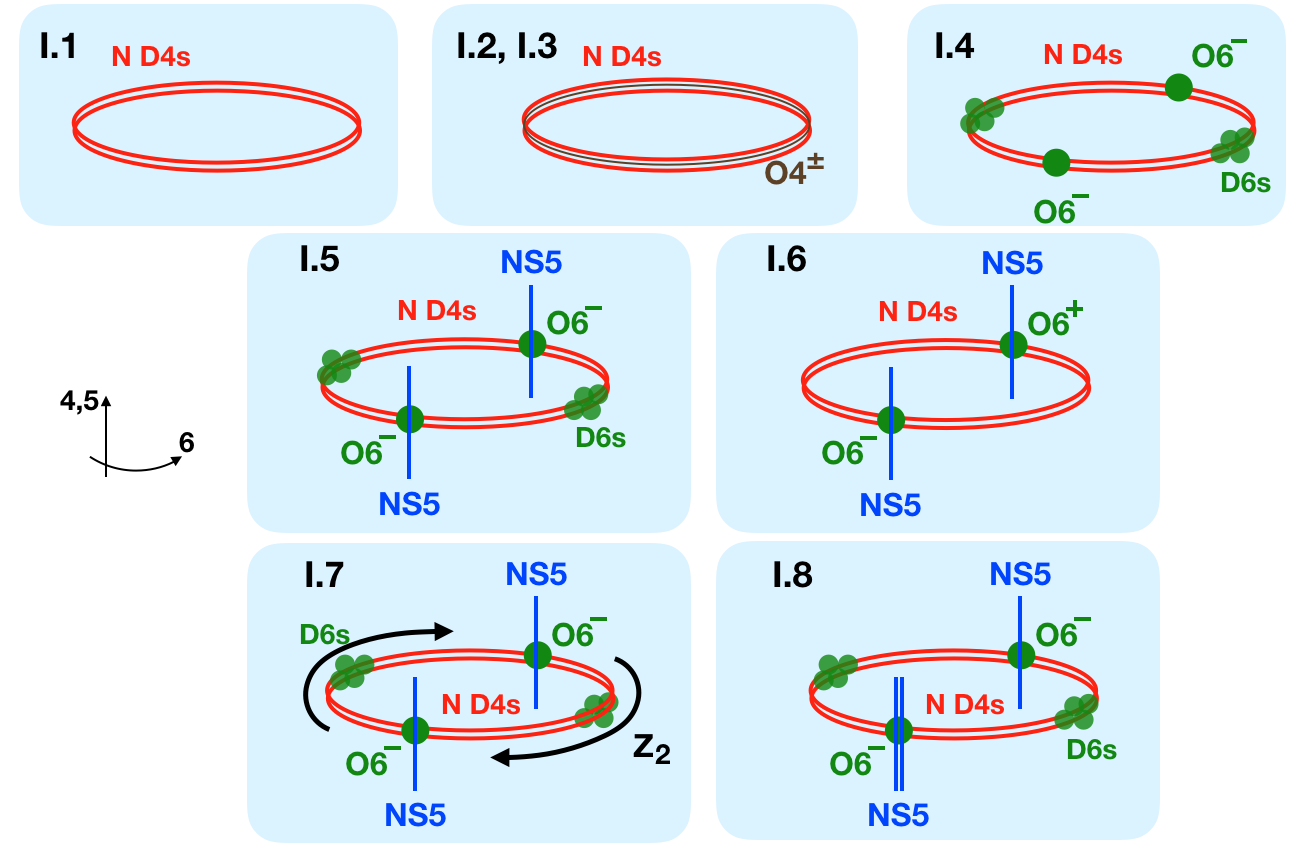}
\caption{\small The type IIA Hanany-Witten brane constructions for theories of class I.}
\label{fig:brane-cooking-classI}
\end{center}
\end{figure}

In this section we describe the brane constructions of the theories in class I in Table \ref{table:SCFTs}. Apart from the type IIA Hanany-Witten construction that we describe, these theories crucially enjoy a type IIB description with D3-branes in orientifold/orbifold quotients of flat space, which we describe as well. As will be further developed in section \ref{sec:strings}, this is tightly related to the infinite-distance limit corresponding to the tensionless limit of a 10d type IIB string in AdS$_5\times\IS^5$ or quotients thereof. We provide the brane construction of the different theories in turn (see Figure \ref{fig:brane-cooking-classI} for a depiction of the corresponding ingredients). 

\subsubsection{The Class I 4d $\NN=4$ theories: I.1, I.2, I.3}
\label{sec:classI-123}

Let us discuss the cases in class I which preserve 4d $\NN=4$ supersymmetry. We discuss them in turns:

\subsubsection*{Theory I.1} 

The gauge theory corresponds to 4d $\NN=4$ $SU(N)$ super Yang-Mills (SYM), and it has a celebrated realization as the low-energy limit of the worldvolume theory on a stack of $N$ D3-branes in type IIB theory. The conformal manifold corresponds to the complex gauge coupling, and the infinite-distance limit is the weak-coupling limit of the gauge theory (see e.g. \cite{Baume:2020dqd,Perlmutter:2020buo}). The structure of the emergent string, to be discussed later on, is that of its gravity dual via holography \cite{Maldacena:1997re}, so the tensionless string limit is weakly-coupled type IIB string theory in AdS$_5\times\IS^5$. Hence, the higher-spin operators that become exponentially conserved at infinite distance in the conformal manifold simply map to higher-spin fields associated with the critical fundamental Type II string.\footnote{A quantitative string-theoretic description of the higher-spin modes becoming massless in the infinite-distance limit is missing. The weakly-coupling limit in the SCFT corresponds to the strong-curvature regime in the worldsheet, thus invalidating the usual $\alpha^\prime$ expansion. See \cite{Gaberdiel:2021qbb,Gaberdiel:2021jrv,Gaberdiel:2022iot} for recent progress in understanding this tensionless string limit using worldsheet techniques.}

For later comparison, it will be useful to provide a type IIA brane realization of this theory, shown in Figure \ref{fig:brane-cooking-classI}. Clearly, it simply corresponds to (the infrared limit of) a stack of $N$ D4-branes, spanning the directions 0123 times a direction 6 compactified on an $\IS^1$. This may be regarded as the simplest example of an elliptic model of the kind described in section \ref{sec:elliptic} in the absence of NS5-branes (i.e. for $k=0$).\footnote{Accidentally, one also gets the same theory for one NS5-brane, $k=1$.\label{one-ns5}} The type IIA realization is directly related to the previous type IIB D3-brane realization via T-duality on the $\IS^1$.

\subsubsection*{Theories I.2 and I.3} 

The gauge theories correspond to 4d $\NN=4$ $SO(N)$ or $USp(N)$ SYM (with $N$ even in the latter case). As studied in \cite{Witten:1998xy}, they are realized as the infrared limit of the worldvolume theory on D3-brane located on top of an O3-plane (an O3$^-$-plane for the $SO(N)$ theory, or an O3$^+$-plane for the $USp(N)$ theory\footnote{There are two possible O planes, that correspond to the $SO(N)$ theory with even or odd $N$. We treat them both simultaneously, as the O3-plane leading to odd $N$ $SO(N)$ is equivalent to an O3$^-$-plane with one stuck D3-brane. There are also two variants of the O3$^+$-plane, which are distinguished by a discrete RR background, i.e. a discrete theta angle in the gauge theory. We refer to \cite{Witten:1998xy} for details.}). The marginal coupling is again the complexified gauge coupling, and the infinite-distance limit corresponds to weak coupling. This tensionless string limit is manifest in the gravitational dual description, and corresponds to weakly-coupled type IIB string theory in the appropriate orientifold of AdS$_5\times\IS^5$, namely AdS$_5\times \IR\IP^5$ \cite{Witten:1998xy}. 

The type IIA brane realization is in terms of a stack of $N$ D4-branes on top of an O4$^\pm$-plane spanning the directions 0123 and wrapped on a direction 6 compactified on an $\IS^1$, see Figure \ref{fig:brane-cooking-classI}. Hence, they are a particular case of the configurations in section \ref{sec:o4-planes} with no NS5-branes ($k=0$). In the T-duality to the type IIB system, we obtain two O3$^\pm$-planes at antipodal points of the T-dual $\IS^1$, but in the infrared limit only the O3$^\pm$-plane on top of which the stack of $N$ D3-branes sits is relevant, thus recovering the previous type IIB realization.

\subsubsection{The Class I 4d $\NN=2$ theories: I.4, I.5, I.6}
\label{sec:classI-456}

We now describe the cases in class I which preserve 4d $\NN=2$. They are realized in terms of brane configurations with O6-planes, see appendix \ref{sec:o6-planes}. We discuss them in turns.

\subsubsection*{Theory I.4} 

It corresponds to a 4d $\NN=2$ $USp(N)$ gauge theory, with hypermultiplets in the $\Yasymm+4\fund$. Using the results in appendix \ref{sec:o6-planes}, this theory is obtained (see in Figure \ref{fig:brane-cooking-classI}) by taking $N$ D4-branes along 0123 and a compact direction 6, adding two O6$^-$-planes along 0123789 and at opposite points in this $\IS^1$, and 8 D6-branes parallel to the O6-plane and at different positions in 6 (4 independent ones and their 4 orientifold images), to cancel the RR charge and attain conformality. 

The same theory is obtained if we include one NS5-brane on top of one of the O6$^-$-planes, which is convenient to make the discussion of the gauge theory content closer to appendix \ref{sec:o6-planes}), as follows. There is one interval, leading to one gauge factor, with the D4-branes mapped to themselves by the O6$^-$-planes, hence producing a $USp(N)$ gauge symmetry. The open strings between the D4-branes across the NS5-brane produce a bifundamental hypermultiplet (in fact, collapsed to an adjoint) in the parent theory, but it is mapped to itself by the O6$^-$-planes, so it projects down to a hypermultiplet in the $\Yasymm$. In addition, the 4 D6-branes (+4 images) lead to 4 hypers in the $\fund$ in the orientifold quotient, completing the spectrum of the gauge theory.

This elliptic model was constructed in \cite{Uranga:1998uj}, and can be T-dualized (see \cite{Park:1998zh} for the general discussion) to a configuration of $N$ type IIB D3-branes in the presence of an O7$^-$-plane with 8 D7-branes on top, a celebrated construction of this CFT, originally considered in \cite{Banks:1996nj,Aharony:1996en,Douglas:1996js,Aharony:1998xz}.

\subsubsection*{Theory I.5} 

It corresponds to a 4d $\NN=2$ $SU(N)$ gauge theory with hypermultiplets in the $2\Yasymm+4\fund$. This model is constructed as an illustrative example in appendix \ref{sec:o6-planes}, namely (\ref{two-o6-ns5-ontop}), but we recap the key concepts of the construction for completeness. This theory is obtained (see Figure \ref{fig:brane-cooking-classI}) by taking $N$ D4-branes along 0123 and a compact direction 6, adding two O6$^-$-planes along 0123789 and at opposite points in this $\IS^1$, with one NS5-brane stuck on top of each O6$^-$-plane. We also have 8 D6-branes at different positions in 6 (4 independent ones and their 4 orientifold images), to cancel RR charge and attain conformality. The two intervals between the NS5-branes are images of each other, so they lead to a $SU(N)$ gauge symmetry in the quotient. Each of the two bifundamental hypermultiplets in the parent theory is mapped to itself under the O6$^-$-plane projection, so it becomes a hyper in the $\Yasymm$ in the quotient. Finally, the 4 D6-branes (+4 images) lead to 4 hypers in the $\fund$ in the orientifold quotient, completing the spectrum of the gauge theory.

This elliptic model was constructed in \cite{Uranga:1998uj}, and its T-dual  \cite{Park:1998zh} is an orientifold of $N$ D3-branes at a $\IC^2/\IZ_2$ singularity by an action $\Omega R (-1)^{F_L}$, with $R$ flipping the two real coordinates transverse to the D3-branes and the $\IC^2/\IZ_2$. To be totally precise, the orientifold action on twisted sectors is chosen without vector structure, in the language of \cite{Polchinski:1996ry}. The model actually corresponds to (a 4d version of) a local $\IC^2/\IZ_2$ singularity in the celebrated 6d orientifold of $\IT^4/\IZ_2$ in \cite{Pradisi:1988xd,Gimon:1996rq}.

\subsubsection*{Theory I.6}  

It corresponds to a 4d $\NN=2$ $SU(N)$ gauge theory with hypermultiplets in the $\Yasymm+\Ysymm$ (and none in the $\fund$). Using the results in appendix \ref{sec:o6-planes}, this theory is obtained by taking $N$ D4-branes along 0123 and a compact direction 6, adding one O6$^-$- and one O6$^+$-plane at opposite points in this $\IS^1$, with one NS5-brane stuck on top of each O6-plane. The O6-plane charges cancel, so there must be no D6-branes in this case to achieve conformality. As in the theory I.5, the two intervals between the NS5-branes are swapped by the orientifold action, and produce a single $SU(N)$ gauge symmetry in the quotient. Each of the two bifundamental hypermultiplets in the parent theory is mapped to itself under the O6-plane projections, but their different signs imply that they project down to $\Ysymm$ at the O6$^+$-plane location and to $\Yasymm$ at the O6$^-$-plane location. Hence we recover the spectrum of the theory.

This elliptic model was constructed in \cite{Uranga:1998uj}, and its T-dual \cite{Park:1998zh} is obtained from $N$ D3-branes at a $\IC^2/\IZ_2$ singularity modded out by an orientifold action (without vector structure, in the language of \cite{Polchinski:1996ry}) $\Omega R\alpha (-1)^{F_L}$, with $R$ flipping the two real coordinates transverse to the D3-branes and the $\IC^2/\IZ_2$. The extra action $\alpha$, is related to the generator $\theta$ of the $\IZ_2$ orbifold by $\alpha^2=\theta$, and its appearance reflects the opposite O6-plane signs in the type IIA picture.  

\subsubsection{The Class I 4d $\NN=1$ theories: I.7, I.8}
\label{sec:classI-78}

These two theories are the only representatives of Class I with 4d $\NN=1$ supersymmetry. They are however rather different from each other, and require different ingredients for their brane constructions. 

\subsubsection*{Theory I.7}

Let us consider theory I.7, which is a 4d $\NN=1$ $USp(N)$ gauge theory with 3 chiral multiplets in the $\Yasymm$ and 12 chiral multiplets in the $\fund$. 
There is a simple type IIB construction of this theory, as an orientifold of a stack of D3-branes at a $\IC^3/(\IZ_2\times\IZ_2)$ singularity with discrete torsion. In fact, this has implicitly appeared in the literature as a local sector in a compact orientifold of $\IT^6/(\IZ_2\times\IZ_2)$ in \cite{Berkooz:1996dw}; the compact model was described as a type I compactification, namely in terms of D9- and D5-branes, but it is straightforward to T-dualize into a model of D7-branes and D3-branes. The model of our interest features a local $\IC^3/(\IZ_2\times\IZ_2)$ singularity at which a set of D3-branes sit. 
Since this construction is standard in the orientifold literature, we quickly review its main ingredients in a language directly adapted to the case of D3-branes, and refer the reader to \cite{Berkooz:1996dw} (see also \cite{Aldazabal:1998mr,Klein:2000tf,Klein:2000qw}) for details.

\medskip

The gauge theory on a stack of $2N$ D3-branes at $\IC^3/(\IZ_2\times\IZ_2)$ with discrete torsion is reviewed in appendix \ref{sec:orbifold-dt}. The generator $\theta$ of one of the $\IZ_2$'s preserves 4d $\NN=2$ supersymmetry and breaks the $SU(2N)$ vector multiplet into $SU(N)\times SU(N)$, and also leads to hypermultiplets in the $(\fund,\antifund)+(\antifund,\fund)$.\footnote{Incidentally, we note that this is the interpolating model of theory II.1 in Table \ref{table:SCFTs}, reviewed in section \ref{sec:gpr}.} The generator $\omega$ of the second $\IZ_2$ exchanges the two $SU(N)$ factors and breaks the supersymmetry down to 4d $\NN=1$, by introducing an extra relative sign (related to the discrete torsion) between the projections on the 4d $\NN=1$ vector and adjoint chiral multiplets in the 4d $\NN=2$ vector multiplet. This results in a 4d $\NN=1$ $SU(N)$ vector multiplet and one adjoint chiral multiplet. In addition, the 4d $\NN=2$ bifundamental hypermultiplets are also exchanged and descend to two 4d $\NN=1$ adjoint chiral multiplets of the final $SU(N)$. Overall we have a 4d $\NN=1$ $SU(N)$ gauge theory with 3 adjoint chiral multiplets (the theory is however only $\NN=1$ because of an extra sign between the two superpotential terms, which is related to the discrete torsion).

We now introduce an O3$^-$-plane orientifold projection, by quotienting by $\Omega R(-1)^{F_L}$, with $R:(z_1,z_2,z_3)\to (-z_1,-z_2,-z_3)$, i.e. flipping all coordinates transverse to the D3-branes, and $F_L$ denoting left-moving worldsheet fermion number. Because of the orbifold quotient, the orientifold group also contains the elements $\Omega R(-1)^{F_L}\theta$, $\Omega R(-1)^{F_L}\omega$ and $\Omega R(-1)^{F_L}\theta\omega$. These introduce O7-planes of thee different kinds, dubbed O7$_i$-branes, $i=1,2,3$, which are located on the 4-planes $z_i=0$. Cancellation of RR charge is not necessary for consistency, but it is rather related to conformality of the resulting gauge theory, so we introduce three sets of 8 D7$_i$-branes at the three 4-planes $z_i=0$. We direct the reader to the literature for the discussion of the precise orientifold action on the Chan-Paton indices, and simply provide the relevant results.

It is illustrative to regard the derivation of the spectrum of the final gauge theory by performing the orbifold and orientifold quotients in different orders. For instance, the 4d $\NN=1$ $SU(N)$ vector multiplet after the $\IZ_2\times\IZ_2$ orbifold is projected  by the O3$^-$-plane down to a $USp(N)$ gauge factor. This may seem puzzling, because O3$^-$-planes usually implement an $SU\to SO$ orientifold projection on D3-branes. But the extra flip is simply due to the discrete torsion, as follows. By performing the orientifold projection before the orbifold, the O3$^-$-plane acting on the parent $SU(2N)$ theory indeed projects it down to an $SO(2N)$ theory. Acting now with $\theta$, we obtain an orientifold of $\IC^2/\IZ_2$ $(\times \IC)$, without vector structure and with one set of 8 D7$_3$-branes, producing a 4d $\NN=2$ theory (\ref{two-o6-ns5-ontop})---which corresponds to theory I.5 in Table \ref{table:SCFTs}, c.f. section \ref{sec:classI-456}---which we copy here for convenience
\beqa
\NN=2 \; {\rm Vector}\quad &SU(N)&\nonumber \\
\NN=2 \; {\rm Hyper}\quad& 2\Yasymm + 4\fund &
\label{z2-withoutvs-again}
\eeqa
We see that the gauge group is broken as $SO(N)\to SU(N)$. A similar projection occurs if one considers the $\omega$ projection instead of the $\theta$ one. Hence, the surviving gauge group is the overlap of the two different $SU(N)\subset SO(2N)$ symmetries defined by the two orbifold generators, which corresponds  to just $USp(N)$. 

The 4d $\NN=1$ chiral matter content can again be understood as the orientifold projection of the 3 $SU(N)$ adjoint chiral multiplets after the $\IZ_2\times\IZ_2$ orbifold quotient, leading to 3 chiral multiplets in the $\Yasymm$. Performing the orientifold and the $\theta$ projections first, the $\Yasymm$'s already show up partially in the 4d $\NN=2$ spectrum (\ref{z2-withoutvs-again}). More specifically, the projection by $\omega$ projects down the 2 hypermultiplets in the $\Yasymm$ to a single one, namely 2 $\NN=1$ chiral multiplets in the $\Yasymm$, and projects the additional $SU(N)$ adjoint chiral multiplet in the 4d $\NN=2$ vector multiplet into the third $\Yasymm$ (i.e. with an extra flip with respect to the above explained action on the gauge group). In addition, the $\omega$ projection on the 4 hypermultiplets in the $\fund$ from the D7$_3$-branes are projected down to 4 chiral multiplets in the fundamental of $USp(N)$. Since the final model is totally symmetric in the three complex directions, a similar comment holds for the flavors from the other D7$_i$-branes. Overall, the 4d $\NN=1$ gauge group and chiral matter content is
\beqa
& USp(N)&\nonumber\\
& 3\Yasymm + 12\fund &
\label{bl-spectrum}
\eeqa
which reproduces the spectrum of the theory I.7.

\medskip
  
Since the model is based on D3-branes at a $\IC^3/(\IZ_2\times\IZ_2)$ singularity, we may expect that there is a T-dual type IIA picture in terms of NS5- and D5-branes. However, this T-duality and such brane constructions have not been considered in the literature in the case of discrete torsion. Although a discussion for general orbifolds is beyond the scope of this work, we now derive the key ingredients of such construction for the $\IC^3/(\IZ_2\times\IZ_2)$ case.

We start with the configuration describing the T-dual of the orientifold of $\IC^2/\IZ_2$ describing the quotient by $\theta$ and the orientifold action, and producing the gauge theory (\ref{z2-withoutvs-again}). It corresponds to an elliptic model with 2 NS5-branes, each stuck on top of an O6$^-$-plane, two stacks of $N$ D4-branes suspended between the NS5-branes in the two intervals, and 4 D6-branes (and their images, see panel I.7. in Figure \ref{fig:brane-cooking-classI}). The theory is 4d $\NN=2$ and has gauge group $SU(N)$. The bifundamental hypermultiplets across each of the NS5-branes project down to hypers in the $\Yasymm$, and the D6-branes produce hypermultiplets in the fundamental, hence reproducing the spectrum (\ref{z2-withoutvs-again})---namely theory I.5, c.f. section \ref{sec:classI-456}.

We would now like to perform the quotient by $\omega$, which corresponds to a $\IZ_2$ orbifold acting as $\omega:(z,w)\to(-z,-w)$ on the coordinates $z=x^4+ix^5$, $w=x^8+ix^9$. Now, contrary to the T-dual of the $\IC^3/(\IZ_2\times\IZ_2)$ without discrete torsion (c.f. appendix \ref{sec:orbifold-nodt}), the $\IZ_2$ generator must act simultaneously as a half-shift on the $\IS^1$, so as to exchange the gauge factors and matter fields in the appropriate way. In order to understand the resulting spectrum, consider the parent theory with the 2 NS5-branes and $N$ D4-branes before the introduction of the O6-planes. The gauge theory is the interpolating model of theory II.1, c.f. (\ref{interpolatingSQCD-again}). Performing now the quotient by $\omega$ times the half-shift on the circle, the two gauge factors become identified into a single $SU(N)$. Similarly, the two $\NN=1$ adjoint chiral multiplets in the $\NN=2$ vector multiplets become identified into a single $SU(N)$ $\NN=1$ adjoint chiral multiplet, and the four $\NN=1$ bifundamental chiral multiplets become identified into two $\NN=1$ chiral multiplets, now also in the adjoint. The whole operation is similar to perfoming the quotient of the theory (\ref{interpolatingSQCD-again}) by the quantum $\IZ_2$ symmetry of the (T-dual) $\theta$ orbifold quotient, but with extra signs preventing coming back to the 4d $\NN=4$ parent theory. Finally, we must reintroduce the O6-planes (and the corresponding flavor D6-branes), which reduce the gauge factor to $USp(N)$ and project down the adjoints to $\Yasymm$'s as explained above for the T-dual type IIB.

\subsubsection*{Theory I.8} 

This is a 4d $\NN=1$ gauge theory with 2 adjoint chiral multiplets and additional chiral multiplets in the $\Yasymm+\bYasymm+2\fund+2\antifund$. The  brane construction for this models turns out to be closely related to those of Class III. Therefore we postpone their discussion to section \ref{sec:classI8-atlast}.

\medskip

As already mentioned and as we will further develop in section \ref{sec:strings}, the infinite-distance limit of all Class I theories corresponds to the tensionless limit of the critical 10d type IIB string in AdS$_5\times\IS^5$ or quotients thereof. 

\subsection{Class II: Models with 2 mobile NS5-branes}
\label{sec:classII}

In this section we consider models in Class II. As we will show, they correspond to constructions involving 2 mobile NS5-branes, i.e., their relative distance can be tuned. Hence, as further developed in section \ref{sec:strings}, the bulk duals of these SCFTs are related to the throat created by these NS5-branes when they are taken close to each other (compared with the energies of any probe). More precisely, the infinite-distance limit in the conformal manifold of all models in this class lead to the same tensionless string, which corresponds to a specific non-critical type IIB string of the kind considered in \cite{Gadde:2009dj}. We study the brane constructions for these models in turns.

\subsubsection{The Class II 4d $\NN=2$ theories: II.1, II.2, II.3}
\label{sec:classII-123}

\begin{figure}[htb]
\begin{center}
\includegraphics[scale=.4]{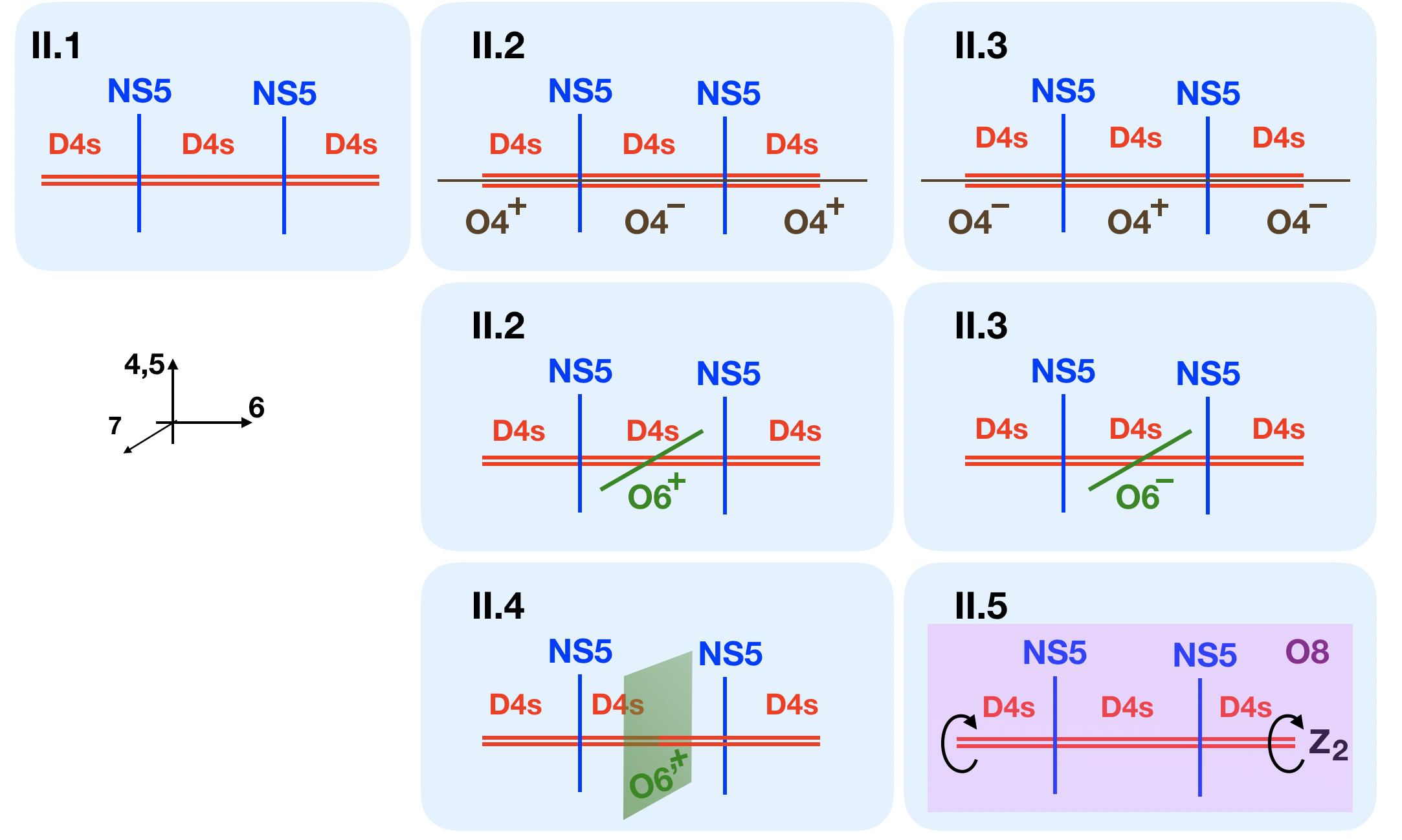}
\caption{\small The type IIA Hanany-Witten brane constructions for theories of class II}
\label{fig:brane-cooking-classII}
\end{center}
\end{figure}

Let us start describing the brane constructions for the Class II theories with 4d $\NN=2$, namely theories II.1, II.2 and II.3, which are sketched in the top three pictures in Figure \ref{fig:brane-cooking-classII}. 

\subsubsection*{Theory II.1} 

It corresponds to a 4d $\NN=2$ $SU(N)$ gauge theory with $2N$ hypermultiplets in the $\fund$, namely the SCQCD theory reviewed in section \ref{sec:gpr}. As discussed there, and specifically in the linear brane model in section \ref{sec:wo-interpolating}, using the rules in appendix \ref{sec:suspended-d4s} (see \cite{Witten:1997sc} for the original reference), this theory is realized (see Figure \ref{fig:brane-cooking-classII}) by considering 2 NS5-branes along 012345, and $N$ D4-branes along 0123 and suspended between the NS5-branes in a non-compact direction 6. We also have $N$ semi-infinite D4-branes stretching from each NS5-brane to infinity in the direction 6, and there are no D6-branes. The configuration preserves 4d $\NN=2$, and the D4-branes in the interval lead to an $SU(N)$ gauge group, while the open strings between these D4-branes and the semi-infinite one on the extremes lead to bifundamental hypermuliplets charged under global $SU(N)$ symmetries. Overall, the theory has the structure (\ref{linear-quiver}) with $k=2$ and $N_i=N$, namely
\beqa
&[SU(N)]\times SU(N)\times [SU(N)] &\nonumber \\
&(\fund,\antifund,1)+(1,\fund,\antifund)&
\eeqa
where square brackets indicate the global symmetry group carried by the semi-infinite D4-branes.\footnote{Despite not being manifest in this construction, this theory has a $U(2N)$ flavor symmetry, not only $SU(N)\times SU(N)$. This can be made manifest in the brane picture by realizing the fundamental hypermultiplets with a stack of $2N$ D6-branes between the two NS5-branes. The configuration with two stacks of $N$ semi-infinite D4-branes can be recovered by separating the D6-branes in two stacks with $N$, and moving each of them to infinity along opposite directions in 6 across the NS5-branes, creating the two stacks of $N$ D4-branes by the Hanany-Witten brane creation effect.}

This theory was considered in \cite{Gadde:2009dj}, as we have reviewed in section \ref{sec:gpr}. For completeness, we also recall the interpolating model also discussed there. It is obtained by simply gluing the above theory into an elliptic model with 2 NS5-branes, or equivalently by gauging the diagonal combination of the two $SU(N)$ global symmetries above. The resulting 4d $\NN=2$ gauge group and hypermultiplet content is
\beqa
&SU(N)\times SU(N) &\nonumber \\
&(\fund,\antifund)+(\antifund,\fund)&
\label{interpolatingSQCD-again}
\eeqa
The theory admits a T-dual description in terms of $N$ D3-branes at a $\IC^2/\IZ_2$ singularity \cite{Douglas:1996sw} (see also \cite{Kachru:1998ys,Lawrence:1998ja,Hanany:1998it}). 

\subsubsection*{Theories II.2 and II.3} 

The two theories are closely related, so they can be  described jointly. They correspond to 4d $\NN=2$ gauge theories with group $G$ and $N\pm 2$ hypermultiplets in the fundamental, with $G=SO(N)$ and $N-2$ flavours for the II.2 case, and $G=USp(N)$ and $N+2$ flavors for the II.3 case. Using the rules in appendix \ref{app:brane-cooking}, there are two possible type IIA brane realizations of each of these theories. 

A first possibility is to consider 2 NS5-branes with $N$ D4-branes suspended between them, and $N\pm 2$ semi-infinite D4-branes at the ends (according to the required number of flavors in each case), and to add an O4-plane, as in appendix \ref{sec:o4-planes} (see Figure \ref{fig:brane-cooking-classII}). The O4-plane flips signs as it crosses each NS5-brane, so for the II.2 theory we choose it to be an O4$^-$-plane in the finite segment between the NS5-branes (and hence, and O4$^+$-plane in the semi-infinite D4-brane regions), and the reverse for the II.3 theory. The $N$ D4-branes between the NS5-branes lead to an $SO(N)$ gauge group in the II.2 case and an $USp(N)$ group in the II.3 case, and the two bifundamental hypermultiplets in the parent theory project down to half-hypermultiplets, giving a total of $N\pm 2$ full hypers in the fundamental, and completing the spectrum of the models. The theories are the two possible $k=2$ models in (\ref{O4-linear}). Notice that the choice of $N\pm 2$ semi-infinite D4-branes ensures that the total D4-brane charge (including the O4$^\pm$-plane contribution) remains constant throughout the whole direction 6, ensuring conformality of the resulting gauge theory.

A second possibility is to consider 2 NS5-branes with $N$ D4-branes suspended between them, and $N\pm 2$ semi-infinite D4-branes at the extremes, and to add an O6-plane in the middle of the finite interval (so that the two NS5-branes are images of each other). We choose this orientifold plane to be an O6$^+$ for the II.2 theory and an O6$^-$ for the II.3 theory (see Figure \ref{fig:brane-cooking-classII}). The $N$ D4-branes suspended between the NS5-branes are mapped to themselves by the O6-plane action, and lead to an $SO(N)$ gauge group in the O6$^+$ case (i.e. the II.2 theory) or an $USp(N)$ group in the O6$^-$ case (i.e. the II.3 theory). The two bifundamental hypermultiplets in the parent theory are swapped by the orientifold action, so they lead to just one copy in the quotient, completing the spectrum of $N\pm2$ fundamental flavors. The theories are the particular $k=2$ case of (\ref{one-O6-ns5-away}).

Let us discuss the interpolating models. In either construction, we can complete the models into elliptic ones, and get a nice surprise: in the elliptic model, one theory of type II.2 combines with a theory of type II.3 to provide a single interpolating model for the two SCFTs, which are then recovered in two different weak-coupling limits.
For instance, in the O4-plane realization, the elliplic model is given by 2 NS5-branes on an $\IS^1$, with an O4-plane of opposite sign in each of the two intervals they define, and with $N'\pm 1$ D4-branes in the interval on top of the O4$^{\mp}$-plane. The resulting 4d $\NN=2$ gauge group and hypermultiplet content is the particular case $k=2$ of (\ref{O4-elliptic}), namely
\beqa
&SO(N'+1)\times USp(N'-1)& \nonumber \\
&(\fund,\fund)&
\label{interpolating-ii-2-3}
\eeqa
where the full bifundamental hypermultiplet arises from two half-hypermultiplets on top of each NS5-brane. Choosing $N'+1=N$ and turning the gauge $USp(N-2)$ symmetry into a global one, we recover the theory II.2, whereas choosing instead $N'-1=N$ and making the $SO(N+2)$ a global symmetry, we recover the theory II.3. 

A similar interpolating model for the two theories can also be constructed in the O6-plane realization, by taking 2 NS5-branes on an $\IS^1$, with an O6-plane of opposite sign in the middle of the two intervals they define, and with $N'\pm 1$ D4-branes in the interval on top of the O6$^{\pm}$-plane. The resulting theory is precisely (\ref{interpolating-ii-2-3}), where now the bi-fundamental hypermultiplet arises from the two bifundamentals in the parent theory, which are swapped by the O6-plane actions. Again we can recover the theories II.2 or II.3 by suitable ungauging limits.

It is similarly easy to build an interpolating model using two theories II.3 in a fully symmetric manner. We take two O6$^-$-planes on the $\IS^1$ parametrized by 6, and locate 2 NS5-branes forming a symmetric pair away from the O6-planes, and we add $N$ D4-branes in each of the two intervals between the NS5-branes. In order to cancel the RR charge, we include 8 D6-branes, distributed in sets of 4 in each interval (i.e. 2 and their orientifold images). Using the familiar rules, the resulting 4d $\NN=2$ gauge group and hypermultiplet content is
\beqa
&USp(N)\times USp(N)&\nonumber\\
& (\fund,\fund)+ 2(\fund,1)+2(1,\fund) &
\eeqa
with the fundamental hypers arising from the D6-banes. By ungauging any of the two factors (i.e. decompactifying the direction 6) the other one reproduces a II.3 theory. The brane realization of the latter contains 2 explicit D6-branes (and their images), and hence it may seem to be different from the brane construction of theory II.3 provided above. However, one may simply send the D6-branes off to infinity in the (now non-compact) direction, creating new semi-infinite D4-branes by the Hanany-Witten effect, and recover the brane construction exactly as above. 

We conclude by noting that it is not possible to provide an interpolating model using two II.2 theories in a symmetric way, because there is no way to attain cancellation of RR charge with two O6$^+$-planes. From a CFT perspective, this is seen from the fact that the II.2 theory has less than $N$ hypermultiplets in the fundamental, which is the minimum amount required to complete a $SO(N)\times SO(N)$ bifundamental hypermultiplet. 

\subsubsection{The Class II 4d $\NN=1$ theories: II.4, II.5}
\label{sec:classII-45}

We now consider the brane constructions for the Class II theories with 4d $\NN=1$.

\subsubsection*{Theory II.4} 

It corresponds to a 4d $\NN=1$ $SO(N)$ gauge theory with a chiral multiplet in the $\Ysymm$ and $2N-8$ chiral multiplets in the fundamental representation. Using the rules in appendix \ref{sec:o6-planes}, we consider a brane construction with 2 NS5-branes along 012345, with $N$ D4-branes along 0123 and suspended in a non-compact direction 6 between the NS5-branes, and $N-4$ semi-infinite D4-branes stretching from each NS5-brane to infinity. We then introduce an O6'-plane (namely a rotated O6-plane, spanning the direction 0123457, see section \ref{sec:o6-planes}) in the middle of the interval, and choose it to be an O6'$^+$-plane (see Figure \ref{fig:brane-cooking-classII}). The configuration preserves 4d $\NN=1$ supersymmetry, and its spectrums is as follows. The $N$ D4-branes in the interval are mapped to themselves, so they experience the O6'$^+$-plane projection, which produces a 4d $\NN=1$ $SO(N)$ vector multiplet and a chiral multiplet in the $\Ysymm$. In addition, the two bifundamental hypermultiplets of the $SU(N)$ gauge and $SU(N-4)$ global symmetries in the parent theory are swapped by the orientifold action, so they descend to one single bifundamental hypermultiplet of the $SO(N)$ gauge and $SU(N-4)$ global symmetries. This descomposes as 4d $\NN=1$ bifundamental chiral multipelts, namely $2N-8$ chiral multiplets in the fundamental of the $SO(N)$ gauge group, thus completing the desired spectrum. The theory is a particular $k=2$ case of (\ref{one-o6p-ns5-away}).

In this case, the natural proposal for an interpolating model would be to turn the above brane configuration into an elliptic model. Namely, to consider 2 NS5-branes at points in an $\IS^1$, and adding two O6'-planes in the middle of the two intervals, with at least one of them being an O6'$^+$-plane. We need the number of D4-branes on top of this O6'$^+$-plane to be $N$, and on top of the other O6'-plane (i.e. in the other interval) to be $N-4$. A natural choice would be to consider the second O6'-plane to be an O6'$^-$-plane, to achieve cancellation of RR charges. The resulting 4d $\NN=1$ gauge group and chiral multiplet content is
\beqa
& SO(N'+2)\times USp(N'-2)&\nonumber \\
& (\Ysymm,1)+2(\fund,\fund)+(1,\Yasymm)&
\label{interpolating-ii4}
\eeqa
with $N'=N-2$ introduced to emphasize the similar structure of the two sectors. This {\em almost} works as interpolating conformal model, because the $SO(N)$ theory with chiral multiplets in the $\Ysymm$ plus $2N-8$ fundamentals is exactly conformal (our original theory II.5), and the $USp(N')$ (with $N'=N-4$) with chiral multiplets in the $\Yasymm$ plus $2N'+8$ fundamental has vanishing one loop beta function,  but is not an exactly conformal theory, i.e. is not in the Table \ref{table:SCFTs}. This serves as a good illustration that the cancellation of RR charges in 4d $\NN=1$ brane constructions only serves as a criterion for 1-loop beta function cancellation, but not of exact conformality. In this vein, choosing the second orientifold plane to be another O6'$^+$-plane leads to an even worse situation, in which not even the 1-loop beta function can be canceled for both gauge factors.
Hence, for the II.5 theory the brane construction does not lead to an exactly conformal interpolating model, at least  within the class of brane constructions we have used. This however is not an obstruction to study the infinite-distance limits of theory II.4, which in the spirit of section \ref{sec:wo-interpolating} can be done without resorting to an interpolating mode.

\subsubsection*{Theory II.5} 

Let us now consider the theory II.5, a 4d $\NN=1$ $SU(N)$ gauge theory with chiral multiplets in the $\Ysymm+\bYasymm+8\antifund$ plus $(2N-4)$ vector-like flavors $\fund+\antifund$. This model is closely related to the elliptic model considered in appendix \ref{sec:orbifold-nodt}. In fact, it is obtained from the theory (\ref{interp-ii5}) by choosing $N_{11}=N$, $N_{21}=N-2$ and ungauging the symmetry $SU(N-2)$ into a global one. The resulting 4d $\NN=1$ gauge group and chiral matter content is
\beqa
& SU(N)&\nonumber \\
& \Ysymm + \bYasymm + 2(N- 4) \fund+2(N +4) \antifund &
\eeqa
This is precisely the theory II.5. 

\begin{figure}[htb]
\begin{center}
\includegraphics[scale=.3]{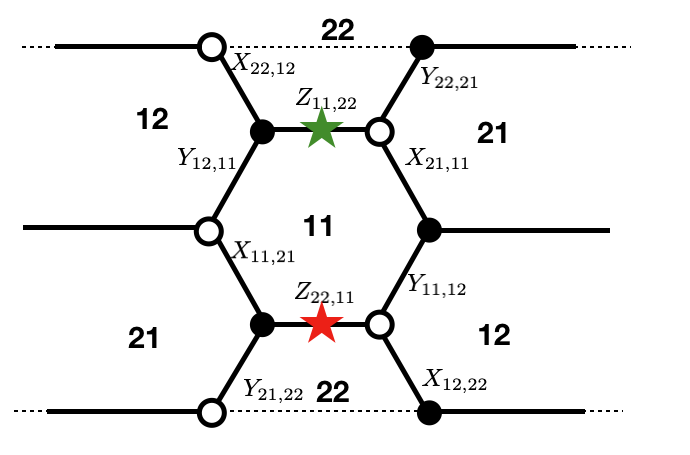}
\caption{\small The non-compact brane tiling obtained as a decompactifying limit of the $\IC^3/(\IZ_2\times\IZ_2)$ theory in Figure \ref{fig:tiling-z2z2}. The dashed horizontal lines are identified. The red and green stars are orientifold fixed points with signs $-$ and $+$, respectively}
\label{fig:tiling-ii5}
\end{center}
\end{figure}

The ungauging corresponds to turning the elliptic model into a linear one, so that the final configuration is given  as follows (see Figure \ref{fig:brane-cooking-classII}). Take 2 NS5-branes along 012345 and at different positions in a non-compact direction 6, with $N$ D4-branes along 0123 and suspended between them in 6, $N-4$ semi-infinite D4-branes stretching out from the NS5-branes to infinity. In addition, we have a $\IZ_2$ orbifold acting as $(z,w)\to (-z,-w)$ on the complex coordinates $z=x^4+ix^5$, $w=x^8+ix^9$, and an orientifold introducing an O8$^-$-plane, as described in appendix \ref{sec:orbifold-nodt}, with a stack of (flavor) D8-branes to achieve charge cancellation. The computation of the spectrum is as explained there, so we skip its detailed discussion. Let us simply mention that it admits a description in terms of a non-compact brane tiling, corresponding to a limit of Figure \ref{fig:tiling-z2z2}, shown in Figure \ref{fig:tiling-ii5}.

Let us finally mention that the elliptic model (\ref{interp-ii5}) in which the theory II.5 is embedded, does not provide a suitable exactly superconformal interpolating model. In fact, in this case the 1-loop beta function of the $SU(N-2)$ gauge group does not vanish exactly, although it does in the large $N$ limit. We will discuss this kind of quasiconformal theories in section \ref{sec:quasiconformal}. We finish by noticing that this elliptic model admits a T-dual type IIB description as an orientifold of $\IC^3/(\IZ_2\times\IZ_2)$, constructed in detail in appendix \ref{sec:orbifold-nodt}.

\subsection{Class III: Models with 3 NS5-branes}
\label{sec:classIII}

\begin{figure}[htb]
\begin{center}
\includegraphics[scale=.32]{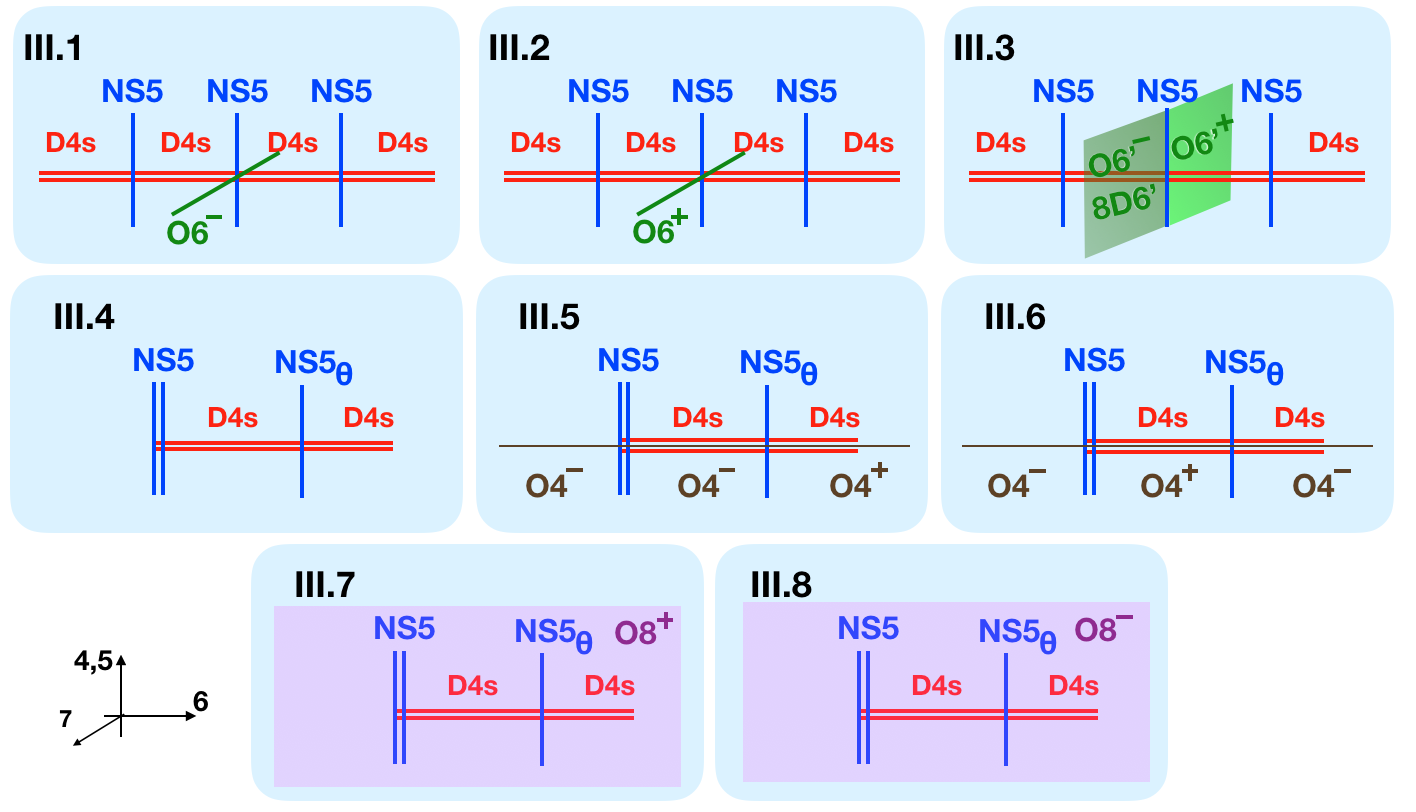}
\caption{\small The type IIA Hanany-Witten brane constructions for theories of class III}
\label{fig:brane-cooking-classIII}
\end{center}
\end{figure}

In this section we consider models in Class III. As we will show, they correspond to constructions involving 3 NS5-branes, but with only one free parameter to tune their relative distances. Hence, as further developed in section \ref{sec:strings}, the bulk duals of these SCFTs are related to the throat created by these 3 NS5-branes when they are taken close to each other (compared with the energies of any probe). More precisely, the infinite-distance limit in the conformal manifold of all Class III theories leads to the same string becoming tensionless. In this case, this string is of a novel kind different from that considered in \cite{Gadde:2009dj}, as anticipated in \cite{Calderon-Infante:2024oed}. We study the brane constructions for these models in turns (see Figure \ref{fig:brane-cooking-classIII}).

\subsubsection{The Class III 4d $\NN=2$ theories: III.1, III.2}
\label{sec:classIII-12}

Let us consider the theories III.1 and III.2 simultaneously. They correspond to a 4d $\NN=2$ $SU(N)$ gauge theory with a hypermultiplets in the $\Yasymm+ (N+2)\fund$ for theory III.1 and in the $\Ysymm+(N-2)\fund$ for theory III.2. Using the rules in appendix \ref{sec:o6-planes}, they admit the following brane construction. Consider 3 NS5-branes along 012345, with one O6-plane along 0123789, with one NS5-brane on top of the O6-plane in the non-compact direction 6, and two other at orientifold image points (see Figure \ref{fig:brane-cooking-classIII}). The configuration with O6$^-$-plane will lead to the theory III.1 and the configuration with an O6$^+$-plane will yield the III.2 theory. We locate $N$ D4-branes along 0123 and suspended in 6 between NS5-branes in the two (orientifold image) intervals, and $N\pm 2$ semi-infinite D4-branes stretching from the outermost NS5-branes to infinity, with the two possible sign choices corresponding to the O6$^\mp$ case. The configuration preserves 4d $\NN=2$ and the choice of numbers of D4-branes is to ensure the no brane bending conditions, i.e. conformality of the resulting theories. 

The spectrum of the resulting theories is as follows. The D4-branes in the two intervals between NS5-branes are exchanged under the orientifold action, so they lead to a 4d $\NN=2$ $SU(N)$ vector multiplet, and the two $SU(N\pm2)$ global symmetry factor from semi-infinite D4-branes are also swapped and lead to one $[SU(N\pm2)]$ global symmetry in the orientifold quotient. The two bifundamentals between the semi-infinite and the finite D4-branes are exchanged and they lead to one bifundamental hypermultiplet under the $SU(N)\times [SU(N\pm 2)]$ symmetry. Finally, the bifundamental hypermultiplet between the two $SU(N)$ factors in the parent theory is mapped to itself by the orientifold action, so it is projected down to a hypermultiplet in the $\Yasymm$ of $SU(N)$ in the O6$^-$-plane case, or in the $\Ysymm$ in the O6$^+$-plane case. This completes the desired spectrum of the III.1 or III.2 theories, which are in fact the particular $k=3$ case of (\ref{one-O6-ns5-ontop}).

Let us quickly note an interesting fact. It is possible to move the stuck NS5-brane off the O6-plane in the directions 789, leaving a configuration with only 2 NS5-branes of the kind considered in the brane construction of theories II.2, II.3. This simply reflects the field theoretic relation 
that a 4d $\NN=2$ $SU(N)$ theory with a hyper in the $\Yasymm$ (like theory III.1) can be Higgsed down to an $USp(N)$ theory (i.e. theory II.3), and similarly that an $SU(N)$ theory with a $\Ysymm$ hypermultiplet (like theory III.2) can be Higgsed down to an $SO(N)$ theory (i.e. theory II.2). This shows that it is possible to flow from theories in one class to theories in another by Higgsing processes. 

It is straightforward to construct elliptic models by combining the above linear quiver models either among themselves, or with other linear brane constructions using O6-planes in section \ref{sec:classII-123}, namely theories II.2 or II.3. When the configurations cancel RR charges, they result in interpolating models for the corresponding theories. Let us describe one simple example, which corresponds to gluing the III.1 and III.2 theories. The configuration contains one O6$^+$- and one O6$^-$-plane, one NS5-brane stuck at each O6-plane, and 2 NS5-branes (forming one orientifold pair) away from the them. By already familiar arguments, the 4d $\NN=2$ gauge group and matter content is
\beqa
&SU(N'-1)\times SU(N'+1)&\nonumber\\
&(\Yasymm,1)+(\fund,\antifund)+(1,\Ysymm)&
\eeqa
where we have introduced $N'$ to emphasize the similar structure of both sectors. Choosing $N'-1=N$ and ungauging the $SU(N'+1)$ factor into a global symmetry one, we recover the theory III.1 from the first gauge factor; choosing instead $N'=N+1$ and ungauging the $SU(N'-1)$ factor, we recover the theory III.2 from the second gauge factor. Hence, the above gauge theory is an interpolating model for both SCFTs.

It is similarly easy to combine two III.1 theories into one elliptic model to obtain a totally symmetric interpolating model. Consider two O6$^-$-planes, one NS5-brane stuck at each O6-plane, and 2 NS5-branes (forming one orientifold pair) away from the them. In order to cancel the RR charge, we add 8 D6-branes, distributed e.g. in sets of 2 in each of the four intervals. Using the by now familiar rules, the 4d $\NN=2$ gauge group and matter content is
\beqa
&SU(N)\times SU(N)&\nonumber\\
&(\Yasymm,1)+(\fund,\antifund)+(1,\Yasymm)+2(\fund,1)+2(1,\fund)&
\eeqa
with the fundamental flavors arising from the D6-branes. By ungauging one of the gauge factors into a global symmetry, we recover the theory III.1. The brane construction contains 4 explicit D6-branes in each of the two (orientifold image) intervals, so the configuration may seem different from that in our above construction of the linear quiver theory. However, they can be moved off to infinity, creating 2 new semi-infinite D4-branes, and matching precisely the brane construction of the III.1 theory presented above.

Let us now provide an example of an interpolating model reproducing theories of the classes II and III in different limits. Consider gluing together a theory III.2 with a theory II.3. Namely, we consider a configuration with one O6$^+$- and one O6$^-$-plane, and a total of 3 NS5-branes, one on top of the O6$^+$-plane and the other two forming an orientifold pair away from the O6-planes. The 4d $\NN=2$ gauge group and matter content is
\beqa
&SU(N'+1)\times USp(N'-1)&\nonumber\\
&(\Ysymm,1)+(\fund,\antifund)&
\eeqa
By choosing $N'+1=N$ and ungauging the second factor we recover a theory III.3 from the first factor, whereas by instead choosing $N'-1=N$ and ungauging the first factor we recover a theory II.3 from the second factor.

\subsubsection{The Class III 4d $\NN=1$ theory with O6'-plane: III.3}
\label{sec:classIII-3}

Let us now consider the theory III.3. It is a 4d $\NN=1$ $SU(N)$ theory with an adjoint chiral multiplet, and extra chiral multiplets in the $\bYasymm+\Ysymm+(N-4)\fund +(N+4)\antifund$. Using the rules in appendix \ref{sec:o6-planes}, we can reproduce it with the following brane construction (see Figure \ref{fig:brane-cooking-classIII}). Consider 3 NS5-branes along 012345 and located at points in a non-compact direction 6. We locate an O6'-plane along 0123457 on top of one of the NS5-branes in the direction 6, and let the other two NS5-branes form a symmetric pair. The O6'-plane is split by the stuck NS5-brane in halves of opposite charge, and we must add 8 half D6'-branes on top of the half O6'$^-$-plane. We finally include $N$ D4-branes in the two (orientifold image) intervals between NS5-branes, and $(N-4)$ semi-infinite D4-branes stretching out of the outermost NS5-branes. 

The computation of the spectrum is as follows. The D4-branes in the two intervals are swapped by the orientifold action, so they produce an $SU(N)$ vector multiplet of the local 4d $\NN=2$ supersymmetry felt in this sector, which decomposes as a 4d $\NN=1$ $SU(N)$ vector multiplet and an adjoint chiral multiplet. The bifundamentals of the parent theory at the NS5-branes away from the O6'-plane are also exchanged by the orientifold action and become a 4d $\NN=2$ bifundamental in the quotient, which decomposes as $(N-4)$ 4d $\NN=1$ chiral multiplets in the $\fund+\antifund$ of $SU(N)$. The bifundamental on top of the NS5-brane stuck at the O6-plane is mapped to itself and projects down to 4d $\NN=1$ chiral multiplets in the $\bYasymm+\Ysymm$ of $SU(N)$, due to the action of the half O6'$^\pm$-planes. Finally, there are 8 extra chiral multiplets in the $\antifund$ of $SU(N)$ due to the half D6'-branes. This brane construction thus reproduces the complete spectrum of the III.3 theory, which we have just shown corresponds to the case $k=1$ of (\ref{one-o6p-ns5-ontop}).

Let us now discuss the construction of an interpolating model. Since the O6'-plane is effectively positively charged (because even the half corresponding to O6'$^-$, with charge $-4$, overlaps with 8 half D6'-brane, turning it to $+4$), the only way to cancel the RR charge is to combine it with an O6'$^-$-plane in an elliptic model. Hence we consider a configuration with 3 NS5-branes in an $\IS^1$ parametrized by 6, with one of them cutting an O6'-plane in two half O6'$^\pm$-planes (with 8 half D6'-branes) and the other two forming an orientifold pair, with an O6'$^-$-plane in the middle. We add $N$ D4-branes in each of the two intervals between the stuck NS5-brane and the other two NS5-branes, and $N-4$ D4-branes in the remaining interval (i.e. where the empty O6'$^-$-plane is located). The resulting 4d $\NN=1$ gauge group and chiral multiplet content is
\beqa
&USp(N-4)\times SU(N) &\nonumber\\
& \Yasymm + (\fund,\antifund)+(\antifund,\fund)+(1,{\rm Adj})+(1,\bYasymm)+(1,\Ysymm)+8(1,\antifund)&
\eeqa
By ungauging the first factor, we recover the theory III.3 from the second. On the other hand, if we instead ungauge the second factor, the first factor is similar to the theory II.4, but with a flip $SO\to USp$, $\Ysymm\to \Yasymm$. This simply reflects the fact that we reach a configuration of 2 NS5-branes away from an O6'$^-$-plane, instead of the O6'$^+$-plane present in the brane configuration of the theory II.4. The theory we obtain has vanishing 1-loop beta function, in agreement from the expectation from RR charge cancellation, but is not exactly conformal (showing again that in 4d $\NN=1$ brane configurations charge cancellation does not imply exact conformal invariance). 

Therefore, the theory III.3 does not admit an embedding into an exactly conformal interpolating model, at least within the class of brane configurations we are considering. However, this is not a problem for the study of the infinite distance points in this SCFT, since, in the spirit of section \ref{sec:wo-interpolating}, it can be carried out without resorting to an interpolating model.

\subsubsection{The Class III 4d $\NN=1$ theories from coincident NS5-branes: III.4, III.5, III.6, III.7, III.8}
\label{sec:classIII-45678}

In this section we argue that there is a further way in which one can build configurations with 3 NS5-branes whose relative positions are controlled by only one parameter, based on the configurations of coincident NS5-branes in appendix \ref{sec:kutasov} (see \cite{Elitzur:1997fh,Elitzur:1997hc} for references). We emphasize that these NS5-branes are formally located at the same position in all directions, even before taking any limit. For convenience, we will refer to them as `glued' NS5-branes, in order to distinguish them from NS5-branes which become coincident in the decoupling limit. In fact, we argue that such constructions underlie the brane realization of the theory III.4, and, via extra orientifolds, theories III.5, III.6, III.7 and III.8, and ultimately even the theory I.8. We discuss them in turns.

\subsubsection*{Theory III.4}

Consider the configuration of $N$ D4-branes suspended between a stack of 2 `glued' NS5-branes along 012345 and 1 rotated NS5$_\theta$-brane, i.e. along 0123 times a 2-plane at angles in the directions 45 and 89, as in section \ref{sec:rotated-branes}. We also add semi-infinite D4-branes attached to the NS5$_\theta$-brane. For a generic angle $\theta$, the theory can be obtained from the rules in appendix \ref{sec:kutasov}).
Namely, we have a 4d $\NN=1$ $SU(N)$ gauge theory, with one chiral multiplet $X$ in the adjoint (with a cubic superpotential $\tr X^3$), and $N_f$ chiral flavors $Q,{\tilde Q}$ in the $\fund+\antifund$, with a quartic superpotential $(Q{\tilde Q})^2$ controlled by the angle $\theta$. As announced, the configuration has 3 NS5-branes, but 2 of them are necessarily coincident, so their relative motion is controlled by a single parameter.

For $N_f=N$, the above gauge theory is a close cousin of the SCFT III.4, except for a missing extra adjoint. Relatedly, it contains 3 NS5-branes, but compared with the previous Class III theories, they are not parallel but rather at some non-trivial angle. In the following we argue that both problems can be solved at once by simply taking $\theta\to 0$. 
This has not been considered in the literature (namely, there is no established gauge theory interpretation of the configuration of `glued' NS5-branes and unrotated NS5-branes). Hence, the following should be considered a heuristic derivation that leads to self-consistent rules for this type of models.

Given the rules in appendix \ref{sec:rotated-branes}, in the limit of small angle $\theta$ the quartic coupling between the fundamental chiral multiplets becomes singular. This simply signals the appearance of an adjoint chiral multiplet for $\theta=0$, which parametrizes the possibility to slide the D4-branes between the parallel NS5-branes in the directions 45. In the standard configurations in appendix \ref{app:basic}, this corresponds to the adjoint chiral multiplet in the 4d $\NN=2$ vector multiplet; in the present setup, however, the configuration keeps memory of its genuine 4d $\NN=1$ nature because of the `glued' NS5-branes (i.e. the pre-existing adjoint $X$ with cubic superpotential). Hence, the final gauge theory is 4d $\NN=1$ $SU(N)$ with 2 adjoint chiral multiplets and $N$ chiral multiplets in the $\fund+\antifund$. The resulting brane setup, depicted in Figure \ref{fig:brane-cooking-classIII}, therefore reproduces the SCFT III.4, and its realization in terms of a configuration of 3 parallel NS5-branes (in the limit) fits perfectly with the  brane realizations of previous Class III theories. In section \ref{sec:strings} we will in fact argue that they all share the same emergent string limit. 

It is interesting to consider the possibility of interpolating models. The natural proposal is to combine two theories III.4 and consider a 4d $\NN=1$ gauge group and chiral matter content
\beqa
& SU(N)\times SU(N) & \nonumber\\
& 2({\rm Adj},1)+(\fund,\antifund)+(\antifund,\fund)+2(1,{\rm Adj})&
\label{interpolating-iii4}
\eeqa
This gauge theory is realized by a brane configuration given by the following elliptic model. Consider 2 `glued' NS5-branes along 012345 and 1 rotated NS5$\theta$-brane along 0123 and a 2-plane at an angle $\theta$ between the 2-planes 45 and 89, at different positions in a direction 6 parametrizing an $\IS^1$, and add $N$ D4-branes in each of the two intervals. Each of the D4-branes in the interval is a theory of the kind considered above, with one adjoint chiral multiplet and $N$ flavors, arising from bifundamentals $(\fund,\antifund)+(\antifund,\fund)$ (from a  bifundamental hypermultiplet of the local 4d $\NN=2$ at the NS5$_\theta$-brane location). In addition, each $SU(N)$ gauge factor should get an extra adjoint chiral multiplet in the limit of vanishing $\theta$, as motivated above. Note that, although there are D4-branes on both sides of the stack of 2 `glued' NS5-branes, there are no corresponding bifundamental multiplets (in contrast with the isolated NS5-brane); this is heuristically due to the presence of an `vanishingly small empty inverval' between the `glued' NS5-branes.\footnote{Note that this argument allows to modify the linear brane model by the addition of semi-infinite D4-branes attached to the `glued' NS5-branes, without modifying the gauge theory. The same holds for the constructions in coming sections. The inclusion of these extra branes is useful for the construction of interpolating models, and we will implicitly keep this possibility in mind in their discussion. \label{foot:hidden-semiD4s}}

In principle it should be possible to perform a T-duality to this elliptic model and find a type IIB description; however, to our knowledge this has not been considered in the literature. Naively, the T-dual corresponds to a stack of $N$ D3-branes at a $\IC^3/\IZ_3$ singularity, but with some additional ingredient which partially freezes the possibility to blow up the singularity, and which breaks the supersymmetry down to 4d $\NN=1$. This is because 2 of the type IIA NS5-branes are `glued', hence frozen to be coincident in their position in 67, and can move in 89 only upon the introduction of extra $\NN=1$ superpotential terms. The nature of this extra ingredient is unclear at this point, and seems different from those arising in other well-known frozen singularities (see e.g. \cite{Landsteiner:1997ei,Witten:1997bs,deBoer:2001wca,Tachikawa:2015wka,Cvetic:2024mtt}). Therefore, we will in general skip the discussion of T-duals of configurations with coincident NS5-branes for this examples, and those in coming sections (except for theory I.8 in section \ref{sec:classI8-atlast}).

In the following sections we discuss adding orientifolds to the above brane construction. Using the rules in appendix \ref{sec:rotated-branes}, we note that only O4- and O8-planes are allowed for generic values of $\theta$, while introducing O6-planes requires taking $\theta=0$ exactly (the NS5$_\theta$-brane would otherwise require the introduction of an image NS5$_{-\theta}$-brane which is not present in the configuration). As we discuss next, adding O4- and O8-planes will lead to the rest of type III theories. Finally, taking $\theta = 0$ exactly and introducing O6-planes will produce the theory I.8, whose brane construction was postponed in Section \ref{sec:classI}. 

\subsubsection*{Theories III.5, III.6}

Theories III.5 and III.6 are easily shown to correspond to including additional O4-planes  to the brane construction of theory III.4 (see Figure \ref{fig:brane-cooking-classIII}). Hence we consider a configuration with 2 `glued' NS5-branes along 012345 and 1 rotated NS5$_\theta$-brane, at different locations in a non-compact direction 6, with $N$ D4-branes along 0123 and suspended along 6 between them, and $N_f$ semi-infinite D4-branes stretching out of the NS5$_\theta$-brane to infinity. We now add an O4-plane, on top of the D4-branes, which using the rules in appendix \ref{sec:o4-planes} flips sign each time it crosses each NS5-brane, i.e. it changes sign across the NS5$_\theta$-brane but does not change sign across the 2 `glued' NS5-branes. There are hence two possible models corresponding to having an O4$^\pm$-plane on top of the $N$ D4-branes suspended in the interval (and on O4$^\mp$-plane on top of the semi-infinite D4-branes).

Consider for instance the configuration with the O4$^-$-plane on top of the interval. The gauge group is projected down to $ SO(N)$, and the 2 bifundamental chiral multiplets from open strings between the D4-branes across the NS5$_\theta$-branes are projected down to a single bifundamental, i.e. $N_f$ chiral multiplets in the $\fund$ of $SO(N)$. We now need to determine the orientifold projection on the adjoint chiral multiplet, which to our knowledge has not been considered in the literature. As explained in appendix \ref{sec:kutasov}, the adjoint coming from the `glued' NS5-branes parametrizes the motion of the $N$ D4-branes in the direction 89 (upon deformation of the adjoint superpotential). This motion is along the NS5$_\theta$-brane, so it is naturally $\NN=1$-like, rather than $\NN=2$-like; this suggests that the orientifold projection will break the 4d $\NN=2$ supersymmetry, leading to a chiral multiplet in the $\Ysymm$.

Finally, we should consider the orientifold projection of the additional chiral adjoint of theory III.4 in the limit of vanishing $\theta$. Contrary to the previous one this is essentially like the adjoint in a system of NS5-and D4-branes, i.e. it is an $\NN=2$-like motion in the directions 45; this suggests that the orientifold projects the adjoint down a $\Yasymm$. Overall, the 4d $\NN=1$ gauge group and chiral multiplet content is
\beqa
& SO(N) &\nonumber \\
& (N-6) \fund + \Yasymm +\Ysymm\, , &
\eeqa
where we have fixed $N_f=N-6$ for conformality. This reproduces theory III.5 in Table \ref{table:SCFTs}. 

One can similarly work out the gauge theory on the same kind of configuration with the O4$^+$-plane on top of the interval. This simply implies a flip of the $SO/USp$ projections for the gauge group and the $\Ysymm/\Yasymm$ matter content. The resulting 4d $\NN=1$ gauge group and chiral multiplet content is
\beqa
& USp(N) &\nonumber \\
& (N+6) \fund + \Ysymm +\Yasymm &
\eeqa
where we have fixed $N_f=N+6$ for conformality. This reproduces theory III.6 in Table \ref{table:SCFTs}.

Let us comment on the possibility of interpolating models. It turns out that there are no interpolating models within our class of theories. Naively one can propose candidate elliptic models, by compactifying the direction 6 (after adding extra semi-infinite D4-branes to the `glued' NS5-branes, c.f. footnote \ref{foot:hidden-semiD4s}). Due to the sign flips of the O4-planes, we must introduce one additional NS5-brane in the elliptic model. Taking it to be an NS5$_\theta$-brane, we end up with an elliptic model with 4 NS5-branes, two of them being `glued', and the other two being NS5$_\theta$-branes, ultimately in the limit of vanishing angle. For instance, for the III.5 theory, the candidate interpolating model  has 3 gauge sectors, two of them corresponding to $N$ D4-branes on the two intervals between the `glued' NS5-brane and each NS5$_\theta$, on top of an O4$^-$-plane (morally, III.5 theories), and a last gauge sector given $N+p$ D4-branes in the interval between the NS5$_\theta$-branes, on top of an O4$^+$-plane (morally a II.3 theory). The 2-index tensor representations arise as explained above, and the leading order (in $1/N$) number of flavors for each gauge factor is reproduced by the bifundamentals across the NS5$_\theta$-branes. However there is no choice of $p$ which makes all the gauge sectors have the right number of flavors to achieve exact conformality. A similar discussion occurs for III.6 theories, using an interpolating models which glues two III.6 theories and a II.2 theory. In other words, the candidate interpolating model is superconformal only in the strict large $N$ limit.

\subsubsection*{Theory III.7, III.8}

Theories III.7 and III.8 correspond to including additional O8-planes to the brane construction of theory III.4. We consider a configuration with 2 `glued' NS5-branes along 012345 and 1 rotated NS5$_\theta$-brane, at different locations in a non-compact direction 6, with $N$ D4-branes  along 0123 and suspended along 6 between them, and $N_f$ semi-infinite D4-branes stretching out of the NS5$_\theta$-brane to infinity. We now add an O8-plane along 012345689, which using the rules in appendix \ref{sec:o8-planes} does not flip sign across the NS5-branes. There are hence two possible models corresponding to the O8$^\pm$-plane cases (see Figure \ref{fig:brane-cooking-classIII}).

Consider fist the case of the O8$^+$ plane. In this case the orientifold projection on D4-branes has an extra sign, compared with the O4-plane case (because of the 4 ND+DN directions), so the gauge group is projected down to $SO(N)$. As in the case of the O4-plane above, the two 4d $\NN=1$ bifundamentals between the finite and semi-infinite D4-branes at the NS5$_\theta$ are projected down to a single bifundamental, i.e. $N_f$ chiral multiplets in the $\fund$ of $SO(N)$. We now have to determine the projection on the adjoint arising from the `glued' NS5-branes. As argued in the case of the O4-plane, it corresponds to a motion of D4-branes in the directions 89, which is $\NN=1$-like; this suggests that the orientifold projection truncates it to a $\Ysymm$. Regarding the adjoint becoming massless in the limit of vanishing $\theta$, it parametrizes motions in the directions 45, so it is $\NN=2$-like. However, in contrast with O4-planes, the O8-plane breaks that supersymmetry, so instead of projecting down to the adjoint of $SO(N)$, we expect that it is projected down to an $\Ysymm$. Combining things together, the resulting 4d $\NN=1$ gauge group and chiral multiplet content is
\beqa
& SO(N) &\nonumber \\
& (N-10) \fund + 2\,\Ysymm &
\eeqa
where we have chosen the number of flavours to get conformality. This corresponds to theory III.7 in the Table \ref{table:SCFTs} of SCFTs.

The case of the O8$^-$-plane is analyzed similarly, and leads to a simple flip of the $SO/USp$ projections for the gauge group and the $\Ysymm/\Yasymm$ matter content. The resulting 4d $\NN=1$ gauge group and chiral multiplet content is
\beqa
& USp(N) &\nonumber \\
& (N+10) \fund + 2\,\Yasymm &
\eeqa
where we have chosen the number of flavours to get conformality. This corresponds to theory III.8 in Table \ref{table:SCFTs}.

Let us consider the possibility of interpolating theories, by making the models elliptic with a compact direction 6 (after adding extra semi-infinite D4-branes to the `glued' NS5-branes, c.f. footnote \ref{foot:hidden-semiD4s}). In this case, since the O8-planes do not flip sign across NS5-branes, it is not necessary to introduce additional NS5-branes. The candidate interpolating models are simply obtained by taking 3 NS5-branes, with two of them being `glued' and the third being an NS5$_\theta$-branes, in the limit of vanishing angle. For the case of an O8$^+$-plane, we almost obtain two copies of the III.7 theory, but there is no choice of numbers of D4-branes which reproduces the right number of flavors to make the two gauge sectors conformal. Hence, within our class of constructions, there is no exactly superconformal interpolating model (except in the strict large $N$ limit) for the theory III.7. For the case of an O8$^-$-plane, we can choose $N$ D4-branes on each interval and introduce 10 D8-branes, and get two copies of the III.8 theory, with the correct number of flavors, thus reproducing an exactly conformal interpolating model, even for finite $N$.

\subsubsection{Grand Finale: The theory I.8}
\label{sec:classI8-atlast}

In this section we apply our previous experience with models with extra adjoints to go back and construct the brane configuration for Theory I.8 (which, with hindsight, was already displayed in Figure \ref{fig:brane-cooking-classI}). Interestingly, even though it corresponds to a theory of Class I, hence with no mobile NS5-branes, the brane construction is closely related to the interpolating model (\ref{interpolating-iii4}) for the theory III.4. Namely, we start with an $\IS^1$ parametrized by the coordinate 6, and we locate 2 `glued' NS5-branes and 1 isolated NS5-brane, at opposite points in the circle. We include 2 stacks of $N$ D4-branes suspended between them in the two intervals. The three NS5-branes span the directions 012345, but they should be regarded as the limit of vanishing $\theta$ of a generic 4d $\NN=1$ configuration, so that the `glued' NS5-branes lead to one extra adjoint for each gauge factor, and no bifundamentals across them (due to the `empty interval' they define). 

We now consider introducing an O6$^-$-plane on top of both the isolated NS5-brane and the stack of 2 `glued' ones (note that this is possible only because we are considering the non-generic case of vanishing angle between the NS5-branes). In addition, in order to cancel the RR charges, we introduce 8 D6-branes, locating 4 of them on top of the 2 `glued' NS5-branes, and the remaining 2 (plus their 2 orientifold images) at generic points in the two intervals.

It is easy to adapt the rules of appendix \ref{sec:o6-planes} and carry out the orientifold projection to the gauge theory content (\ref{interpolating-iii4}), as follows. The two stacks of $N$ D4-branes are exchanged, leading to a single $SU(N)$ gauge group and, similarly, to a single copy of the 2 adjoint chiral multiplets. On the other hand, the 2 bifundamental chiral multiplets at the location of the isolated NS5-brane are each mapped to itself by the O6$^-$-plane, and project down to chiral multiplets in the $\Yasymm+\bYasymm$. Finally, the 2 D6-branes (and their images) in the intervals lead to 2 chiral multiplets in the $\fund+\antifund$, while those on top of the `glued' NS5-branes lead to no additional flavors (due to the `empty interval' defined by the NS5-branes). Overall, the 4d $\NN=1$ gauge group and chiral matter content is
\beqa
& SU(N)&\nonumber \\
& 2\,{\rm Adj}+\Yasymm+\bYasymm+2(\fund+\antifund)
\eeqa
which reproduces Theory I.8 as announced. 

As already mentioned, the elliptic version of models of `glued' NS5-branes has not been considered in the literature, so there are no results on the T-dual type IIB realization of such constructions. However we can easily identify the key ingredients for the realization of theory I.8. Since the type IIA realization has 3 NS5-branes with $N$ suspended D4-branes, we expect that the T-dual type IIB picture contains a stack of $N$ D3-branes at a $\IC^2/\IZ_3$ singularity. An important observation is that, because there are 2 `glued' NS5-branes in the type IIA picture, whose relative motion in the directions 6, 7 is effectively frozen, the corresponding blowup mode in the $\IZ_3$ singularity must be considered to be frozen, even though the microscopic mechanism does not seem to correspond to any of the known ones (see e.g. \cite{Landsteiner:1997ei,Witten:1997bs,deBoer:2001wca,Tachikawa:2015wka,Cvetic:2024mtt}). In addition, the presence of O6$^-$-planes and 8 flavor D6-branes in the original type IIA picture implies that the type IIB side contains an orientifold of the previous system, which introduces an O7$^-$-plane and 8 flavor D7-branes. 

\medskip

This completes our construction of brane realizations for the SCFTs in Table \ref{table:SCFTs}. We now turn to their application to the understanding of the universality classes of infinite-distance limits, and the corresponding emergent tensionless strings.

\section{Universality Classes and the Emergent String Theories}
\label{sec:strings}

In this section we use the brane constructions in the previous section to explain why the infinite-distance limits of the SCFTs in Table \ref{table:SCFTs} fall into three universality classes, as predicted in \cite{Calderon-Infante:2024oed}, and characterize the nature of the tensionless strings arising in these three different limits. We show that the resulting ``flat space'' worldsheets (see Footnote \ref{foot:flat-space}) are the 10d critical Type IIB string for class I; the non-critical string in \cite{Gadde:2009dj} (associated to the closed string background generated by two NS5-branes becoming coincident) for class II; and a new kind of string, which we describe as associated to three NS5-branes becoming coincident, for class III.

\subsection{Class I}
\label{sec:strings-classI}

These correspond to theories whose infinite-distance limit is a weakly-coupled 10d critical Type IIB string theory. In the brane constructions this simply follows because the gravity dual is obtained as the near-horizon limit of a stack of $N$ D3-branes in flat space or some orbifold or orientifold thereof, as we now derive explicitly.

The prototype is theory I.1, which is 4d $\NN=4$ $SU(N)$ SYM, realized on the worldvolume of a stack of $N$ D3-branes in flat space. The infinite-distance limit in the conformal manifold is the weak-coupling limit of the gauge theory, which corresponds, in the holographic gravity dual, to taking the string weak-coupling limit of  10d type IIB string theory in AdS$_5\times\IS^5$. Hence, the HS operators map to HS fields arising from the fundamental (critical) Type IIB closed string. 

As discussed in section \ref{sec:classI-123}, the theories I.2 and I.3 arise on the worldvolume of $N$ D3-branes on top of an O3$^-$- or an O3$^+$-plane, respectively. The dual gravitational string background is hence an orientifold of 10d type IIB on AdS$_5\times\IS^5$, studied in \cite{Witten:1998xy}. In particular, the action of an O3-plane is $\Omega R(-1)^{F_L}$, where $R:(z_1,z_2,z_3)\to(-z_1,-z_2,-z_3)$ on the $\IC^3$ transverse to the D3-branes. This action corresponds to an antipodal identification on the angular $\IS^5$, so we obtain 10d type IIB string theory on AdS$_5\times\IR\IP_5$. The difference between theories I.2 and I.3 (namely, the O3$^\pm$-plane) is a different background value of the NSNS 2-form on $\IR\IP_2\subset \IR\IP_5$.

An important observation is that these theories actually correspond, in the large $N$ limit, to the same tensionless string in the infinite-distance limit in the conformal manifold. This simply follows because the worldsheet theories of parent and orientifold models are locally identical, and only differ in the inclusion of unorientable worldsheets in the genus expansion of the latter, i.e. the inclusion of crosscaps. Some of the states of the parent theory are projected out by the orientifold, which leads to modifications in the spectrum that reflect the different backgrounds in which the string propagates. However, the very energetic part of the spectrum (as compared to the string scale, the AdS scale, and the scale of the internal manifold) is expected to be mostly insensitive to these global aspects of the model, such that its coarse-grained properties should only depend on the local dynamics in the worldsheet. This explains from a string-theoretic perspective why the large-$N$ Hagedorn temperature of these SCFTs coincides, as pointed out in \cite{Calderon-Infante:2024oed}. As emphasized also in \cite{Calderon-Infante:2026rkj}, the large-$N$ Hagedorn temperature precisely encodes coarse-grained information about the high energy spectrum, namely its exponential growth. Furthermore, as emphasized in the context of holographic duality in e.g. \cite{Kakushadze:1998tr}, correlators of operators that survive the projection are identical to those in the parent theory at leading order in the large $N$ limit with fixed 't Hooft coupling $\lambda=g_s N$, since each crosscap contributes $+1$ to the Euler characteristic of the worldsheet and is hence weighted by an extra $g_s$ factor. 

Let us note that the above large $N$ limit is not directly related to the infinite distance limit in the conformal manifold of the CFT. The former corresponds to an expansion in $1/N$, for an arbitrary fixed 't Hooft coupling $\lambda$ ($\sim g_{YM}^2 N$), while the latter  corresponds to a weak-coupling limit ($g_{YM}\to 0$) for large $N$ theories. The interplay between both is that the large $N$ limit provides a valid expansion for any value of $\lambda$, and in particular holds also for $\lambda\to 0$, which reproduces the limit in the conformal manifold picture. In this limit the gravitational dual does not corresponds to semiclassical gravity (which holds for large $\lambda$), rather it corresponds to a weakly coupled string, as befits the emergent string limit of the CFT. Note that, from a complementary perspective, the large $N$ limit with $\lambda$ fixed effectively zooms into the weak-coupling region which corresponds to the universality class related to the weakly-coupled string of interest. 

As we will use in later examples, the same argument of the previous paragraphs apply to any theories related by orientifolding. Similarly, the presence of $N_f$ flavor D-branes in the holographic dual model is also subleading in $1/N$ (in the quenched approximation, $N_f\ll N$), because the extra branes are only felt via the introduction of worldsheets with boundaries. Since each boundary contributes $+1$ to the Euler characteristic of the worldsheet, it is weighted by $1/N$ in the large $N$ expansion. In what follows, we will apply this to many other examples with orientifolds, and with flavour D-branes (when their number is subleading in $1/N$). 

For instance, as mentioned in section \ref{sec:classI-456} the theory I.4 is obtained from a stack of $N$ D3-branes on top of an O7$^-$-plane, with 8 D7-branes on top. The emergent string corresponds to 10d type IIB on an $\Omega R_3(-1)^{F_L}$ orientifold of AdS$_5\times\IS^5$, where the action of $R_3$ on the $\IS^5$ is inherited from its action $R_3:(z_1,z_2,z_3)\to (z_1,z_2,-z_3)$ on the $\IC^3$ transverse to the D3-branes. The presence of an explicit O7$^-$-plane along a fixed AdS$_5\times \IS^3$ implies that the model must contain 8 explicit D7-branes along that geometry.\footnote{The fact that the D7-branes are stable (even supersymmetric) despite the $\IS^3$ being trivial is because the corresponding cycle is a generalized calibration \cite{Gutowski:1999iu,Gutowski:1999tu,Townsend:1999nf,Gauntlett:2001ur,Gauntlett:2002sc,Gauntlett:2003cy,Martelli:2003ki,DallAgata:2003txk,Cascales:2004qp} (similar comments apply to the flavor D-branes in examples below).} This configuration was considered in \cite{Banks:1996nj,Aharony:1996en,Douglas:1996js}, and in \cite{Aharony:1998xz} in the context of holography. By the argument above, the orientifolding and the addition of flavor branes are subleading in $1/N$, so the emergent string theory is essentially the same as in 4d $\NN=4$ SYM theory. 

Theories I.5 and I.6 correspond to an orientifold of the system of $N$ D3-branes at an $\IC^2/\IZ_2$ ($\times\IC)$, generated by $\theta:(z_1,z_2,z_3)\to (-z_1,-z_2,z_3)$. The precise orientifold in the type IIB side can be derived from their different type IIA realization in terms of NS5- and D4-brane configurations, as described in section \ref{sec:classI-456}, and corresponds to orientifolds considered in \cite{Park:1998zh}. Theory I.5 corresponds to an orientifold by $\Omega R_3(-1)^{F_L}$, with $R_3:z_3\to-z_3$ as above, while theory I.6 corresponds to an orientifold by $\Omega R_3\alpha (-1)^{F_L}$, with $\alpha^2=\theta$. The emergent string theories in the infinite-distance limit of the conformal manifolds of these theories correspond to the critical 10d type IIB string on the corresponding orientifold of AdS$_5\times S^5/\IZ_2$. The actions of the $\IZ_2$ and the orientifolds, and in particular their fixed point sets, on the $\IS^5$ are directly obtained from those on the $\IC^3$ transverse to the D3-branes. In the theory I.5, there is an explicit O7$^-$-plane (with 8 D7-branes on top) on AdS$_5\times \IS^3$, while in theory I.6 there are no O7-planes (neither D7-branes).

These theories again correspond to the same emergent 10d Type IIB string, in the large $N$ limit. This follows from the results in \cite{Kachru:1998ys,Lawrence:1998ja,Bershadsky:1998mb} in the context of holography for general orbifolds of $\IC^3$, combined with our previous observation for the addition of orientifolds.  Clearly, the same comment applies to any theories related by orbifolding, as we will see in several later examples.

In particular, as described in section \ref{sec:classI-78}, the theory I.7 is obtained from an $\Omega R(-1)^{F_L}$ orientifold of a system of D3-branes at a $\IC^3/(\IZ_2\times\IZ_2)$ singularity (with discrete torsion), formally T-dual to that considered in \cite{Berkooz:1996dw} (in a compact setup). The emergent string is the critical 10d type IIB string on the orientifold AdS$_5\times \IR\IP_5/(\IZ_2\times\IZ_2)$. In this case the combination of orbifold and orientifold actions implies the presence of three different kinds of O7$_i$-planes (each with 8 D7$_i$-branes on top) on different AdS$_5\times\IS^3$ subspaces (actually, quotiented by the extra $\IZ_2$ actions) corresponding to the angular part in the $\IR^4$'s defined $z_i=0$ in $\IC^3$.

Finally, theory I.8 is slightly different from the previous ones. As mentioned in section \ref{sec:classI8-atlast}, the theory would be realized on an $\Omega R_3(-1)^{F_L}$ orientifold of a stack of $N$ D3-branes at a partially frozen $\IC^2/\IZ_3$ singularity. Even though the microscopic physics responsible for the freezing of the singularity is not known, it is reasonable to expect from the standard near horizon limit that the emergent string is a critical 10d type IIB string on the corresponding orientifold of AdS$_5\times \IS^5/\IZ_3$, where the geometric actions of the orientifold and orbifold are those inherited from $\IC^3$. The $\IS^5/\IZ_3$ geometry contains an $\IS^1$ of $\IC^2/\IZ_3$ fixed points, which should be partially frozen in the sense explained above. The geometry contains an explicit O7$^-$-plane (with 8 D7-branes on top) on (an orbifold quotient of) AdS$_5\times \IS^3$. The fact that the string theory corresponds to the same as in previous class I theories suggests that the additional ingredients partially freezing the singularity are subleading in the large $N$ limit. We will use this fact in some of the Class III examples discussed later.

\medskip

The conclusion is that all theories in Class I share the same emergent string theory in the infinite-distance limits of their conformal manifolds, which is the critical 10d type IIB theory on AdS$_5\times\IS^5$. This explains why all these theories share similar properties in the weak-coupling limit, as for instance the value of the large-$N$ Hagedorn temperature in the free limit as well as the value of the exponential rate of the HS tower at large $N$ \cite{Calderon-Infante:2024oed}. We emphasize again that our brane realization of all Class I theories has been the key ingredient for this important lesson. We turn to  a similar analysis for Class II and Class III theories. 

\subsection{Class II}
\label{sec:strings-classII}

Let us consider the structure of the emergent string for class II theories, which will correspond to a sub-critical closed string. 
We first revisit the origin of this string in the context of the theory II.1 (as studied in \cite{Gadde:2009dj} and summarized in Section  \ref{sec:gpr}) and then explain why it is the same emergent string for all theories of class II.

The gravitational dual of the theory II.1 is given by the near horizon geometry of D4-branes in a non-trivial background generated by 2 NS5-branes in the double scaling limit, as proposed in \cite{Gadde:2009dj}. We summarize the key ingredients of the construction in Section \ref{sec:gpr} so we refer the reader to that section for further details. In summary, one can start with the Type IIA background of Figure \ref{fig:interp} and take the low energy limit such that the string coupling goes to zero at the same time than the NS5-branes approach each other (see \eqref{eq:decoupling-limit}), which is claimed to reproduce $\mathcal{N}=2$ SQCD at finite gauge coupling. Before taking into account the backreaction of the D4-branes on the geometry, the closed string background generated by these approaching NS5-branes admits the exact $\mathcal N= (2,2)$ worldsheet description \cite{Giveon:1999px}
\begin{equation} \label{Liouville-k=2}
    \IR^{5,1}\times \IR_\rho \times U(1)_\varphi/\IZ_2 \ , \quad \Phi(\rho)=-\rho/2 \ , \quad W = \tau_0^2 \, e^{-(\rho - i \varphi)} \, .
\end{equation}
As indicated by the linear dilaton background and superpotential above, $\rho - i \varphi$ is the chiral superfield of a $\mathcal N=2$ superLiouville theory. The bottom component of this chiral superfield\footnote{As it is customary, we denote the superfield and its scalar bottom component by the same symbol.} parametrizes a cylinder with radius $2$ in string units, i.e. $\varphi \sim \varphi + 4 \pi$. The Liouville superpotential is related to the separation between the NS5-branes, $\tau_0$, and has the important role of shielding the strongly-coupled region $\rho \to - \infty$. The $\IZ_2$ orbifold acts as $\varphi \to \varphi + 2 \pi$, hence shortening the radius of the cylinder by half. We will not take this into account explicitly, as it would obscure the relation to the Hanany-Witten setup that we discuss now for later convenience. The $\mathcal N=2$ superLiouville theory describes the 6789 hyperplane transverse to the two NS5-branes. The coordinate $\rho$ goes in the radial direction in this hyperplane and around the center of mass of the two NS5-branes (that are slightly separated in the direction 6), while the angular coordinate $\varphi$ parametrizes a great circle inside the remaining three-sphere. For concreteness, we take this great circle to be the one in the 67 plane. More precisely, we take $\varphi=0,2\pi$ to correspond to $x_7=0$ with $x_6>0$ and $x_6<0$, respectively. As recalled in section \ref{sec:gpr}, the other two directions inside the three-sphere are gapped in the case of two NS5-branes, and hence no longer contribute to the central charge \cite{Gadde:2009dj,Dei:2024frl}. This is why this is considered a non-critical string background, even though the string theory is critical in the sense of the worldsheet having vanishing central charge.

Upon a further T-duality along $\varphi$, \eqref{Liouville-k=2} is equivalent to the worldsheet theory \eqref{eq:KS-coset}, namely
\beqa
    \IR^{5,1}\times \frac{SL(2,{\bf R})_2}{U(1)}\mathlarger{\mathlarger{\mathlarger{/}}}\IZ_2\, ,
\label{cigar-ii}
\eeqa
where $SL(2,{\bf R})_2/U(1)$ is a Kazama-Suzuki coset, also known as the cigar CFT. Morally, the Liouville superpotential is replaced by a geometric cap-off of the cylinder, such that the string coupling is now bounded from above by a value of the order of the gauge coupling of $\mathcal N=2$ SCQCD (c.f. \eqref{eq:decoupling-limit}). In this Type IIB frame, the D4-branes turn to D3-branes and flavour D5-branes; whose backreaction at large $N$ is expected to generate an AdS$_5$ throat times a three-dimensional internal space (containing a circle of string size). The infinite-distance limit in the conformal manifold of theory II.1 corresponds to the weak-coupling limit of said gauge theory, and therefore maps to the weak (string) coupling limit of the above non-critical string background. The emergent string is therefore this subcritical superstring that become tensionless in Planck units. 

\medskip

As derived in Section \ref{sec:classII}, the brane configuration for the remaining Class II theories are obtained as (orientifold or orbifold) quotients of the configuration for theory II.1, and therefore, they inherit a similar weak-coupling limit. The result is a quotient of the type IIA or type IIB backgrounds in \eqref{Liouville-k=2} and \eqref{cigar-ii}, respectively. As in the discussion in section \ref{sec:strings-classI}, orientifold and orbifolds do not change the local dynamics on the worldsheet and imply only subleading $1/N$ corrections on the correlators of operators that survive the projection.\footnote{For orientifolds, it just follows from the same argument about the Euler characteristic of crosscaps. For orbifolds, even though we no longer have orbifolds of flat space, general field theory arguments imply the subleading behaviour in their orbifold quotients (see e.g. \cite{Schmaltz:1998bg}).} Hence, the infinite-distance limits of all class II theories is controlled by the same non-critical emergent string. In the following we quickly describe the concrete quotients arising for each theory. 

Theories II.2 and II.3 can be obtained from II.1 in two ways, either introducing an O4- or an O6-plane. Let us first consider reproducing these theories using an O4$^\pm$-plane. Since the O4-plane flips all directions transverse to the D4-branes (i.e. directions 45789) we have an orientifold of the type IIA background (\ref{Liouville-k=2}) flipping the directions 45 in $\IR^{5,1}$, acting as $\varphi \to -\varphi$ on $U(1)_\varphi$ (i.e. turning the circle into a segment), and leaving the radial direction $\IR_\rho$ invariant.\footnote{We note that $\varphi \to -\varphi$ is a symmetry of \eqref{Liouville-k=2}, as required by consistency of the orientifold action. This transformation amounts to changing the superpotential $W$ by its complex conjugate, which indeed leaves the theory invariant.} After the T-duality, we expect the O4-plane to T-dualize exactly as the D4-branes, namely we obtain a quotient of (\ref{cigar-ii}) by orientifold action flipping the directions 45 in $\IR^{5,1}$ and corresponding to a combination of O3-plane at a point and an O5-plane wrapped on the cigar geometry, with opposite signs so as to reproduce the sign flips of the original O4-planes across the NS5-branes. The configuration also includes flavor branes, which are subleading in $1/N$ and can be included in a probe approximation. We will skip their detailed discussion here and in the following examples.

Let us also quickly consider the O6-plane realization, where the orientifold flips the directions 456. This corresponds to the type IIA background (\ref{Liouville-k=2}) with an orientifold action flipping the directions 45 in $\IR^{5,1}$, acting as $\varphi \to 2 \pi - \varphi$ on $U(1)_\varphi$, and leaving $\IR_\rho$ invariant. The action on the superfield $\varphi$ looks different to that in the O4-plane realization. However, both realizations in fact lead to the same orientifold of \eqref{Liouville-k=2} upon taking into account the $\IZ_2$ orbifold $\varphi \to \varphi + 2\pi$. It is satisfactory to see how two different orientifold Hanany-Witten realizations of the same SCFT lead to the same worldsheet theory in the double-scaling limit. Morally, the difference between the two realizations differ in their behaviour on fields which get gapped.

For the theory II.4, the brane configuration is a quotient of that of theory II.1 by introducing and O6'-plane, flipping the directions 689. After the double-scaling limit, this corresponds to a type IIA background (\ref{Liouville-k=2}) with an orientifold action leaving invariant both $\IR^{5,1}$ and $\IR_\rho$, while acting as $\varphi \to 2 \pi -\varphi$ on $U(1)_\varphi$, or equivalently as $\varphi \to -\varphi$ on $U(1)_\varphi/\IZ_2$. Note that this action on the internal CFT is as in the O4-plane or O6-plane cases above, but differs from it in the action on $\IR^{5,1}$. Performing again a T-duality, the type IIB background we obtain is a quotient of (\ref{cigar-ii}) by an orientifold action leaving $\IR^{5,1}$ invariant, and acting on the internal coordinates as the previous O3/O5-plane. Namely, we obtain a combination of an O5-plane along $\IR^{5,1}$ and an O7-plane along the $\IR^{5,1}$ times the cigar geometry. The different dimensionality of the orientifold planes explains that (upon introducing the D3/D5-branes and their backreaction) there is only 4d $\NN=1$ supersymmetry.

Finally, for the theory II.5, the brane configuration is a quotient of that of theory II.1 by a $\IZ_2$ orbifold flipping all the directions 4589 and an O8-plane orientifold flipping the direction 7. After the double-scaling limit, this corresponds to a type IIA background (\ref{Liouville-k=2}) with a $\IZ_2$ orbifold and an orientifold. The generator of the orbifold group acts by flipping the directions 45 in $\IR^{5,1}$. The orientifold acts as $\varphi \to -\varphi$ on the $U(1)_\varphi$ factor, while leaving invariant both $\IR^{5,1}$ and the radial component $\IR_\rho$. Since the orbifold does not act on $\varphi$, it stays an orbifold after T-dualizing. The type IIB background is thus a $\IZ_2$ orbifold of (\ref{cigar-ii}) with an extra orientifold action leaving $\IR^{5,1}$ invariant, and corresponding to a combination of an O7- and an O5-plane in the cigar geometry. 

Note that this construction is different from previous orientifolds, because its orientifold group is $\IZ_2\times \IZ_2$, with an orientifold action $\Omega R$ (with $R$ acting geometrically as $\varphi\to -\varphi$) and an orbifold action (generated by $\Theta:(x_4,x_5)\to -(x_4,x_5)$). This is in contrast with the previous configurations, where the only quotient is a $\IZ_2$ orientifold action. We also note that, due to the underlying symmmetry of the NS5/D4-brane system under rotations in 789, it would have been similarly possible to choose the O8-plane transverse to e.g. the direction 9, and the orbifold acting on 4578. In that case, after the limit the orientifold would be just $\Omega$, i.e. with trivial geometric action, also on the direction $\varphi$ (still defined in terms on the angle in the 2-plane 67), because the direction 9 is `gapped'. On the other hand, the orbifold would be acting as $\Theta':(x_4,x_5,\varphi)\to -(x_4,x_5,\varphi)$. One may fear that this would produce a different model. However, there is no problem, because the full orientifold group would be generated by $\Omega$ and $\Theta'$, and is equivalent to the group generated by $\Omega R$ and $\Theta$ (because $\Theta'=R\Theta$). It is very satisfactory that the models obtained via diferent choices of actions related by a symmetry in the initial configuration end up corresponding to the same construction in the limit.

\subsection{Class III}
\label{sec:strings-classIII}

In this section we describe the tensionless string limit of Class III theories and argue that it is related to a string probing the background of 3 NS5-branes becoming coincident in the decoupling limit. We first consider a parent configuration in the absence of additional orientifolds or orbifolds, and introduce those ingredients subsequently to describe the limit of all Class III SCFTs.

\subsubsection{The limit from a 3 NS5-brane parent theory}
\label{sec:strings-classIII-limit}

Class III brings a new interesting feature we would like to highlight. As emphasized in the introduction, our general perspective is that the infinite-distance limits of large classes of CFTs arising from a parent one (via orientifold or orbifold quotients, etc) fall in the same universality class in the sense that their bulk dual is controlled by the same tensionless string. For the above cases of Class I and Class II theories, the underlying parent theory belongs to the corresponding class itself, namely theories I.1 and II.1. For Class III theories, in contrast, there is no theory within Class III which can act as parent theory for the others. This however does not contradict our general perspective, and simply means that one should apply it to the full set of SCFTs which can be built using brane configurations. Indeed, the parent theory of the Class III models does not lie within Class III, but rather corresponds to a theory with two gauge factors within a general class of models considered in \cite{Calderon-Infante:2026rkj}. This is an interesting demonstration that the classification of limits of SCFTs in universality classes should be regarded as a statement in the whole set of SCFTs (at least, within those realized using brane configurations).

In fact, one can already see from the construction of Class III theories in section \ref{sec:classIII} in terms of type IIA brane configurations, that they are all closely related (mainly by orientifold and/or orbifold quotients) to an underlying parent theory, which we now discuss\footnote{The Class III theories in section \ref{sec:classIII-45678} have the special feature of involving `glued' NS5-branes with only 4d $\NN=1$, and require some modifications of the discussion. Still, as we later argue, their infinite-distance limit is still related to this same 4d $\NN=2$ parent theory, so we skip this subtlety momentarily. \label{foot:kutasov}} (see appendix \ref{app:basic} for the basic rules). The underlying theory is a 4d $\NN=2$ linear quiver SCFT realized in a type IIA configuration with 3 NS5-branes along 012345 and stacks of $N$ D4-branes along 0123 and stretched in the two intervals among them in 6, and with $N$ semi-infinite D4-branes at the ends, extended in the non-compact direction 6. Explicitly, the 4d $\NN=2$ symmetry groups and hypermultiplet content is 
\beqa
&[SU(N)]\times SU(N)_1\times SU(N)_2\times [SU(N)] &\nonumber\\
& N(\antifund_1,1)+(\fund_1,\antifund_2)+N(1,\fund_2)\, , &
\label{classIII-parent}
\eeqa
where as usual the factors in brackets correspond to global symmetries. This 4d $\NN=2$ SCFT corresponds to the $k=3$ case of (\ref{linear-quiver}), and as recently studied in \cite{Calderon-Infante:2026rkj} and reviewed below, admits infinite-distance limits in its conformal manifold passing through overall weak coupling. The theory is not listed in Table \ref{table:SCFTs} because it has two gauge factors; however, upon diverse operations e.g. orientifold projections, it projects down to a gauge theory with a single gauge factor and reproduces theories in Class III. Therefore it is interesting to identify the infinite-distance limit in this parent theory that corresponds to the infinite-distance limits of Class III theories, so as to identify their common tensionless string in the bulk. As we will see, this common string is the one arising from the double-scaling limit in which 3 NS5-branes become coincident. Furthermore, the infinite distance limit in the SCFT maps to the weak string coupling limit of this string theory.

It is easy to carry out a discussion of this limit in terms of an interpolating model, in analogy with the discussion in \cite{Gadde:2009dj}, reviewed in section \ref{sec:gpr}. In the same way that the latter provides the emergent string limit of all Class II theories, see section \ref{sec:strings-classII}, the following discussion provides the emergent string limit of all Class III theories. The interpolating theory we consider is obtained by a type IIA elliptic model with 3 NS5-branes and $N$ D4-branes in each of the 3 intervals they define. The 4d $\NN=2$ gauge group and hypermultiplet content is 
\beqa
&SU(N)_0\times SU(N)_1\times SU(N)_2 &\nonumber\\
& (\fund_0,\antifund_1,1)+(1,\fund_1,\antifund_2)+(\antifund_0,1,\fund_2) &
\label{classIII-parent-interpolating}
\eeqa
This theory is the $k=3$ elliptic model (\ref{elliptic-N2}). It can be regarded as obtained from (\ref{classIII-parent}) by gauging the diagonal subgroup of the two $SU(N)$ global symmetries; conversely, the theory (\ref{classIII-parent}) is recovered by turning the gauge factor $SU(N_0)$ into a global one.

The type IIA realization of the interpolating model with a compact direction 6 allows the construction of a T-dual type IIB realization. It is given by a stack of $N$ D3-branes at a $\IC^2/\IZ_3\times\IC$ singularity. The D3-branes are in the regular representation of $\IZ_3$, i.e. there are no fractional branes, matching the fact that the number of D4-branes is the same in all the intervals in the type IIA picture. The type IIB realization leads easily to a holographic dual realization of the interpolating model (\ref{classIII-parent-interpolating}), given by type IIB theory on AdS$_5\times\IS^5/\IZ_3$, where the $\IZ_3$ generator acts as $\theta:(z_1,z_2,z_3)\to (e^{2\pi i/3}z_1,e^{-2\pi i/3}z_2,z_3)$ on the $\IS^5$ defined as a unit ball in $\IC^3$, $|z_1^2+|z_2|^2+|z_3|^2=1$. As emphasized in \cite{Kachru:1998ys,Hanany:1998it,Gukov:1998kk}, there is an $\IS^1\subset \IS^5$ of $\IC^2/\IZ_3$ singularities, which leads to additional twisted fields. They are associated to two independent $\IP_1$'s collapsed to zero size at the local $A_2$ singularity (whose affine 3-node Dynkin diagram is an explicit depiction of the 3 intervals between the 3 NS5-branes in the type IIA T-dual). In particular, the NSNS 2-form backgrounds control the ratios of the individual gauge couplings with the overall one, fixed by the type IIB dilaton. More explicitly, we have

\beqa
&& \frac{1}{g_0^2}+\frac{1}{g_1^2}+\frac{1}{g_2^2}=\frac{1}{2\pi g_s} \, ,\nonumber \\
&& \frac{g_1^2}{g_0^2}=\frac{\beta_1}{1-\beta_1-\beta_2}\quad , \quad \frac{g_2^2}{g_0^2}=\frac{\beta_2}{1-\beta_1-\beta_2} \, ,
\eeqa
where $\beta_1$ and $\beta_2$ correspond to the period of the NSNS 2-form over the above mentioned two independent $\IP_1$'s.

We are interested in the limit $g_0\to 0$, while keeping the couplings $g_1$, $g_2$ finite. As in section \ref{sec:gpr}, this can be conveniently studied in the type IIA T-dual, where the relation among parameters is
\beqa
&&\frac{1}{g_1^{\,2}}=\frac{\beta_1 R}{2\pi g_s^A l_s}\quad ,\quad \frac{1}{g_2^{\,2}}=\frac{\beta_2 R}{2\pi g_s^A l_s}\quad ,\quad \frac{1}{g_0^{\,2}}=\frac{(1-\beta_1-\beta_2) R}{2\pi g_s^A l_s} \, .
\label{eq:couplings-k=3}
\eeqa
As explained above, the gauge theory \eqref{classIII-parent} is recovered in the limit of sending $g_0\to 0$, while keeping $g_1,g_2$ finite, and upon decoupling the free sector. In other words, one of the gauge groups is turned into a global one.

The NS5-branes end up arranged in a line with a separation between consecutive ones given by
\beqa
\tau_1=2\pi\beta_1 R \quad ,\quad \tau_2=2\pi \beta_2 R \, ,
\label{NS5-distances-k=3}
\eeqa
namely, a 3 NS5-branes version of (\ref{doublescalingLST}). Analogously, the above weak-coupling limit corresponds to a double-scaling limit which sends this distance and the IIA string coupling $g_s^A$ simultaneously to zero, so that $g_0\to 0$ while keeping fixed the individual gauge couplings $g_1, g_2$ in (\ref{eq:couplings-k=3})
\beqa
\tau_1,\tau_2\to 0\quad ,\quad g_s^A\to 0 \quad ,\quad g_1,g_2={\rm fixed}
\label{double-scaling-k=3}
\eeqa

Moreover, in order to reproduce Class III theories with two gauge factors identified by orientifold and/or orbifold actions, we are interested in restricting to $g_1=g_2$, so we can set $\beta_1=\beta_2\equiv \beta$, hence $\tau_1=\tau_2\equiv \tau$. On the other hand, for Class III theories with glued NS5-branes, we have e.g. $\beta_1\equiv 0$, $\beta_2\equiv \beta$, hence $\tau_1\equiv 0$, $\tau_2\equiv\tau$. The Class III configurations are specified by a single effective parameter $\tau$.

The above brane construction allows to provide the bulk interpretation of the infinite-distance limit of Class III theories. Their overall weak-coupling limit corresponds to the tensionless string limit of the above stringy background, whose worldsheet we describe more explicitely in the next subsection. Thanks to the interpolating model, the string can be argued to be `made' of the fundamental Type IIB string and the D3-branes wrapped on the two collapsed 2-cycles in AdS$_5\times\IS^5/\IZ_3$. We recall that the extra ingredients (e.g. the orientifold/orbifold identifications) reduce these two 2-cycles to effectively an independent one in the quotient. This situation is analogous to the $k=2$ case reviewed in section \ref{sec:gpr}. In the following section we discuss additional information regarding the worldsheet theory of the emergent tensionless string.

\subsubsection{The ``flat space'' worldsheet description of the tensionless string} 
\label{ss:worldsheet}

The discussion of the limit in the brane configuration is close in spirit to that in \cite{Gadde:2009dj}, reviewed in section \ref{sec:gpr}, while the construction of the ``flat space'' string worldsheet theory brings in several novel aspects. 

Let us start discussing the limit of the brane configuration. Ignoring the color D4-branes for the time being, the background for the corresponding emergent string in our case is given by that of 3 NS5-branes in a scaling limit in which they all become coincident. 
The configurations of $k$ coincident type IIA NS5-branes has been considered e.g. in \cite{Aharony:1998ub,Giveon:1999zm,Giveon:1999px,Giveon:1999tq} from different viewpoints, and is described by the CHS worldsheet theory \cite{Callan:1991at}
\beqa
\IR^{5,1} \times \IR_\rho \times SU(2)_k \, ,
\label{wzw-iii}
\eeqa
where $SU(2)_k$ is a supersymmetric level-$k$ WZW model, and $\IR_\rho$ describes the radial coordinate $\rho$ with a linear dilaton $\Phi=-\rho/\sqrt{2k}$. However, to take the double-scaling limit in \eqref{double-scaling-k=3} we first need to consider a deformation of this background for which the NS5-branes are separated. These worldsheet deformations can be understood from the holographic duality between this string theory with linear dilaton background and the worldvolume theory on the NS5-branes \cite{Aharony:1998ub}. From the viewpoint of the latter, the positions of the NS5-branes are controlled by the VEV of four worldvolume scalars. As in AdS/CFT holographic dualities, giving non-zero VEVs to these scalars corresponds to turning on normalizable modes in the bulk, which in turn corresponds to deforming the worldsheet CFT with the corresponding normalizable vertex operator. While referring to \cite{Aharony:2003vk,Aharony:2004xn} for more details, we will now introduce the main ingredients of this duality for the case we are interested in.

To describe the vertex operators of \eqref{wzw-iii}, it is more convenient to use the equivalent description in which WZW model is replaced by
\begin{equation} \label{chs-other-frame}
    \IR^{5,1}\times \IR_\rho \times \left(U(1)_\varphi \times \frac{SU(2)_k}{U(1)} \right)/\IZ_k \, .
\end{equation}
The $U(1)_\varphi$ factor describes a circle of radius $\sqrt{2k}$, as fixed by the action of the $\IZ_k$ that we will discuss momentarily. As the notation suggests, it involves a chiral field $\varphi$ whose bottom component parametrizes the circle, i.e., $\varphi \sim \varphi + 2 \pi \sqrt{2k}$.\footnote{As usual, we are denoting the superfield and its bottom component by the same symbol.} The $SU(2)_k/U(1)$ theory is a $\mathcal N=2$ minimal model that can be described as the IR fixed point of a Landau-Ginzburg model with a chiral superfield $\chi$ and superpotential $W = \chi^k$. The $\IZ_k$ acts as a rotation by an angle of $2 \pi/k$ on $U(1)_\varphi$ (i.e. as $\varphi \to \varphi + 2 \pi \sqrt{2k}/k$) and as $\chi \to \exp(2\pi i/k) \chi$ on the $SU(2)_k/U(1)$ factor. As described in \cite{Aharony:2003vk} (see also \cite{Fotopoulos:2007rm}), a general distribution of NS5-branes on a complex plane is described by deforming the worldsheet Lagrangian by
\begin{equation} \label{deformations}
    \delta \mathcal L = \sum_{j} \lambda_j \int d^2 \theta \, \chi^{k-2(j+1)} e^{-\sqrt{\frac{2}{k}} (j+1) (\rho - i \varphi)} + c.c. \, ,
\end{equation}
where $d^2 \theta$ denotes integration over half of superspace and $j=0,1/2,1,\ldots,(k-2)/2$. Notice that, for generic $j$, the vertex operator in the integral is well-defined thanks to the $\IZ_k$ orbifold. As advanced above, the couplings $\lambda_j$ are related to the VEV of a complex scalar $A$, morally in the adjoint representation of $SU(k)$, which in general can be written as 
\beqa
\left< A \right> = \text{diag}(a_1,\cdots,a_k) \quad {\rm with} \quad \sum_n a_n=0\,. 
\label{traceless-positions}
\eeqa
Taking string units and ignoring an overall normalization factor, we have
\begin{equation} \label{map-distances-couplings}
    \lambda_j \sim \left< \Tr (A^{2j+2}) \right> + \cdots \, ,
\end{equation}
where the ellipsis represents multi-trace terms (see \cite{Aharony:2003vk,Aharony:2004xn}). Notice that these extra terms are absent for $j=0,1/2$ by virtue of $\Tr(A)=0$. 

Before going on, let us recall the most studied deformation of this type and that leads to the double-scaling limit in \cite{Giveon:1999px,Giveon:1999tq}. It corresponds to placing the NS5-branes homogeneously around a circle of radius $R$, such that $\left< \Tr (A^{2j+2}) \right>=0$ for $j<(k-2)/2$. Thanks to this highly symmetric arrangement, only the $j=(k-2)/2$ deformation in \eqref{deformations} is turned on:\footnote{For the same reason, there are no multi-trace contributions to $\lambda_j$ with $j=(k-2)/2$.}
\begin{equation}
    \delta \mathcal L = \mu \int d^2 \theta \, e^{-\sqrt{\frac{k}{2}} (\rho - i \varphi)} + c.c. \, , \quad \mu \sim R^k \, .
\end{equation}
We see that this deformation corresponds to turning on a $\mathcal N=2$ Liouville superpotential on the cylinder parametrized by $\rho-i \varphi$. Upon T-dualizing this theory to the $SL(2,{\bf R})_k/U(1)$ Kazama-Suzuki model \cite{Hori:2001ax}, we get the oftentimes discussed worldsheet theory
\begin{equation} \label{circle-theory}
    \IR^{5,1}\times \left(\frac{SL(2,{\bf R})_k}{U(1)} \times \frac{SU(2)_k}{U(1)} \right)\mathlarger{\mathlarger{\mathlarger{/}}}\IZ_k \, ,
\end{equation}
This reproduces \eqref{cigar-ii} for $k=2$ as the factor $SU(2)_2/U(1)$ disappears and no longer contributes to the central charge.\footnote{The total central charges vanishes (see e.g. \cite{Dei:2024frl}),
\beq
c_{\rm tot}=c_{\rm matter}+c_{\rm ghosts}+c\left[\frac{SL(2,{\bf R})_k}{U(1)} \right]+c\left[\frac{SU(2)_k}{U(1)} \right]=0
\eeq
with $c\left[\frac{SL(2,{\bf R})_k}{U(1)} \right]=3\left(1-\frac2k\right)$ and $c\left[\frac{SU(2)_k}{U(1)} \right]=3\left(1-\frac2k\right)$. Hence, for $k=2$, the last contribution vanishes.}

\medskip

For our purposes, we now focus on the case of $k=3$ NS5-branes distributed along a line, since this is the configuration relevant for the Class III SCFTs. Notice that this differs from the highly symmetric arrangement that yields \eqref{circle-theory}, as we need the NS5-branes to be distributed along a line rather than homogeneously around the circle\footnote{For $k=2$ both arrangements are equivalent, and that is why \eqref{circle-theory} provides the accurate worldsheet description for the closed string spectrum dual to Class II SCFTs. However, this is no longer the case for $k\geq 3$.}. Following our previous notation (\ref{NS5-distances-k=3}) with NS5-branes located at positions $-\tau_1,0,\tau_2$ (with $\tau_1,\tau_2\geq 0$), we can shift the origin by $(\tau_1-\tau_2)/3$ and put it in the form of the above traceless combination (\ref{traceless-positions}) as
\beqa
a_1=-\frac 23\tau_1 - \frac 13 \tau_2\;,\; a_2=\frac 13\tau_1-\frac 13 \tau_2\; ,\; a_3=\frac 13 \tau_1+\frac 23 \tau_2
\eeqa
or equivalently
\begin{equation}
    \left< A \right> = \tau \, \text{diag}(\gamma_1,\gamma_2,-\gamma_1-\gamma_2) \, ,
\end{equation}
with $\tau\equiv \tau_1+\tau_2$ the distance between the outmost NS5-branes, and $\gamma_i=a_i/\tau$.

Hence, we are interested in the theory in \eqref{chs-other-frame} deformed by \eqref{deformations} with
\begin{equation}
    \lambda_0 \sim \tau^2 \left( \gamma_1^2 + \gamma_2^2 + (\gamma_1+\gamma_2)^2 \right) \, , \quad \lambda_{1/2} \sim \tau^3 \left( \gamma_1^3 + \gamma_2^3 - (\gamma_1+\gamma_2)^3 \right) \, .
\end{equation}
Let us focus on the two different arrangements that appeared in our brane realization of Class III theories. The first one corresponds to 3 equally spaced NS5-branes, namely $\tau_1=\tau_2$ or $\gamma_1=-1/2$ and $\gamma_2=0$. This leads to $\lambda_{1/2}=0$, so that the deformation simplifies to
\begin{equation} \label{eq:Liouville-like}
    \delta \mathcal L \sim \tau^2 \left( \int d^2\theta \, \chi \, e^{-\sqrt{\frac{2}{3}} \, (\rho - i \varphi)} + c.c \right) \, .
\end{equation}
Notice that this deformation turns on a superpotential similar to that in $\mathcal N=2$ superLiouville theory, but with the crucial difference that it also couples the cylinder theory parametrized by $\rho-i \varphi$ to the $\mathcal N=2$ minimal model containing the chiral field $\chi$. For this reason, the deformed theory will no longer take the nice form of a direct product as happens for the worldsheet describing the NS5-branes at a point \eqref{chs-other-frame} or homogeneously distributed along a circle \eqref{circle-theory}. The second type of arrangement we found corresponds to two of the three NS5-branes being glued, namely $\tau_1=0$ or $\gamma_1=\gamma_2=-1/3$ above. In this case, both $\lambda_0 \sim \tau^2$ and $\lambda_{1/2} \sim \tau^3$ are turned on in \eqref{deformations}.

The discussion above provides a complete description of the worldsheet CFT. Something to note is that, unlike in the case of NS5s distributed homogeneously around a circle, the deformation in \eqref{eq:Liouville-like} couples the cylinder theory with the linear dilaton background to the $\mathcal N=2$ minimal model. To the best of our knowledge, the effect of this type of Liouville-like superpotential has not been considered in the literature, and we will not consider it further here. We will also not address the question of T-dualizing to a type IIB description, which is not expected to take the simple form of the product of two coset CFTs as in \eqref{circle-theory}. We leave these interesting open questions for future work.

Notice that, in general, the worldsheet theory with $\lambda_0 \sim \tau^2$ and $\lambda_{1/2} \sim \tau^3$ is invariant under $\rho \to \rho + c$, $g_s \equiv \exp \Phi \to \exp(c / \sqrt{6}) g_s$ and $\tau \to \exp(c/ \sqrt{6}) R$. Hence, the string perturbative expansion is controlled by the combination $g_{\rm eff}=g_s/\tau$, which is precisely the ratio kept fixed in the double-scaling limit (\ref{double-scaling-k=3}) (see \cite{Giveon:1999px} for the analogous statement in the $k=2$ case). In fact, this ratio also controls the gauge couplings $g_1 \sim g_2$, with $\gamma_1$ and $\gamma_2$ fixed, of the parent SCFT (c.f. \eqref{eq:couplings-k=3}). As advertised, the infinite-distance limit associated to the overall weak-coupling limit of the dual SCFT corresponds to taking $g_{\rm eff} \to 0$ in the worldsheet, namely the tensionless string limit.

It is worth emphasizing that the two possible NS5-brane geometries we have considered actually correspond to the {\em same worldsheet CFT} up to deformations by marginal couplings, described by the 2d superpotential deformations above. This is analogous to the familiar situation of emergent strings in infinite distance limits in CY compactifications, where the actual worldsheet theory depends on marginal couplings controlled by the CY moduli. We adhere to this notion of equivalent worldsheet CFTs in our claim that all Class III theories share a common emergent string in their infinite distance limits.

\medskip

The above configuration describes the worldsheet theory of the emergent string in the infinite-distance limit of the parent theory. As anticipated, this also controls the corresponding limit of all Class III SCFT. This follows from our general argument that the extra ingredients necessary to turn the parent theory into the Class III ones (orientifold and orbifold quotients, inclusion of flavor branes, and inclusion of `gluing' of NS5-branes) do not change the local dynamics on the worldsheet. Furthermore, they only lead to modifications to the correlators of operators that survive the projection that are subleading in the $1/N$ expansion, that corresponds to the weak string coupling limit in the bulk. It is possible to translate the effect of these operations in the Type IIA brane configuration to gaugings in the worldsheet theory described above, much in the spirit of section \ref{sec:strings-classII} for Class II theories. A key difference is however that we lack a simple Type IIB dual worldsheet theory in this Class III case, which complicates the story. This important open question is left for future work.

\medskip

This concludes our discussion of the emergent tensionless strings for all large $N$ SCFTs with simple gauge group. We again highlight the important lessons from their brane constructions: the overall weak-coupling limits of these SCFTs fall in universality classes; these classes are characterized by the corresponding infinite distance limits in some parent theories, of the kinds considered in \cite{Calderon-Infante:2026rkj}; and the corresponding emergent tensionless strings are associated to the number of NS5-branes becoming coincident in the double-scaling limit. From our general perspective explained in the introduction, these lessons tested in detail in the case of simple gauge group SCFTs have far more general applicability, to much large class of SCFTs with multiple gauge factors, or even limits of non-exactly conformal theories, as we explore in the next section.

\section{Hitchhiking Beyond: A Guide to Generalizations}
\label{sec:generalizations}

In this section we come back to the general perspective motivating this work, and argue that the brane cooking toolkit that we have introduced can be applied in a far more general manner. After considering interpolating models, we discuss how our previous results on universality classes and tensionless string limits in the bulk generalize to superconformal gauge theories with more than one gauge factor and to theories with simple gauge group and vanishing 1-loop beta function but without a conformal manifold. 

\subsection{Network of SCFTs and interpolating models}
\label{sec:network}

\begin{figure}[htb]
\begin{center}
\includegraphics[scale=.41]{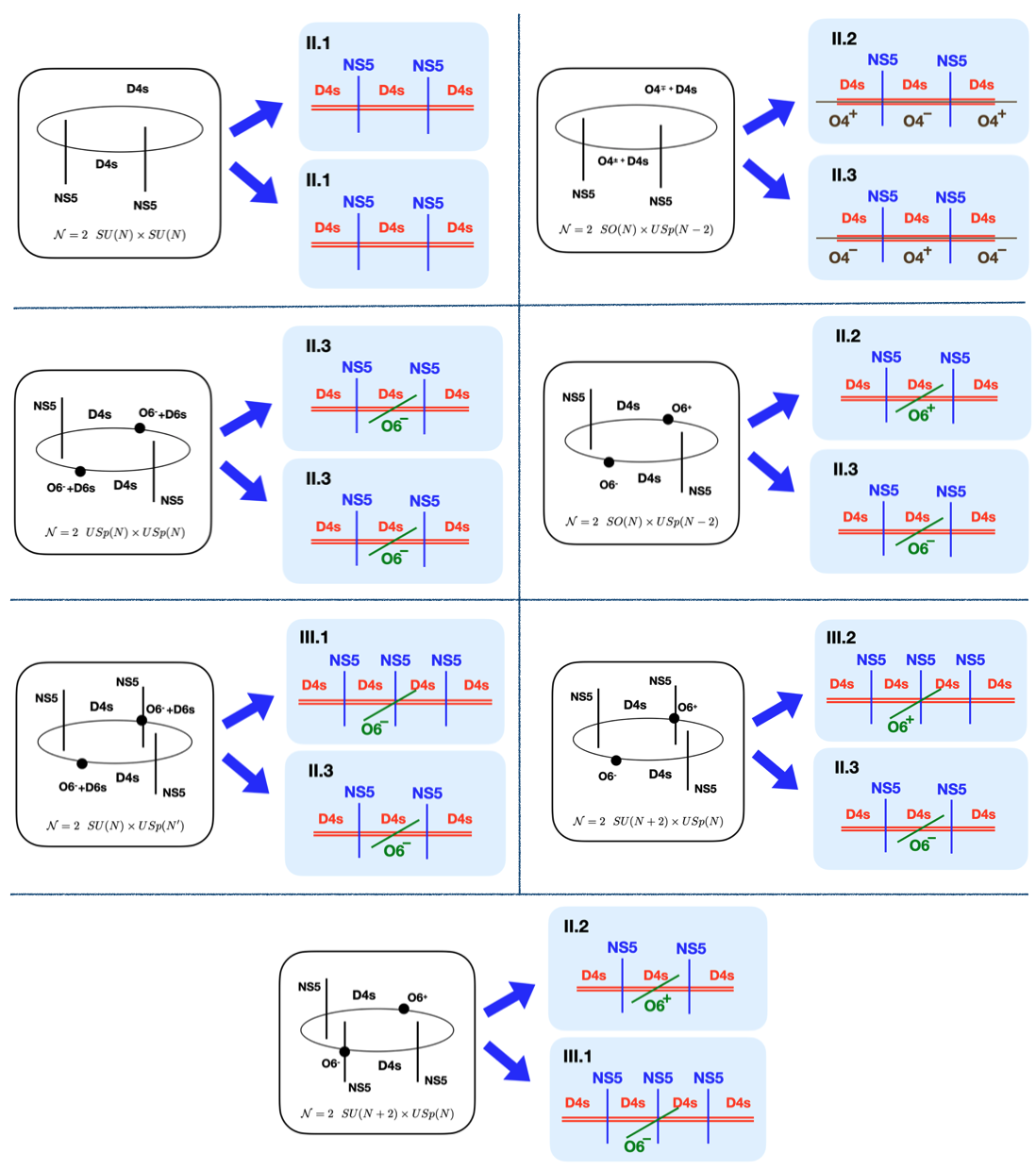}
\caption{\small Interpolating models for $\mathcal{N}=2$ SCFTs whose Hanany-Witten brane model contains two or three NS5-branes.}
\label{fig:interpolating}
\end{center}
\end{figure}

We begin by highlighting the connections between the SCFTs studied in this paper, which will pave the way to posterior generalizations. When constructing the brane configuration dual to each of the SCFTs with simple gauge groups of Table \ref{table:SCFTs}, we have seen that they can also be recovered from taking a partial weak-coupling limit of a larger SCFT with more than one gauge factor.\footnote{As stressed in section \ref{sec:wo-interpolating}, we recall that, strictly speaking, the partial weak-coupling limit of the interpolating model recovers the flavor singlet of the SCFT of interest.} These weak-coupling limits are associated to exactly marginal deformations in the case of $\mathcal{N}=2$ theories, although are in general not marginal for $\mathcal{N}=1$ theories. These interpolating models therefore yield a network of SCFTs related by marginal and RG flow deformations. They were discussed in detail for each SCFT in Section \ref{sec:constructions}, but we summarize those relating $\mathcal{N}=2$ SCFTs and containing two and three NS5-branes in their Hanany-Witten realization in Figure \ref{fig:interpolating}.

\subsubsection{Interpolating Models and the CFT Distance Conjecture}

As shown for several illustrative examples in Figure \ref{fig:interpolating}, starting from SCFTs with two gauge factors, one always flows to one of the SCFTs listed in Table \ref{table:SCFTs} upon taking a weak-coupling limit. This limit is implemented by sending one of the gauge couplings to zero while keeping the other fixed, and subsequently restricting to the interacting degrees of freedom while decoupling those that become free. Since the original theory involves two gauge couplings, this procedure naturally establishes a connection between two possibly distinct SCFTs with simple gauge groups. From the perspective of the brane configuration, taking a gauge coupling to zero while keeping the others fixed corresponds to two NS5's approaching each other while going to weak string coupling, in the double-scaling limit explained in Section \ref{sec:strings}. For the interpolating models in Figure \ref{fig:interpolating}, the mobile NS5's can approach each other on either side of the circle, yielding therefore two SCFTs (with a single gauge factor) in each case. In particular, the first interpolating model in Figure \ref{fig:interpolating} reproduces that of \cite{Gadde:2009dj} summarized in Section \ref{sec:gpr}. 

As the number of coincident NS5's (and therefore, the nature of the emergent tensionless string) can be different when taking different limits, the resulting SCFTs can belong to different universality classes, and also differ from the original one. Nevertheless, it is interesting to notice that the exponential rate of the HS tower  in the initial and resulting theory will be related as follows. Consider the original theory to have gauge group $G_1\times G_2$ and central charge $c$. When taking either $g_1\to$ or $g_2\to 0$, there will be a tower of HS operators whose anomalous dimension decreases exponentially in the Zamolodchikov distance with exponential rate given respectively by (see \eqref{eq:alpha-value}),
\beq
\alpha_{i}=\sqrt{\frac{2c}{\textrm{dim}\, G_i}} \ ,\quad i=1,2 \, .
\eeq
For concreteness, let us consider first the limit $g_1\to 0$. The resulting interacting theory will have a gauge group $G_2$ with central charge $c_2< c$. When taking the remaining gauge coupling $g_2\to 0$ there will be again a HS tower becoming conserved with exponential rate
\beq
\alpha^{(1)}_{2}=\sqrt{\frac{2c_2}{\textrm{dim}\, G_2}} \, .
\eeq
Notice that this tower is the same than the one arising in the original theory when sending $g_2\to 0$, but now $\alpha^{(1)}_{2}$ is the exponential rate measured in the new theory resulting upon taking the first weak-coupling limit $g_1\to 0$ (with central charge $c_2$) while $\alpha_{2}$ was measured in the original theory (with central charge $c$). Both exponential rates are related by
\beq
\alpha^{(1)}_{2}=\alpha_2\sqrt{\frac{c_2}{c}} \, ,
\label{alpha12}
\eeq
so the exponential rate always decreases when measured in the new theory, as the latter always has a smaller central charge since some degrees of freedom have been decoupled. For instance, in the interpolating model of \cite{Gadde:2009dj} to obtain SQCD, one starts with the quiver $SU(N)\times SU(N)$ with $c=N^2/2$ and $\textrm{dim}\, G_1=N^2$, which yields $\alpha_2=1$. Upon taking $g_1\to 0$ and decoupling the free sector one recovers $\mathcal{N}=2$ SQCD with $c=N^2/3$, yielding $\alpha^{(1)}_{2}=\sqrt{2/3}$ using \eqref{alpha12}, which matches with the exponential rate associated to this theory (see Table \ref{table:SCFTs}).
Hence, in this type of interpolating models, the smallest possible values of the exponential rates are precisely those classified for gauge theories with simple gauge factors in \cite{Razamat:2020pra}, namely $\alpha=1/\sqrt{2},\sqrt{2/3},\sqrt{7/12}$. Interestingly, even if $\alpha=1/\sqrt{2}$ is the smallest possible one, it never appears as the resulting theory of an interpolating model. We believe that this is related to the fact that $\alpha=1/\sqrt{2}$ is associated to holographic Einstein-gravity bulk duals, and it seems pathologic to decouple a sector of an Einstein-gravity theory while keeping another interacting Einstein-like theory. Whenever taking a decoupling limit, the resulting theory always seems to keep some light higher-spin fields (with mass of order AdS) that makes the theory non-Einstein (see e.g. \cite{Mantegazza:2026spd}).

In general, the analogous of the taxonomy rules studied in flat space \cite{Etheredge:2024tok} for how to combine different towers in the same moduli space are quite trivial in conformal manifolds, as the HS vectors are orthogonal \cite{Calderon-Infante:2024oed,Calderon-Infante:2026rkj}. However, it would be interesting to explore whether the above relations between the exponential rates can be leveraged to derive some rules constraining the network of SCFTs related by marginal and other RG flow deformations.

Clearly, these interpolating models are not unique since there might be several ways to recover the same theory. We could also imagine starting with a model including a larger number of gauge factors (and NS5-branes) so that it allows for many different weak-coupling limits; connecting in this way many more SCFTs. The exponential rates will still be related by the ratio of the central charges, as in \eqref{alpha12}. This naturally brings us to the generalization studied in the next subsection.

\subsubsection{Interpolating models and Hanany-Witten configurations}

Given the key role played by brane constructions in this work, we would now like to discuss how the various limits in the conformal manifold of a gauge theory with several gauge factors are reflected in this setup. A large class of these theories is easily engineered by a generalization of the previously discussed type of Hanany-Witten models with larger numbers of NS5-branes. In fact, we have already encountered examples of this kind. In section \ref{sec:strings-classIII-limit} we considered the configuration with 3 NS5-branes and stacks of $N$ D4-branes suspended in the different intervals, leading to the theories (\ref{classIII-parent}) in the linear quiver case (non-compact direction 6) and (\ref{classIII-parent-interpolating}) in the elliptic model case (compact direction 6). As we will see next, the latter has an overall weak-coupling limit of type I, and furthermore serves as interpolating models for theories with limits of type II and III.

\begin{figure}[htb]
\begin{center}
\includegraphics[scale=.35]{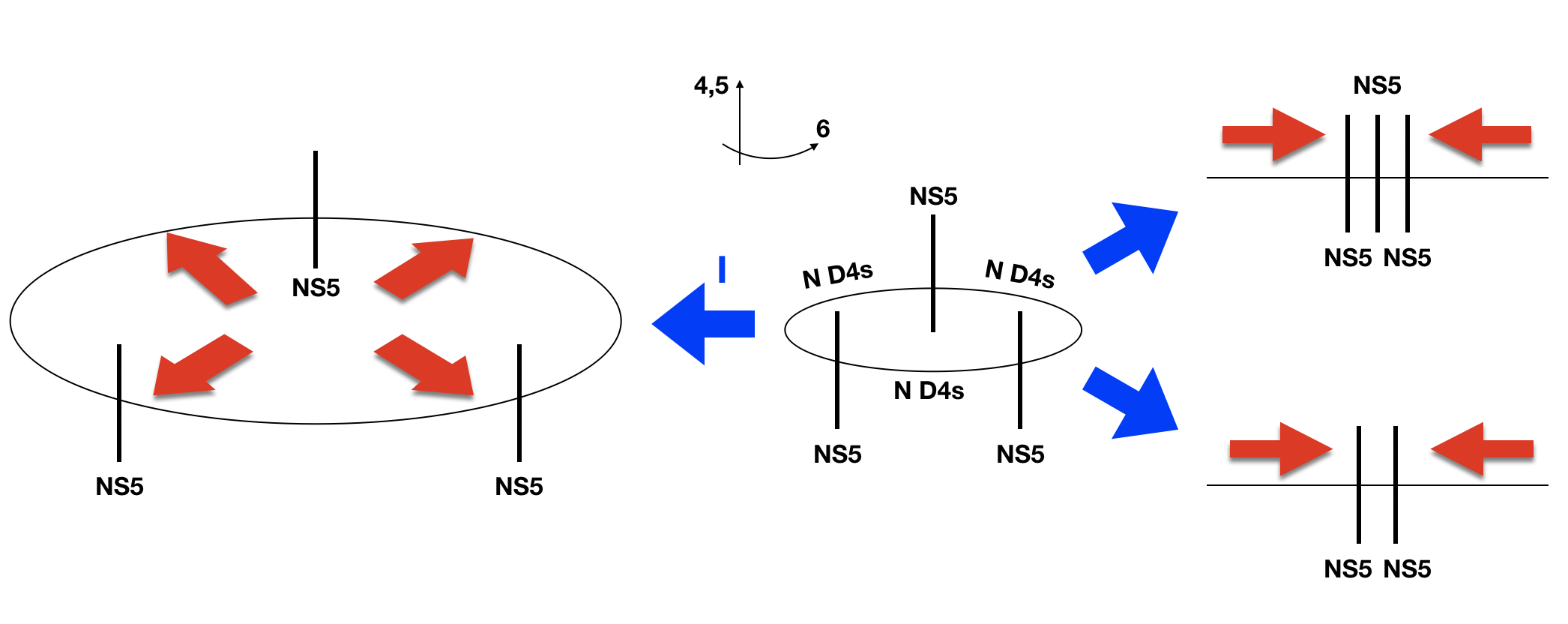}
\caption{\small Depiction of the brane configuration for the interpolating model (\ref{classIII-parent-interpolating}) and its three types of limits which, as explained in the main text, are related to theories of class I, II and III, respectively. The scale of these figures is set by $g_s l_s$, which is implicitly being sent to zero to implement the double-scaling limit. 
}
\label{fig:3ns5-limits}
\end{center}
\end{figure}

The brane model of interest is depicted in the central panel in Figure \ref{fig:3ns5-limits}. In the double-scaling limit in which the radius of the direction 6, $R$, the string length, $l_s$, and the string coupling, $g_s$, are sent to zero with the ratio $R/(g_s l_s)$ and the relative distances between NS5-branes kept fixed, this setup recovers the $\mathcal N=2$ circular gauge quiver with gauge group $SU(N)^3$. As it is already familiar, the squared gauge coupling of each gauge factor is proportional to $g_s l_s$ divided by the distance between adjacent NS5-branes. In what follows, we would like to combine the double-scaling limit above with a further limit in the conformal manifold of this gauge theory in which one, two or three of the gauge factors are decoupled. How these different limits affect the brane construction is depicted in Figure \ref{fig:3ns5-limits}. We warn the reader that all distances in this figure are measured relative to $g_s l_s$, which is being sent to zero implicitly to implement the double-scaling limit in which the gauge theory is isolated from the rest of the string theory spectrum. In what follows, we discuss the interpretation of each of these three infinite-distance limits in the conformal manifold:

\begin{itemize}

\item Overall weak-coupling limit: This corresponds to sending the three gauge couplings to zero while keeping their ratios fixed. This corresponds to a Type I limit; the large-$N$ Hagedorn temperature of this theory in this limit is the same as that of $\mathcal N=4$ SYM \cite{Calderon-Infante:2026rkj}. As depicted in the left part of Figure \ref{fig:3ns5-limits}, this limit is achieved by sending $R\to\infty$ in units of $g_s l_s$ (which we again recall is being sent to zero in the double-scaling limit) in the Hanany-Witten configuration. It is however more illuminating to consider the Type IIB dual picture, that leads to the weak string coupling limit of the critical 10d type IIB theory on AdS$_5\times \IS^5/\IZ_3$, as befits a Type I limit.

\item Decoupling two gauge factors: We can also send two of the gauge couplings to zero, while keeping the other one fixed. After getting rid of the decoupled sector, we end up with the flavor singlet sector of $\mathcal N=2$ SCQCD, which has a weak-coupling limit of Type II in its conformal manifold. In the Hanany-Witten setup, this limit corresponds to keeping fixed the distance between a pair of NS5-branes in units of $g_s l_s$ while the radius $R$ is sent to infinity and the remaining NS5-brane goes off to infinity. As depicted in upper right part of Figure \ref{fig:3ns5-limits}, this effectively recovers the double-scaling limit of the Hanany-Witten setup that engineers $\mathcal N=2$ SCQCD. In fact, this is a simple generalization of how this theory is recovered from the interpolating model in \cite{Gadde:2009dj}.

\item Decoupling one gauge factor: Finally, we can consider sending only one gauge coupling to zero while keeping the other two fixed. This limit was discussed in detail in section \ref{sec:strings-classIII-limit}. It leads to the $SU(N)^2$ linear quiver gauge theory whose overall weak-coupling limit is of Type III and that plays the role of parent theory for all the gauge theories with simple gauge factor in Class III. In analogy to the previous case, this leads to the double-scaling limit in which 3 NS5-branes are brought together (see the lower right part of Figure \ref{fig:3ns5-limits}).

\end{itemize}

It is amusing to recover a type I overall weak-coupling limit as well as partial weak-coupling limits leading to theories with type II and III limits from a single interpolating model, furthermore illustrating the powerful intuitions from the brane configuration.

We note that the introduction of extra ingredients such as orientifold planes in this model relates the gauge couplings of different factors and prevents the realization of some of the limits of the parent theory in the descendant theory (relatedly, these extra ingredients in general change the number of gauge factors in the theory, as already explained in section \ref{sec:strings-classIII-limit}). On the other hand, we do emphasize that all the infinite-distance limits of the descendant theory can be realized in terms of the parent theory, and therefore fall in the same universality class. This is precisely what happens with theories of Class III, whose tensionless string limit corresponds to 3 NS5-branes becoming coincident, and therefore all of them lie in the same universality class of the similar limit in the parent theory. 

The previous discussion generalizes straightforwardly to an arbitrary number of NS5-branes. We can focus on a general elliptic model given by $k$ NS5-branes with stacks of $N$ D4-branes on the $k$ intervals they define, with no D6-branes. This system leads to the gauge theory (\ref{elliptic-N2}) with equal ranks for all the gauge factors $N_i\equiv N$. This is a 4d $\NN=2$ SCFT which has $k$ marginal couplings, corresponding to the gauge couplings for the $k$ gauge factors. We may also consider the linear quiver theory SCFT obtained by ungauging one of the factors, namely a brane configuration with non-compact direction 6, $k$ NS5-branes with $N$ D4-branes suspended in the $(k-1)$ intervals, as well as two sets of semi-infinite D4-branes at the ends.\footnote{It is straightfoward to construct other SCFTs with the same quiver structure of gauge factors, but possibly non-equal ranks, compensated by extra D6-branes. All these possibilities generically involve modifications that are subleading in $1/N$, so they do not modify the discussion of the infinite-distance limit structure, and we will not consider them explicitly, and refer the reader to \cite{Calderon-Infante:2026rkj} for more details on those cases.}

\begin{figure}[htb]
\begin{center}
\includegraphics[scale=.42]{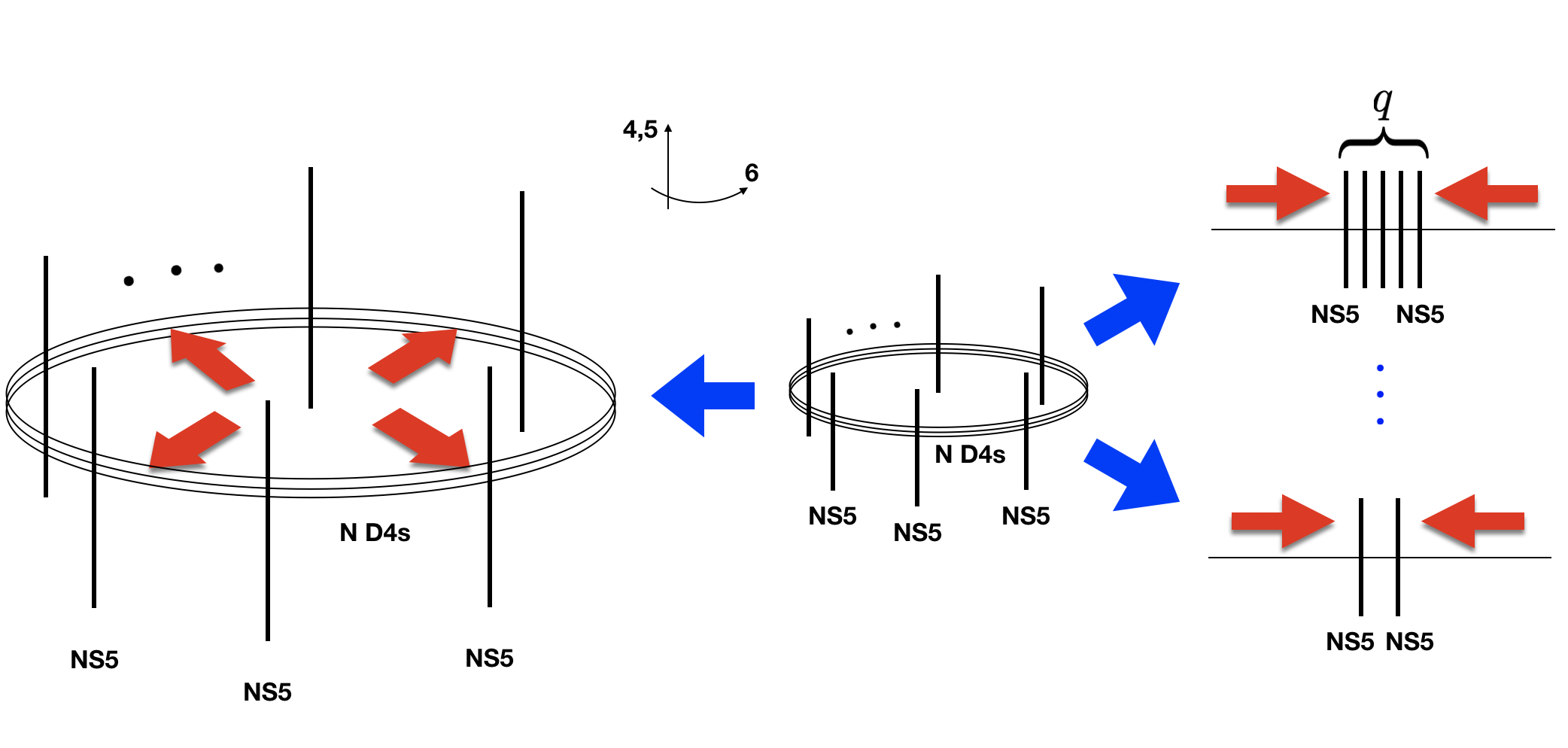}
\caption{\small The brane configuration of the elliptic model with $k$ NS5-branes and $k$ stacks of $N$ D4-branes suspended among them (center of the image) admits several limits, including the overall weak-coupling limit (left part), and limits in which $q$ NS5-branes become coincident (right part) for $q=2,\ldots, k$. The distances in this figures are measured in units of $g_s l_s$, which we remind is sent to zero in the double-scaling limit in which the SCFT is decoupled from the rest of the string theory spectrum. 
}
\label{fig:general-k}
\end{center}
\end{figure}

Let us consider starting from the elliptic model, see the center part of Figure \ref{fig:general-k}. The behavior of different infinite-distance limits, and a convex hull formulation for their combination, has been recently studied in \cite{Calderon-Infante:2026rkj} from the SCFT perspective. It is straightforward to see how these limits are translated to the brane picture. In the overall weak-coupling limit, the radius of the direction 6 is send to infinity in units of $g_s l_s$ while keeping the relative NS5-brane positions fixed (left part of Figure \ref{fig:general-k}). On the other hand, partial weak-coupling limits are such that $q$ of these NS5-branes (for $q=2,\ldots,k$) are kept at finite distance (again in units of $g_s l_s$) while the remaining ones go off to infinity sending $R/(g_s l_s)$ to infinity (see right part of Figure  \ref{fig:general-k}). This leads to the double-scaling limit in which $q$ NS5-branes are brought together. We will discuss this type of limit and its associated ``flat space'' worldsheet in the next section.

\subsection{Universality classes for SCFTs with multiple gauge factors\label{sec:multiple}}

After having explained how theories with more than one gauge factor lead to network of SCFTs connected by partial decoupling limits and how this is reflected in the brane constructions, let us now focus on what the latter can tell us about the appearance of universality classes and bulk tensionless strings in the overall weak-coupling regime. In a simple generalization of our results for theories with simple gauge group, the type of tensionless string emerging in the bulk is controlled by the number of coincident NS5-branes in the double-scaling limit. This provides a string-theoretic explanation to the universality classes discussed in \cite{Calderon-Infante:2026rkj} and characterized by the large-$N$ Hagedorn temperature. Within each universality class, various SCFTs as those discussed in \cite{Calderon-Infante:2026rkj} play the role of parent theories for several others. Theories in this finer classification are also guaranteed to share the value of the exponential rate of the HS currents in the overall weak-coupling limit. This paves the way towards an ultimate general classification, in the spirit of that for emergent string limits in CY compactification. In this section, we briefly sketch some of these generalizations and interesting open directions for future research.

\subsubsection{Worldsheet theory in the double-scaling limit of $k$ NS5-branes} \label{ss:worldsheet-general}

We start by considering our basic Hanany-Witten configuration with $k$ NS5-branes spanning the directions 012345 and separated along a non-compact direction 6. Adding $N_i$ D4-branes along 01236 stretched between the $i$-th and $(i+1)$-th NS5-branes and $K_i$ D6-branes along 0123789 and placed in the interval between the $i$-th and $(i+1)$-th NS5-branes, this setup engineers a 4d $\mathcal N=2$ gauge theory with gauge group $\prod_i SU(N_i)$, with one hypermultiplet in the bifundamental representation of each pair of adjacent $SU(N_i)$, and with $K_i$ hypermultiplets in the fundamental representation of $SU(N_i)$. More precisely, this gauge theory is isolated from the rest of the system in the double-scaling limit
\begin{equation} \label{eq:decoupling-limit-general}
\begin{split}
	g_s \to 0 \, , \quad \tau_i\to 0\, , \quad l_s \to 0 \, , \\
	\text{with} \quad \frac{1}{g_{i}^2} \sim \frac{\tau_i}{g_s l_s} \quad \text{fixed} \, ,
\end{split}
\end{equation}
where $\tau_i$ is the distance between the $i$-th and $(i+1)$-th NS5-branes along the direction 6, and the second line ensures that the gauge couplings $g_i$ remain constant in the limit. In the overall weak-coupling regime, the large-$N_i$ Hagedorn temperature of these $\mathcal N=2$ linear quivers only depends on $k$, i.e., on the number of NS5-branes \cite{Calderon-Infante:2026rkj}. A straightforward generalization of the results we developed for theories with simple gauge group above also provides a string-theoretic understanding for the appearance of these universality classes.

As in the previously studied class II and III cases, by removing the D-branes and studying the backreaction of the NS5-branes we can obtain the ``flat space'' worldsheet for the bulk string that emerges in the overall weak-coupling limits of the gauge theory. In this generalized setup, we are led to studying the near-horizon geometry of $k$ parallel NS5-branes separated along a line in the double-scaling limit above. In fact, the resulting worldsheet theory was already introduced in section \ref{ss:worldsheet}. It is described by \eqref{chs-other-frame} supplemented by the superpotential deformations in \eqref{deformations} that implement the separations of the NS5-branes along the direction 6. We will not attempt to write down the mapping between the separation of the NS5-branes $\tau_i$ and the worldsheet couplings $\lambda_j$, which turns out to be more involved for $k \geq 3$ due to the multi-traces contributions represented by the ellipsis in \eqref{map-distances-couplings}. Generically, all deformations in \eqref{deformations} will be turned on, while for specially symmetric arrangements some of the couplings $\lambda_j$ may vanish. As we did in the $k=3$ case in section \ref{ss:worldsheet}, using $\tau$ to denote the distance between the outermost NS5-branes, then equation \eqref{map-distances-couplings} leads to the power-law scaling $\lambda_j \sim \tau^{2j+2}$. Similarly to the $k=2$ and $k=3$ case discussed in \cite{Giveon:1999px} and section \ref{ss:worldsheet} respectively, the effective string coupling is controlled by $g_s/\tau$, which is precisely the combination kept fixed in the double-scaling limit in \eqref{eq:decoupling-limit-general} (recall that in the worldsheet we are taking string units). This effective string coupling is related to the overall gauge coupling of the SCFT in such a way that, as expected, the overall weak-coupling infinite-distance limit is mapped to the tensionless and weakly-coupled string limit in the bulk.

The discussion in the previous paragraph is only sensible to the number of NS5-branes in the Hanany-Witten setup, but not to distribution of D-branes that controls the ranks of the gauge group and the number of fundamental hypermultiplets. This fits with the result in \cite{Calderon-Infante:2026rkj} regarding the large-$N$ Hagedorn temperature of this type of theories, that suggested that they share the same type of bulk tensionless string in the overall weak-coupling limit. Indeed, even though the particular AdS$_5$ background generated by the backreaction of the D-branes will be different, we do find that the ``flat space'' worldsheet coincides for all theories sharing the same Hagedorn temperature.

Our brane model perspective also sheds some light on another observation made in \cite{Calderon-Infante:2026rkj}, namely that there are theories with more than one gauge factor that share the same value of $T_H$ but not of the exponential rate $\alpha$. As we explained in our detailed study of the cases with simple gauge group, that the large $N$ value of $\alpha$ coincides between the different theories in a given universality class is guaranteed by the fact that they all descend from a single parent theory. This however is no longer the case when we allow for semi-simple gauge group. For instance, focusing on the case with three NS5s---or two gauge factors---there are infinitely many distributions of D-branes corresponding to SCFTs, each of them leading to a different value of $\alpha$ and serving as parent theory for many others SCFTs.\footnote{As discussed in section 4.2 of \cite{Calderon-Infante:2026rkj}, any pair of positive integers $N_1$ and $N_2$ satisfying $1 \leq N_1/N_2 \leq 2$ leads to a superconformal gauge quiver with gauge group $SU(N_1) \times SU(N_2)$, a hypermultiplet in the bifundamental representation, and $K_1=2N_1-N_2$ and $K_2=2N_2-N_1$ hypermultiplets in the fundamental representation of $SU(N_1)$ and $SU(N_2)$, respectively. Using the equations therein, one can also see that the large $N$  value of the exponential rate $\alpha$ in the overall weak-coupling limit is controlled by the ratio $N_1/N_2$.} In the next section, we take a particularly simple distribution of D-branes in a setup with $k$ NS5-branes to illustrate how it plays the role of parent theory upon introducing orientifold planes.

\subsubsection{Hitchhiking with extra orientifolds: the case study of 4d $\NN=2$ SCFT}
\label{sec:multiple-orientifolds}

As we have explained, our brane realization approach explains that there are large classes of SCFTs whose overall weak-coupling infinite-distance limits fall into universality classes that reflect the type of string becoming tensionless in the bulk. Furthermore, a subset of theories within a universality class also share the value of the exponential rate of the HS currents, $\alpha$, which is guaranteed when they descend from a single parent theory. This paves the way to, starting from the theories in \cite{Calderon-Infante:2026rkj} as parents, provide a finer classification of larger classes of 4d $\NN=2,1$ SCFTs with infinite-distance limits with the same features. Although a full classification is beyond the scope of the present work (because of its richness, and because of lack of guidance given the absence of general classifications of SCFTs from the field theory perspective), we now illustrate the idea by restricting to 4d $\NN=2$ theories. 

As a case study, we start with the set of parent theories given by a brane configuration with non-compact direction 6, $k$ NS5-branes and $N$ D4-branes in all intervals and the semi-infinite endpoints, and consider the modifications producing new classes of $\NN=2$ SCFTs. The bulk tensionless string emerging in the overall weak-coupling limit of these SCFTs and the large-$N$ value of the exponential rate $\alpha$ are the same one as that of the parent theory by construction. The restriction to $\NN=2$ is motivated because it greatly simplifies the possible modifications which can be introduced in the parent theories, in particular we may only introduce O4- or O6-planes. It also simplifies the realization of the conformality conditions, which amount to RR charge cancellation conditions, as explained in section \ref{sec:constructions}.

\subsubsection*{$\NN=2$ theories with O4-planes}

Let us consider starting with the addition of an O4-plane to parent configurations of NS5- and D4-branes, i.e. the SCFTs in \cite{Calderon-Infante:2026rkj}, using the rules in appendix \ref{sec:o4-planes} (see \cite{Landsteiner:1997vd} for original reference). As explained, we focus on the case with no flavor D6-branes. For the case of linear quivers, the resulting theories are condensed in the expression (\ref{O4-linear}), where we recall that the alternating structure of $SO/USp$ gauge factors is due to the orientifold plane charges changing sign when they cross an NS5-branes. Making (\ref{O4-linear}) more explicit, and particularizing to the conformal case, when the number of NS5-branes in the parent theory is even $k=2p$, there are two possible classes of theories, according to whether the semi-infinite orientifold planes are O4$^-$- or O4$^+$-planes. The theories respectively have the $\NN=2$ gauge groups and hypermultiplet contents
\beqa
{\rm O4}^-:\; & [SO(N+2)]_0\times USp(N)_1 \times\ldots \times SO(N+2)_{2p-2}\times USp(N)_{2p-1}\times [SO(N+2)]_{2p}&\nonumber\\
& \frac 12 (\fund_0,\fund_1)+\frac 12(\fund_1,\fund_2)+\ldots +\frac 12(\fund_{2p-2},\fund_{2p-1})+\frac 12 (\fund_{2p-1},\fund_{2p})&\\
{\rm O4}^+:\; & [USp(N)]_0\times SO(N+2)_1 \times\ldots \times USp(N)_{2p-2}\times SO(N+2)_{2p-1}\times [USp(N)]_{2p}&\nonumber\\
& \frac 12 (\fund_0,\fund_1)+\frac 12(\fund_1,\fund_2)+\ldots +\frac 12(\fund_{2p-2},\fund_{2p-1})+\frac 12 (\fund_{2p-1},\fund_{2p})&
\eeqa
where squared brackets indicate global symmetries, as usual.
On the other hand, for odd number of NS5-branes $k=2p-1$, there is a single kind of theory with $\NN=2$ gauge groups and hypermultiplet content
\beqa
& [SO(N+2)]_0\times USp(N)_1 \times\ldots \times SO(N+2)_{2p-2}\times [USp(N)]_{2p-1}&\nonumber\\
& \frac 12 (\fund_0,\fund_1)+\frac 12(\fund_1,\fund_2)+\ldots +\frac 12 (\fund_{2p-3},\fund_{2p-2})+\frac 12 (\fund_{2p-2},\fund_{2p-1})&
\eeqa
The models are easily embedded into elliptic interpolating models, by closing the linear quiver into a circular one. In the even $k=2p$ case, this simply requires gauging the diagonal global symmetry of the endpoints, whereas in the $k=2p+1$ case it requires adding one extra NS5-branes. The resulting theories have the structure (\ref{O4-elliptic}).

By our general arguments, the bulk string becoming tensionless in overall weak-coupling limits of these theories is inherited by the parent theory. That is, its ``flat space'' worldsheet theory is an orientifold of that in section \ref{ss:worldsheet-general}. Additionally, these theories admit other types of infinite-distance limits that lead to the decoupling of part of the theory. Hence, they serve as interpolating models for theories with other types of overall weak-coupling limits whose bulk tensionless string is related to the configuration of $q<k$ NS5-branes becoming coincident in the orientifolded brane configuration. A similar comment applies to the constructions below. 

\subsubsection*{$\NN=2$ theories with O6-planes}

Let us consider the addition of O6-planes to parent configurations of NS5- and D4-branes, i.e. the SCFTs in \cite{Calderon-Infante:2026rkj}, using the rules in appendix \ref{sec:o6-planes} (see \cite{Landsteiner:1997ei} for original reference), again for the case with no flavor D6-branes. For the case of linear quivers, we again must distinguish between the case of even or odd number $k$ of NS5-branes. In the case of even $k=2p$, the theories are of the kind in (\ref{one-O6-ns5-away}). Recall that the NS5-branes arrange in $p$ pairs symmetric with respect to the origin in the direction 6, where the O6-plane sits. The middle interval produces an $SO$ or $USp$ symmetry for the case of O6$^+$- or O6$^-$-plane, respectively. Focusing on the conformal case, the resulting theories have the $\NN=2$ gauge groups and hypermultiplet contents
\beqa
{\rm O6}^+ :\;& [SU(N-2p)_0]\times SU(N-2(p-1))_1\times \ldots \times SU(N-2)_{p-1}\times SO(N)_p &\nonumber\\
& (\fund_0,\antifund_1)+\ldots+(\fund_{p-2},\antifund_{p-1})+(\fund_{p-1},\fund_p) &\label{linearo6plus-wons}\\
{\rm O6}^-:\; & [SU(N+2p)_0]\times SU(N+2(p-1))_1\times \ldots \times SU(N+2)_{p-1}\times USp(N)_p &\nonumber\\
& (\fund_0,\antifund_1)+\ldots+(\fund_{p-2},\antifund_{p-1})+(\fund_{p-1},\fund_p) & \label{linearo6minus-wons}
\eeqa
Considering now the case of odd $k=2p+1$, the theories are of the kind in (\ref{one-O6-ns5-ontop}). Recall that there must be one NS5-branes stuck at the O6-plane, while the remaining ones organize in $p$ symmetric pairs. The O6-plane does not map any interval onto itself, so all gauge factors are unitary groups, but it maps one bifundamental of the parent theory to itself, so it is projected down to a 2-index tensor in a representation determined by the O6-plane sign. In the conformal case, the resulting theories have the $\NN=2$ gauge groups and hypermultiplet contents
\beqa
{\rm O6}^+:\; & [SU(N_0)]\times SU(N_1)\times \ldots \times SU(N_{p}) &\nonumber \\
& (\fund_0,\antifund_1)+\ldots+(\fund_{p-1},\fund_p)+ \Ysymm_p & \label{linearo6plus-wns}\\
{\rm O6}^-:\; & [SU(N_0)]\times SU(N_1)\times \ldots \times SU(N_{p}) &\nonumber \\
& (\fund_0,\antifund_1)+\ldots+(\fund_{p-1},\fund_p)+ \Yasymm_p & \label{linearo6minus-wns}
\eeqa
These linear theories can be easily completed to elliptic interpolating models, morally by combining pairs of such theories (see \cite{Uranga:1998uj} for original reference). We note that, when making the direction 6 compact, the configuration contains two O6-planes, whose sign can be chosen in different ways. If they have different sign, the resulting theories are combination of (\ref{linearo6plus-wons}) with (\ref{linearo6minus-wons}), (\ref{linearo6plus-wons}) with (\ref{linearo6minus-wns}), or (\ref{linearo6minus-wons}) with (\ref{linearo6plus-wns}). On the other hand, if both are O6$^-$-planes, the theories must necessarily contain 8 D6-branes to cancel the RR orientifold charges, so we resort to the general forms (\ref{one-O6-ns5-away}) or (\ref{one-O6-ns5-ontop}), with coefficients fixed by conformality (equivalently, equal linking numbers for all NS5-branes). We refer the reader to \cite{Uranga:1998uj} for a more explicit discussion of these specific examples.

Again, from our general arguments, the infinite-distance limits in which all gauge couplings become small are in the same universality class and those of the parent theory. That is, the ``flat space'' worldsheet for the bulk string corresponds becoming tensionless in this limit is given by an orientifold of that discussed in section \ref{ss:worldsheet-general}. Furthemore, other infinite-distance limits in which some gauge couplings remain fixed reveal that these theories again serve as interpolating models for other SCFTs whose overall weak-coupling limits are related to $q<k$ NS5-branes becoming coincident (possibly with an orientifold action). 

\medskip

This discussion illustrates a particular instance of our main point. Namely, that upon the introduction of orientifolds (or other modifications in more general cases), one can get large classes of SCFTs with conformal manifolds passing through weak coupling and leading to the same type of bulk tensionless string as in the parent theory. We hope that this suffices to give a flavor of the potential of using brane configurations and the techniques we have developed to construct new large classes of SCFTs with larger number of gauge factors, and to explain that their tensionless string limits fall in universality classes controlled by the number of NS5-branes becoming coincident in the double-scaling limit. We leave the systematic characterization of these interesting generalizations in the more involved $\NN=1$ case for future work.

\subsection{Other large $N$ SCFTs / Quasiconformal theories}
\label{sec:quasiconformal}

The class of theories we have considered so far are exactly conformal SCFTs admitting a large $N$ limit, which provide a well-defined setup to address the implications of the Distance Conjecture in an exactly AdS gravity dual. The limits we have considered correspond to infinite distance points at weak coupling, where we have shown that they are classified by universality classes according to the resulting tensionless string in the bulk.

In this section we would like to start the exploration of an interesting (and conceptually non-trivial) extension of the above story. We consider 4d supersymmetric gauge theories with simple gauge group, admitting a large $N$ limit, and which have vanishing 1-loop beta function but are not exactly conformal theories. One motivation to consider this class of theories is that the one-to-one correspondence between the exponential rate of the HS tower and the large-$N$ Hagedorn temperature found in \cite{Calderon-Infante:2024oed} also holds for them, even though they are not exactly conformal away from the free point. We refer to such theories as quasiconformal theories (or QSCFTs). The picture is that the theories can be regarded as perturbations of an UV fixed point defined by the free theory, which subsequently have a non-trivial RG flow in the IR. Correspondingly, one may regard the putative gravity duals of these theories as gravitational backgrounds whose asymptotics (but not their interior) are morally the same kind of AdS$_5$ backgrounds arising in the weak coupling infinite-distance limits of exact SCFTs. In other words, we may expect that the universality classes of emergent strings in the infinite-distance limits in SCFTs may also include the limiting behaviour of large classes of quasiconformal theories. 

In order to test these ideas, we explore several illustrative examples. The results suggest that there is a very rich set of possibilities, including QSCFTs whose weak coupling regime falls in universality classes of the kind discussed in \cite{Calderon-Infante:2024oed}, c.f. sections \ref{sec:constructions} and \ref{sec:strings}, but also other examples which correspond to new universality classes different from those in \cite{Calderon-Infante:2024oed} and \cite{Calderon-Infante:2026rkj}. These could correspond to new kinds of emergent tensionless strings, which may arise in limits of other SCFT universality classes. A systematic exploration is beyond the scope of this work and left for future research.

\subsubsection{QSCFTs limits in known universality classes}
\label{sec:QSCFTs-easy}

Let us begin by considering large $N$ QSCFTs whose weak-coupling limits fall in the universality classes we have already studied. For concreteness, we focus on cases with simple gauge group, so the resulting limits should be those of class I, II or III theories. By our general arguments, these can be easily built by applying our model building rules (orientifold and/or orbifold quotients, addition of flavor branes, etc) to the brane configurations realizing the parent theories of class I, II, III, such that the brane configuration has cancellation of RR charges. As already explained, this guarantees the cancellation of the 1-loop beta function, but the result in general is not an exactly conformal theory, hence results in a QSCFT. By our general arguments, the type of bulk string becoming tensionless in the weak-coupling limit of the QSCFT will be teh same as that in the underlying parent theory.

It is straightforward to build examples following this procedure. In fact we have already encountered some of these examples in our discussion in previous sections. For instance, in the attempt to construct an interpolating model for theory II.4, we ended up considering the theory (\ref{interpolating-ii4}). In the limit of decoupling the $SO$ gauge factor, we obtain a 4d $\NN=1$ theory with gauge group and matter content (upon redefining the rank notation)
\beqa
& USp(N) &\nonumber \\
& \Yasymm+(2N+8)&\fund
\eeqa
As mentioned there, this has vanishing one-loop beta function, but is not an exactly superconformal theory. In fact, it becomes conformal only in the strict large $N$ limit. This is manifest using the brane construction, where the theory arises as a set of $N$ D4 branes suspended between 2 NS5-branes, in the presence of an O6'$^-$-plane. This is just an orientifold quotient of the brane construction of the II.1 theory, the parent theory of all Class II theories. 

By the familiar arguments in section \ref{sec:strings}, the orientifold does not change the local dynamics of the bulk ``flat space'' worldsheet, which is thus identical to that of type II.1 theory. Hence, even though the gauge coupling is not exactly marginal, the behavior of the theory in the weak-coupling limit leads to the same bulk tensionless string as the theories in class II. It would be interesting to provide a precise formulation of this kind of tensionless string limits for non exactly conformal theories, a topic which we leave for future research.

\subsubsection{QSCFTs in new universality classes}
\label{sec:QSCFTs-hard}

There is a classification of theories with simple gauge group and vanishing 1-loop beta function \cite{Razamat:2020pra}. It is easy to realize that some of them admit a possible embedding in brane construction using the techniques reviewed in Appendix \ref{app:brane-cooking}, including the familiar type IIA brane configurations or more general brane tiling techniques \cite{Hanany:2005ve,Kennaway:2007tq}. In the following we provide brane constructions of some illustrative examples for some of these QSCFTs. These constructions show there exist sets of theories which are related by operations such as orientifolding, addition of flavor branes, etc, hence defining universality classes. They moreover show that these universality classes in general are not associated to parallel NS5-branes becoming coincident, and therefore the possible tensionless strings are different from those in class I, II, II theories in this work, namely in \cite{Calderon-Infante:2024oed}, and from those in the more general class in \cite{Calderon-Infante:2026rkj}.

\begin{figure}[htb]
\begin{center}
\includegraphics[scale=.3]{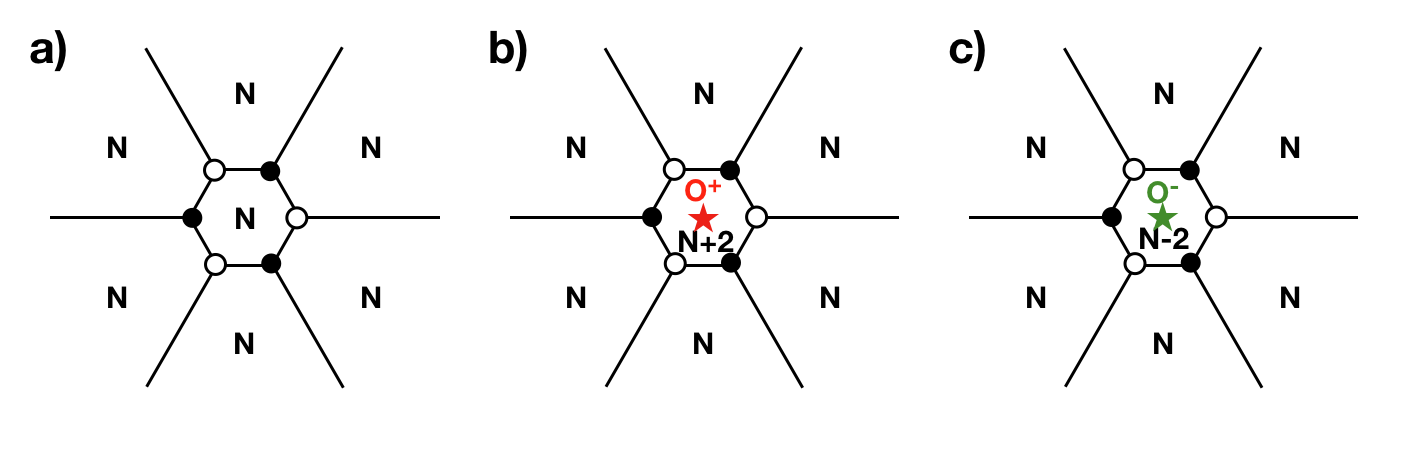}
\caption{\small The brane tilings of $\IR^2$ leading to a class of QSCFTs. Figure a) shows the parent theory, while figures b) and c) correspond to orientifolded versions.}
\label{fig:1face}
\end{center}
\end{figure}

Consider the 4d $\NN=1$ theory with the following gauge symmetry and matter content
\beqa
&SU(N)&\nonumber \\
& 3N\,(\fund+\antifund)
\label{3nfund}
\eeqa
This theory has vanishing 1-loop beta function but it is not exactly conformal. It is easy to realize this gauge theory in terms of a brane tiling. In fact, it corresponds to a particular decoupling limit of the brane tiling associated to D3-branes at a $\IC^3/(\IZ_2\times\IZ_2)$ singularity, described in section \ref{sec:orbifold-nodt} (see Figure \ref{fig:tiling-z2z2}a). The gauge theory is given in (\ref{z2z2}). Consider now taking the decoupling limit of three of the gauge factors, and sending the overall coupling to zero, while keeping the gauge coupling of one gauge factor fixed. This is analogous to turning an elliptic interpolating model into a linear one in the context of IIA brane configurations, c.f. section \ref{sec:gpr}. In the brane tiling picture, this corresponds to decompactifying the $\IT^2$ tiled by the dimer graph, sending the area of three of the facets to infinity, while keeping the area of one face finite. The result is a tiling of $\IR^2$ with non-compact faces, except for a single finite one, see Figure \ref{fig:1face}a. 

We should note that the brane tiling model is more than a powerful graph technique: it really corresponds to an actual IIB configuration of D5-branes spanning the faces, suspended between NS5-branes which span the edges, hence providing a realization of the gauge theory. The corresponding ``flat space'' worldsheet for the bulk theory can be morally obtained by analogy with the IIA brane configurations, namely as the near-horizon geometry generated by the backreaction of the NS5-branes in Figure \ref{fig:1face}a (without the D5-branes) in the appropriate double-scaling limit in which the NS5-branes bounding the finite size face become coincident (and the face formally collapses to zero size).

This description of the limit seems to be very different from those of parallel IIA NS5-branes becoming coincident, encountered in previous SCFTs, suggesting it may lead to a new kind of emergent tensionless string in the gravity dual. However, the brane tiling picture is very different from the type IIA Hanany-Witten descriptions, so a direct comparison is difficult. Therefore, in order to provide a direct comparison, we will now describe a translation of the $\IC^3/(\IZ_2\times\IZ_2)$ interpolating model to a type IIA Hanany-Witten construction, and describe the limit of interest in that picture, showing it indeed corresponds to a new universality class.

\begin{figure}[htb]
\begin{center}
\includegraphics[scale=.3]{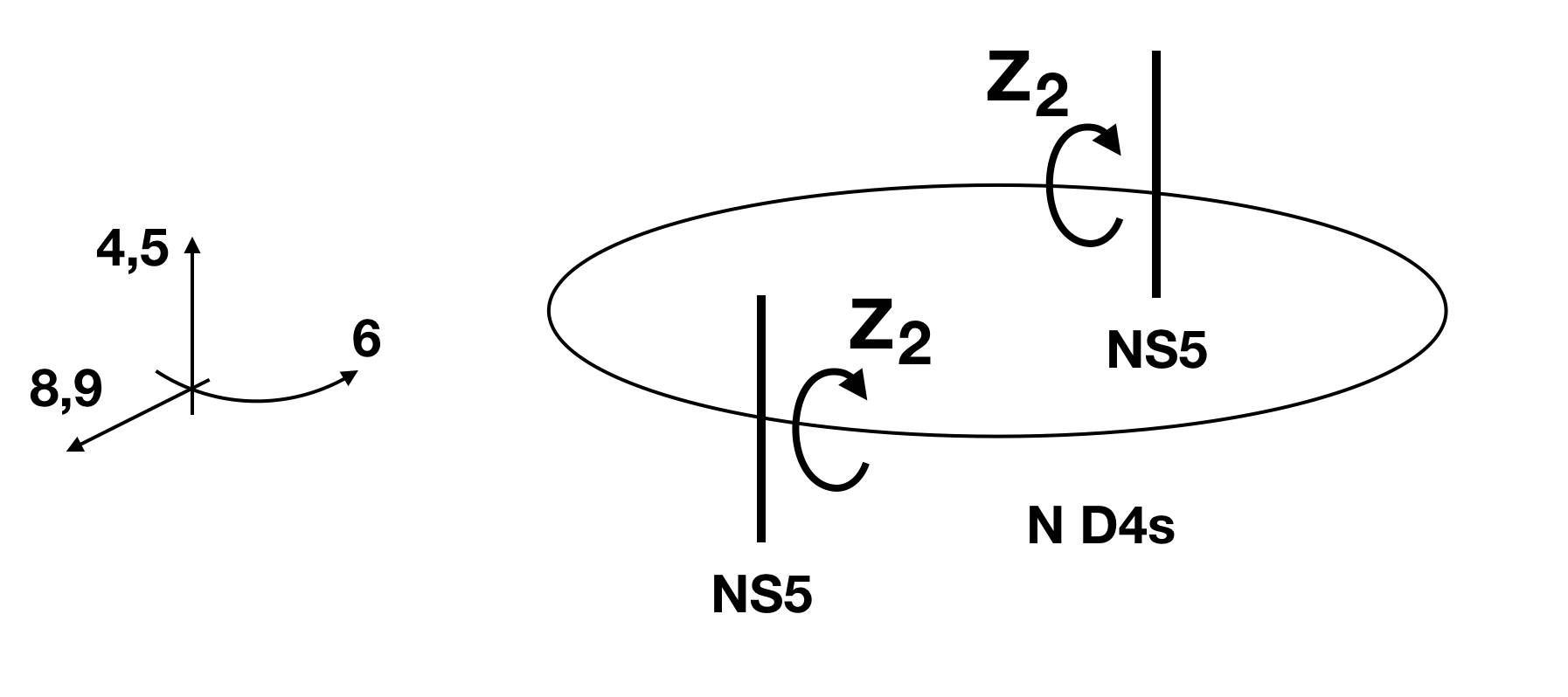}
\caption{\small The elliptic brane configuration with a $\IZ_2$ orbifold providing the T-dual of D3-branes at a $\IC^3/(\IZ_2\times\IZ_2)$ singularity.}
\label{fig:elliptic-z2}
\end{center}
\end{figure}

The type IIA configuration can be obtained by acting with a T-duality on the system of D3-branes at $\IC^3/(\IZ_2\times\IZ_2)$, as described in the brane construction of theory II.5 in section \ref{sec:classII-45} (equivalently, by a different T-duality on the NS5/D5-brane tiling configuration). The resulting IIA configuration in an elliptic model with the direction 6 parametrizing an $\IS^1$, with 2 NS5-branes along 012345, a stack of D4-branes in the directions 0123 and suspended in 6 in the two intervals between the NS5-branes, and a $\IZ_2$ orbifold acting on the coordinates $45$ and $89$, see Figure \ref{fig:elliptic-z2}. The theory is an orbifold quotient of the $k=2$ elliptic model in section \ref{sec:elliptic} (i.e. of the interpolating model studied in section \ref{sec:gpr}), and the spectrum (\ref{z2z2}) follows from the rules in \cite{Lykken:1997gy}, as reviewed in section \ref{sec:orbifold-nodt}. We recall that each stack of D4-branes suspended in an interval bounded by 2 NS5-branes leads to an $SU(N)\times SU(N)$ theory, due to the splitting of the D4-branes into fractional branes because of the $\IC^2/\IZ_2$ orbifold at whose tip they are sitting. Since there are two such intervals in the elliptic model, the complete theory has four gauge factors, in agreement with the picture of D3-branes at $\IC^3/(\IZ_2\times\IZ_2)$, and with the brane tiling construction.

Now we would like to follow in this type IIA Hanany-Witten construction the weak-coupling limit leading to the theory with simple gauge group (\ref{3nfund}). One might think that, because the configuration is an orbifold of the interpolating model of the theory II.1 (c.f. section \ref{sec:gpr}), the weak-coupling limit would be controlled by that of II.1 theory, and we would recover a class II universality class. This would however not be correct. The key point is that the limit in the brane tiling picture shows that all gauge factors except are turned into global symmetries.

Now we should recall the origin of the four gauge factors from the combination of intervals between NS5-branes and fractional D4-branes due to the $\IZ_2$ orbifold. This implies that the limit corresponds to decompactifying the $\IS^1$, while letting one of the intervals become infinity (which decouples the two gauge factors from that interval), keeping the other interval finite, and simultaneously sending the NSNS 2-form field in the collapsed $\IS^2$ at the $\IC^2/\IZ_2$ singularity to zero (which decouples the third gauge factor), while keeping the gauge coupling of the last gauge factor finite. Clearly, this combined limit leads to a fairly singular configuration, with strong non-perturbative dynamics arising from the orbifold singularity with vanishing $B$-field. Also, it inherently exploits the existence of an orbifold singularity and of its twisted sector scalars which control the necessary gauge couplings. Hence, even though it is an orbifold theory, its behaviour in the limit of interest cannot be understood without that orbifold, and therefore it is not related to the limit of the II.1, which therefore does not provide a suitable parent. In other words, even though we get two NS5-branes becoming coincident in the double-scaling limit, they are not placed in flat space but in a singular flat space orbifold with vanishing $B$-field. This is expected to change the near-horizon geometry generated by the backreaction of these NS5-branes. The theory thus displays a new kind of universality class not encountered before, as signaled by the new value for $\alpha=\sqrt{3/4}$ and of the large-$N$ Hagedorn temperature (see general result for theories with $SU(N)$ gauge group in \cite{Calderon-Infante:2024oed}).

On the other hand, the theory (\ref{3nfund}) can itself acts as a parent theory of others in this new universality class. By our general arguments in earlier sections, QSCFTs related to (\ref{3nfund}) by e.g. orientifold operations will have limits in this new unniversality class. In Figure \ref{fig:1face}b, c we provide two such examples. The picture shows the same brane tiling, but with additional orientifold projection with a fixed point, indicated with a red/green star, depending on the orientifold sign, see section \ref{sec:orbifold-nodt}. Using the rules therein to compute the resulting spectrum, the gauge factor in the finite size face is projected down to an $SO/USp$ factor, while the flavors arising from the finite edges around it are identified in pairs. The gauge group and chiral multiplet content for the resulting 4d $\NN=1$ theories for Figures \ref{fig:1face}b, c are respectively
\beqa
&O^+\,:\;\;SO(N+2)\quad, &\quad O^-\,:\;\; USp(N-2)\nonumber \\
& \quad\quad 3N\,\fund\quad & \quad\quad\quad\quad\quad 3N\,\fund
\eeqa

The fact that this QSCFTs built from brane tilings leads to a new universality class is not an exception, but rather seems to be the rule. It is in fact easy to build other examples of SQCFTs from brane tilings with no orientifolds, which enjoy a variety of seemingly new limits, which are shared by all its orientifolded descendant theories. 

\begin{figure}[htb]
\begin{center}
\includegraphics[scale=.35]{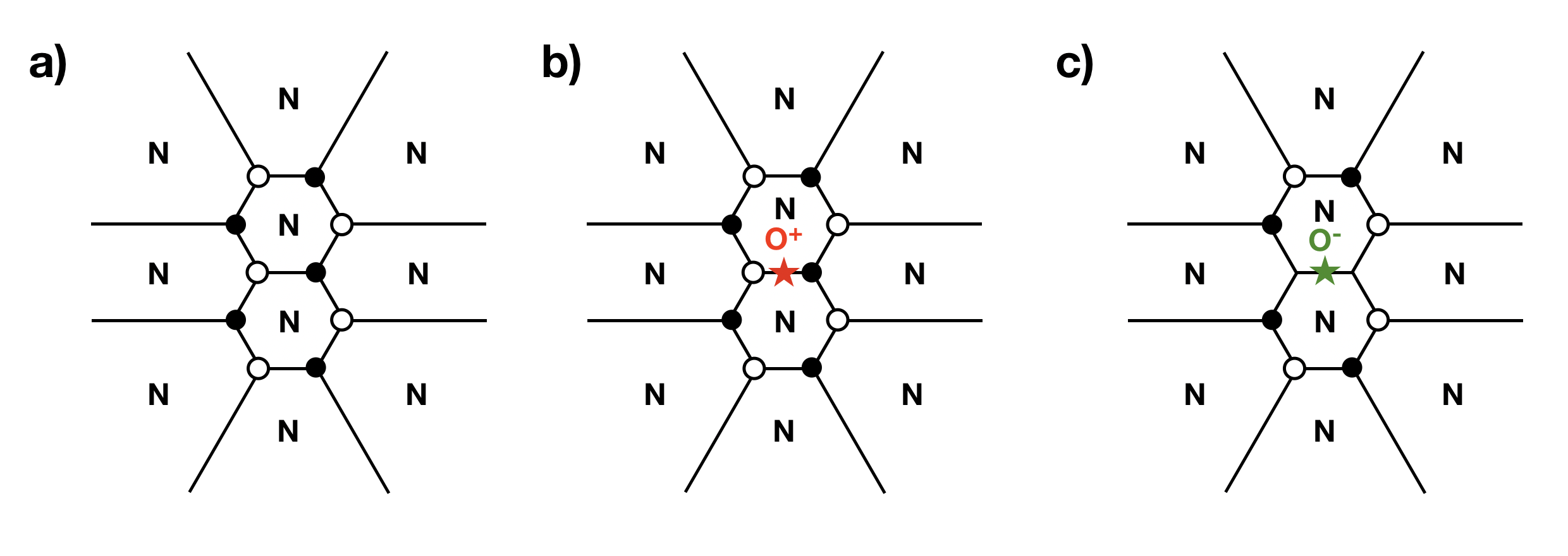}
\caption{\small The brane tilings of $\IR^2$ leading to a further class of QSCFTs. Figure a) shows the parent theory, while figures b) and c) correspond to orientifolded versions. For simplicity, in the latter we are neither showing the ${\cal O(1/N)}$ modifications in the ranks in the non-compact faces, nor including the necessary extra flavor branes.}
\label{fig:2face}
\end{center}
\end{figure}

As further illustration of this phenomenon, in Figure \ref{fig:2face}a we provide the brane tiling of a QSCFT with two gauge factors, with the 4d $\NN=1$ gauge group and chiral matter content
\beqa
&SU(N)\times SU(N) &\nonumber \\
& (\fund,\antifund) + 2N (\fund,1)+3N(\antifund,1)+3N(1,\fund)+2N(1,\antifund)&
\eeqa
This field content can be obtained using the general rules for brane tilings in section \ref{sec:orbifold-nodt}.\footnote{Although this configuration does not admit an embedding into that of the $\IC^3/(\IZ_2\times\IZ_2)$ theory, it is easy to find other theories which can play the role of interpolating models, for instance the $\IC^3/(\IZ_2\times \IZ_3)$ theory (equivalent to $\IC^2/\IZ_6'$, where the $\IZ_6'$ is generated by $z_i\to e^{2\pi i v_i}z_i$, $i=1,2,3$, with $v=(1,2,-3)/6$).} 

This theory has a limit in which the two finite size faces shrink in a $\IZ_2$ symmetric manner. This provides the parent model for the limits of descendant quasiconformal theories obtained by including an orientifold quotient. These are shown in Figures \ref{fig:2face}b, c, which respectively have the resulting 4d $\NN=1$ vector and chiral multiplets:
\beqa
&O^+\,:\;\quad \quad SU(N) \quad \quad \quad\quad, \quad \quad & O^-\,:\;\quad\quad SU(N) \nonumber\\
& \quad\quad \Ysymm + (2N-3)\fund+(3N+1)\antifund \quad \quad \quad& \quad\quad\Yasymm+(2N+3)\fund+(3N-1)\antifund 
\eeqa
where we have adjusted the ${\cal O}(1/N)$ numbers of flavors to achieve vanishing 1-loop beta function.

Note that, although the parent theory has two gauge factors, the orientifolded theories have simple gauge group due to the non-trivial orientifold identifications. This is analogous to a similar phenomenon encountered for class III theories, although it arises here in the context of QSCFTs and in a completely different universality class.

We hope these examples suffice to illustrate the rich structure of QSCFTs and their possible limits, and we postpone to future works their more detailed exploration, as well as the conceptual aspects of the proper definition of emergent string limits for QSCFTs.

\section{Conclusions} 
\label{sec:conclusions}

In this work we have provided a string-theoretic explanation for the universality classes of infinite-distance weak-coupling limits in conformal manifolds observed in large classes of four-dimensional SCFTs \cite{Calderon-Infante:2024oed,Calderon-Infante:2026rkj}. Focusing first on the complete set of superconformal gauge theories with simple gauge group and admitting a large-$N$ limit, we have constructed the brane setup realizing each of those SCFTs in a double-scaling limit. It was observed in \cite{Calderon-Infante:2024oed} that this set of theories can be divided in three classes at large $N$ according to the properties of their weak-coupling limit, such as the Hagedorn temperature and the exponential rate at which higher-spin operators saturate the unitarity bound. We have shown that all these SCFTs can be realized in terms of Type IIA Hanany--Witten brane configurations, and that each of them can be traced back to three underlying parent configurations through explicit operations including orientifold and orbifold quotients, the addition of small numbers of flavor branes, and suitable combinations thereof. This provides a sharp explanation as of why the weak-coupling limits of these theories fall into three universality classes. Even though some operators in the parent theory are projected out, this reduction is not strong enough to modify the large-$N$ Hagedorn temperature that captures the leading exponential behavior of the density of states at very high energies. Furthermore, since the operations done to the parent theories do not modify the leading large-$N$ behavior of correlators between operators that survive the projection, this explains why the towers of higher-spin operators become conserved at the same rate. 

Our brane constructions also provide a direct explanation to the bulk interpretation of these universality classes proposed in \cite{Calderon-Infante:2024oed}, namely that they are related to the type of string becoming tensionless at the infinite-distance limit. The key result is that the relevant universality class is controlled simply by the number of coincident NS5-branes in the double-scaling limit in which the brane configuration reduces to the SCFT. Studying the backreaction of these coincident NS5-branes and their near-horizon limit, we obtain what we dubbed the ``flat space'' worldsheet (see Footnote \ref{foot:flat-space}) for the corresponding bulk string becoming tensionless and weakly-coupled in the infinite-distance limit. In this way, only Class I theories have an Einstein-gravity bulk dual as they can be realized as D4-branes probing a string background without NS5-branes becoming coincident in the limit, and which can be related through suitable operations to that of AdS$_5\times S^5$ (the parent theory). The tensionless string in the Class I weak-coupling limit is therefore simply the fundamental (critical) type IIB string. In contrast, theories with Class II and Class III overall weak-coupling limits are non-Einstein and the emergent string will be different. In particular, we found that all theories with a Class II limit are controlled by two coincident NS5-branes and therefore share the same tensionless string, namely the sub-critical string previously proposed for $\mathcal{N}=2$ superconformal QCD in \cite{Gadde:2009dj} (which acts as the parent theory of this class). Such string propagates only in 8 space-time dimensions, one of them of string size. Likewise, all theories with a Class III limit are controlled by three coincident NS5-branes, which leads to another type of string whose woldsheet description we identify to be given by a class of marginal deformations of the CHS worldsheet \cite{Callan:1991at} with level $k=3$. These deformations generalize the more well-studied one related to $\mathcal N=2$ superLiouville in Type IIA or the cigar CFT in Type IIB. In this case, the string propagates in 10 space-time dimensions but 3 of them are of string size. 

For many of these theories we have provided an embedding into interpolating models, either in terms of elliptic IIA Hanany-Witten constructions or T-dual IIB systems of D3-branes probing orbifold and/or orientifold singularities. The latter are useful to identify the nature of the emergent strings in the limit as combinations of the fundamental string and wrapped D3-branes. Hence the emergent string may be regarded as a (highly non-trivial) sector in a standard holographic setup. 

It is also clear now why there were only three universality classes for 4d SCFTs with a single gauge factor, since their brane construction cannot involve more than 3 NS5's, as otherwise would yield more than one gauge factor.\footnote{Notice that, at best, we can only reduce by 2 the number of gauge groups by adding one O6-plane in the linear model, as described in Appendix \ref{sec:o6-planes}.} Clearly, new universality classes can then appear when considering more general SCFTs with multiple gauge factors.

A further outcome of our analysis is that the brane toolkit is not restricted to theories with simple gauge group. Rather, it extends naturally to theories with multiple gauge factors, allowing one to relate broad classes of SCFTs to parent models whose infinite-distance limits can be easily understood. In this way, the classification of weak-coupling limits can be organized in terms of the same underlying tensionless strings. We have illustrated this in several classes of $\mathcal{N}=2$ theories, extending the results in \cite{Calderon-Infante:2026rkj}, and also discussed extensions to quasiconformal theories, where new types of tensionless strings seem to arise.

Taken together, our results show that the seemingly diverse landscape of four-dimensional large-$N$ Lagrangian SCFTs simplifies drastically in the neighborhood of infinite-distance limits. At leading order, what matters is not the detailed open-string sector, nor the precise choice of orientifold or orbifold quotient, but rather the closed-string background that controls the tensionless limit. This explains why theories within the same class share universal weak-coupling data, such as the large-$N$ Hagedorn temperature and the exponential rate at which the higher-spin towers become light. More broadly, our analysis supports the idea that infinite-distance limits in conformal manifolds admit a top-down classification in terms of emergent string backgrounds, much as happens for infinite-distance limits in moduli spaces of string compactifications. 

\medskip

There are, however, several important questions that remain open. Although our original motivation was the classification of tensionless strings arising at weak-coupling/infinite-distance limits, the results should also be relevant more broadly for top-down holography. Indeed, our constructions provide many new putative AdS/CFT pairs beyond the Einstein gravity regime. While we have identified the ``flat space'' worldsheet description, it would be very interesting to study in greater detail the brane configurations that engineer the SCFTs and their backreaction on the geometry leading to the expected AdS near-horizon throat, as done in \cite{Gadde:2009dj,Dei:2024frl} for $\mathcal{N}=2$ for SCQCD. All these theories are expected to contain low-lying higher-spin fields. It would be interesting to explore their role in more detail to shed new light on holography for non-Einstein theories.

The brane constructions are mostly based on well-established rules, which explain the right pattern and the many features of the resulting gauge theories. On the other hand, we have been driven to improve some of the rules for configurations involving glued NS5-branes, i.e. coincident NS5-branes with extra adjoint chiral multiplets and preserving only 4d $\NN=1$. Such configurations are related by T-dualities to partially frozen singularities in the IIB side, or to glued M5-branes in M-theory. It would be interesting to use these complementary perspectives to provide further evidence for the resulting gauge theory results we have proposed for these configurations, to provide alternative realizations for the corresponding SCFTs, or to learn about the mechanisms underlying these novel frozen singularities. 

Regarding the ``flat space'' worldsheet backgrounds, those related to $k>2$ coincident NS5-branes involve the superpotential deformations in \eqref{deformations} that generalize the usual Liouville superpotential by not only turning an exponential superpotential on the cylinder, but also coupling it to other sector of the 2d SCFT. To the best of our knowledge, this type of superpotential terms have not been studied in detail in the literature and it would be interesting to do so in the future. Applying T-duality to the $k=2$ case leads to an exact Type IIB worldsheet description involving a direct product between the cigar CFT and the $SU(2)/U(1)$ minimal model that oftentimes is more tractable than the Type IIA description. Even though the analog Type IIB description for the $k>2$ case is not expected to take the simple form of a product of coset CFTs due to the non-trivial coupling between the cylinder and $\mathcal N=2$ minimal model in Type IIA, it would be valuable to study the effect of T-duality on these backgrounds.

A further natural direction is to push the classification beyond the cases considered here. Our techniques already apply to theories with multiple gauge factors, and we have discussed some cases in this paper, but a more systematic exploration of weak-coupling limits in general gauge theories would be very valuable. In particular, it would be interesting to classify all possible emergent tensionless strings that can arise in the bulk duals of exactly conformal theories, in analogy to the classification program of infinite-distance limits that have taken place in flat space string compactifications. So far, all the limits studied in this paper, including those beyond simple gauge groups, are ultimately controlled by the background of $k$ coincident NS5-branes. This raises the question of whether other types of new tensionless strings can arise in weak-coupling limits of exactly conformal theories, and if so, in which ones. Quasiconformal theories already indicate that new string backgrounds are possible, but it remains to be understood how far this can be extended within the landscape of CFTs.

Another interesting class of limits concerns infinite-distance points in the conformal manifold that are not the standard weak-coupling limits of the gauge field theory. For instance, we can also have strongly coupled limits at infinite distance which, after an appropriate duality transformation, are again described as weak-coupling limits of the same theory---when the theory is self-dual---or of a different gauge theory. In particular, it would be especially interesting to analyze limits related by Argyres--Seiberg-type dualities \cite{Argyres:2007cn} and whether they can provide new types of bulk tensionless strings. Although these typically involve weak-coupling limits of finite-$N$ gauge theories and therefore do not lead to a weakly-coupled gravitational dual, they should still provide a valuable testing ground for the general structure of infinite-distance limits in conformal manifolds.

Our results also highlight an intriguing contrast with the situation in flat-space string theory compactifications. In all known flat-space examples, infinite-distance limits associated with tensionless strings lead to critical fundamental strings, in agreement with the Emergent String Conjecture  \cite{Lee:2019wij}. By contrast, in AdS/CFT one finds a much richer set of possibilities, including non-critical strings propagating effectively in fewer than ten dimensions, even if they also have vanishing worldsheet central charge. It would be very interesting to understand whether this proliferation of possibilities is an intrinsically AdS phenomenon, or whether some analogue could also arise in flat space. Clarifying this point may shed light on the scope of the Emergent String Conjecture and on the extent to which AdS backgrounds enlarge the possible UV completions in terms of different string backgrounds.

Finally, flat-space compactifications admit not only tensionless-string limits but also decompactification limits, signaled by towers of Kaluza--Klein modes. No analogous decompactification limits are known in  conformal manifolds of 4d CFTs. Establishing their absence in full generality would provide the missing ingredient needed to complete the proof of the CFT Distance Conjecture \cite{Perlmutter:2020buo}. From this perspective, understanding why conformal manifolds in $d>2$ appear to allow only the higher-spin/tensionless-string type of infinite-distance limit remains one of the most fundamental open problems in the subject.

We hope that the brane perspective developed in this paper will provide useful tools to address these questions and to extend the top-down classification of infinite-distance limits in conformal manifolds to much broader classes of quantum field theories. More generally, our results suggest that AdS/CFT provides a natural arena in which Swampland constraints emerge as intrinsic features of the conformal manifold, hinting at a deep interplay between quantum gravity consistency conditions and CFT dynamics.

\section*{Acknowledgments}

We would like to thank Florent Baume, Jacques Distler, Matteo Lotito, Amineh Mohseni, Miguel Montero and Ethan Torres for useful discussions. This work is supported by the Spanish Agencia Estatal de Investigaci\'on through the grant ``IFT Centro de Excelencia Severo Ochoa'' CEX2020-001007-S, the grants PID2021-123017NB-I00 and PID2024-
156043NB-I0, funded by MCIN/AEI/10.13039/501100011033 and by ERDF A way of making Europe. This work was performed in part at Aspen Center for Physics, which is supported by National Science Foundation grant PHY-2210452. We also thank the Simons Center for Geometry
and Physics for hospitality during the Summer Workshop ’25. The work of JC is supported by the Walter Burke Institute for Theoretical Physics at Caltech and by the U.S. Department of Energy, Office of Science, Office of High Energy Physics, under Award Number DE-SC0011632. The work of I.V. is supported by the ERC Starting Grant QGuide101042568 - StG 2021, and the Project ATR2023-145703 funded by MCIN/AEI/10.13039/501100011033.

\appendix

\section{Brane Constructions}
\label{app:brane-cooking}

In this appendix we overview the construction of Hanany-Witten configurations \cite{Hanany:1996ie} of NS5-branes with D4-branes suspended between them \cite{Witten:1997sc}, possibly in the presence of D6-branes, and orbifolds and orientifolds thereof, to describe 4d $\NN=2$ and $\NN=1$ SCFTs. For a general review of these ingredients and the model building rules, see \cite{Giveon:1998sr}. We will provide original references of these ingredients and constructions as needed.

\subsection{The basic construction}
\label{app:basic}

In this section we review the configurations of NS5-branes with D4-branes suspended between them, as a realization of 4d $\NN=2$ gauge theories providing large classes of SCFTs. They are also  the starting point for the more elaborated models with orientifolds in later sections. 

\subsubsection{D4-branes suspended between NS5-branes}
\label{sec:suspended-d4s}

The basic configurations of D4-, NS5- and D6-branes were studied in \cite{Witten:1997sc}, see \cite{Elitzur:1997hc} for a review. We will follow the conventions in that literature, and consider D4-branes along the directions 0123 and 6, which in the latter coordinate are suspended between NS5-branes along 012345, and with D6-branes (if present) along 0123789. Because the D4-branes stretch on intervals between NS5-branes in the direction 6, their 5d gauge theory is reduced to 4d at low-energies. The system of D4- and NS5-branes preserves 8 supercharges, so we get 4d $\NN=2$ theories. The introduction of D6-branes does not break any additional supersymmetries and preserves this $\NN=2$. 

Consider a configuration of $k$ NS5-branes at different positions in the direction 6, so that they define $k-1$ intervals. We locate $N_i$ stacks of D4-branes suspended between NS5-branes at the $i^{th}$ interval, $i=1,\ldots, k-1$, and we locate $N_0$ and $N_k$ semi-infinite D4-branes extending from the NS5-branes at the ends. We also include $w_i$ D6-branes in the $i^{th}$ interval (in non-compact models, one can move them to infinity by using Hanany-Witten transitions \cite{Hanany:1996ie}, and bring the configuration to a pure NS5/D4-brane configuration. The general class of theories corresponds to a 4d so-called linear $\NN=2$ quiver gauge theory, with symmetry group and hypermultiplet matter content
\beqa
&[SU(N_0)]\times SU(N_1)\times\ldots SU(N_{k-1})\times [SU(N_k)]&\nonumber \\
& (\fund_0,\antifund_1)+(\fund_1,\antifund_2)+(\fund_2,\antifund_3)+\ldots +(\fund_{k-1},\antifund_k)&\nonumber \\
& w_1\fund_1+\ldots+ w_{k-1}\fund_{k-1} &
\label{linear-quiver}
\eeqa
Here the groups in brackets correspond to global symmetries on the semi-infinite D4-branes, while others correspond to gauge factors. The $U(1)$ factors disappear as gauge symmetries by brane bending \cite{Witten:1997sc}, and remain as global baryonic symmetries. Regarding matter, the open strings stretched between the D4-branes on the two sides of an NS5-brane lead to a massless 4d $\NN=2$ hypermultiplet in the bifundamental of the corresponding symmetry groups (including the global symmetries at the extremes). Finally, the open strings between the D4- and D6-branes in the same interval lead to fundamental hypers of the corresponding gauge factor.

One simple class of conformal theories corresponds to choosing all $N_i$ equal, i.e. $N_i\equiv N$ $\forall i=0,\ldots,k$, and including no D6-branes. In these $\NN=2$ cases, this condition is related to the absence of brane bending, i.e. the cancellation of the backreaction of D4-branes ending on the NS5-branes \cite{Witten:1997sc}. The simplest example of this kind is obtained for $k=2$, leading to the theory
\beqa
& [SU(N)]_0\times SU(N)_1\times [SU(N)]_2 & \nonumber\\
& (N,{\overline N},1)+(1,N,{\overline N}) &
\eeqa
where we have introduced subindices to distinguish the different $SU(N)$ factors. Namely, in 4d $\NN=1$ terms we obtain an $SU(N)$ gauge theory with one adjoint and $2N$ chiral multiplets in the $\fund+\antifund$. This is precisely the SCFT II.1 in Table \ref{table:SCFTs}, as already studied in \cite{Gadde:2009dj}. Models with additional NS5-branes lead to analogous linear quiver theories with more gauge factors, recently studied in \cite{Calderon-Infante:2026rkj}, which we will revisit in section \ref{sec:multiple}.

It is possible to construct other conformal cases with different $N_i$'s, if they are suitably compensated by appropriate numbers of D6-branes, to ensure that all NS5-branes have the same linking number (equivalently, the number of D4-branes on each interval is equal once all the D6-branes have been moved off using Hanany-Witten transitions). These theories have been recently considered in 
\cite{Calderon-Infante:2026rkj}.

\subsubsection{Elliptic models and T-duality to D3-branes on $\IC^2/\IZ_k$}
\label{sec:elliptic}

We now turn to elliptic models \cite{Witten:1997sc}, which are useful to construct interpolating models, with a tractable T-dual in terms of D3-brane systems.

The starting point are the NS5/D4-brane configurations in the previous section, for general $N_i$. Elliptic models are built by taking the direction 6 to be an $\IS^1$. From the brane perspective, this means that the stacks of semi-infinite D4-branes at the ends of the linear quiver become finite, such that the global symmetry groups $SU(N_0)$ and $SU(N_k)$ are identified (so we must require $N_0=N_k$) and become an extra gauge factor, which we choose to denote by $SU(N_k)$. From the QFT viewpoint, this operation corresponds to a gauging of the diagonal subgroup of the $[SU(N_0)\times SU(N_k)]$ global symmetry (with $N_0=N_k$).

The resulting 4d $\NN=2$ gauge theory has gauge group and hypermultiplet content
\beqa
&SU(N_1)\times\ldots SU(N_k)&\nonumber \\
& (\fund_1,\antifund_2)+(\fund_2,\antifund_3)+\ldots +(\fund_k,\antifund_1)&
\label{elliptic-N2}
\eeqa
The picture is that the NS5-brane locations split the $\IS^1$ of the coordinate 6 into $k$ intervals. The $N_i$ D4-branes suspended between two parallel NS5-branes in the $i^{th}$ interval lead to a 4d $\NN=2$ vector multiplet of $SU(N_i)$. The conformal case corresponds again to choosing all $N_i$ equal, i.e. $N_i\equiv N$ $\forall i=1,\ldots,k$. For the special cases of $k=0$ or $k=1$, we obtain the 4d $\NN=4$ $SU(N)$ SYM theory. The case $k=2$ was used in \cite{Gadde:2009dj} to build an interpolating model for the SCFT II.1 in Table \ref{table:SCFTs}, as we generalize for large $k$ in section \ref{sec:multiple} (see also \cite{Calderon-Infante:2026rkj}). 

For elliptic models, we can T-dualize the configurations along the direction 6 to get type IIB configurations (see e.g. \cite{Karch:1998yv,Hanany:1998it}), as follows. The $k$ NS5-branes are geometrized into a $k$-center Taub-NUT in the T-dual, which locally correspond to a $\IC^2/\IZ_k$ orbifold singularity, and the D4-branes map to 
D3-branes at the orbifold fixed point. The D3-branes in each  $SU(N_i)$  gauge factor map to fractional branes (with the $U(1)$'s being removed by $BF$ couplings \cite{Ibanez:1998qp}). The $(k-1)$ independent positions of the NS5-branes in the directions 789 are mapped to geometric blow-ups of the orbifold singularity, while the separations of the NS5-branes in the direction 6 are mapped to the values of the $B$-fields in the $(k-1)$ non-trivial 2-cycles, controlling the gauge couplings of the fractional D3-brane gauge factors (with the overall gauge coupling being controlled by the dilaton). In case the original configuration contains D6-branes, they map to D7-branes, as we will discuss in explicit examples in later sections.

\subsection{Including orientifold planes}
\label{app:oplanes}

A standard procedure to generate new examples of gauge theory realizations using brane configurations is to add orientifold planes to the previous setup. There are several kinds of orientifold quotients that can be performed, preserving 4d $\NN=1$ supersymmetry, and which correspond to adding O4-, O6- or O8-planes.\footnote{Adding more than one kind of orientifold planes actually implies quotienting by additional orbifold actions. This possibility will be considered separately in section \ref{app:orbifolds}.} We study them in turns.

\subsubsection{Adding O4-planes}
\label{sec:o4-planes}

The simplest kind of orientifold quotient that can be imposed on the configuration of NS5/D4-branes corresponds to adding an O4-plane along the direction 01236 and located at the origin in the directions 5789. In other words, we quotient the configuration by $\Omega R (-1)^{F_L}$, where $\Omega$ is worldsheet parity, $R:(x^5,x^6,x^7,x^8,x^9)\to (-x^5,-x^6,-x^7,-x^8,-x^9)$ and $F_L$ is left-moving fermion number. The orientifold preserves the same supersymmetries as the D4-branes, hence we obtain 4d $\NN=2$ theories (which may accidentally enhance to $\NN=4$ in special cases). The brane configurations with these O4-planes were considered in \cite{Landsteiner:1997vd} (see \cite{Evans:1997hk,Elitzur:1997hc} for earlier 4d $\NN=1$ models), to which we refer the reader for details. We note that, in order for the parent configuration to be invariant under the orientifold symmetry, the D4-branes must be located at orientifold image locations in the directions 45 along the NS5-branes, while the NS5-branes may be located at arbitrary positions in the direction 6. For simplicity, we do not consider introducing D6-branes, which is an straightforward generalization of the coming construcions.

There are two choices of orientifold plane,\footnote{Actually, there are two other variants \cite{Gimon:1998be,Hori:1998iv}, the O4$^0$-plane (corresponding to an O4$^-$ with a stuck D4-brane on top, so we include it as a simple modification of our models), and a variant of the O4$^+$-plane with an extra RR flux, which leads to same models up to a discrete $\theta$ parameter.} denoted by O4$^{\pm}$, which differ by the RR charge they carry ($\pm 1$ in D4-brane charge units, as counted in the covering space). They differ in their orientifold action on gauge group and matter. An important observation \cite{Evans:1997hk,Landsteiner:1997vd} is that when an O4-plane crosses an NS5-brane, it flips its RR charge, so an O4$^+$-plane becomes an O4$^-$-plane, and viceversa. This has an important impact on the pattern of gauge groups and matter content in the resulting 4d $\NN=2$ theories.

The effect of the O4-plane in the brane configuration is as follows. Each D4-brane interval is mapped to itself, so each $SU(N_i)$ gauge factor is projected down to an SO or USp factor, depending on whether the corresponding interval corresponds to an O4$^-$- or O4$^+$-plane, respectively. On the other hand, each bifundamental hypermultiplet $(\fund_i,\fund_{i+1})$ is mapped to itself under the orientifold action, so it is projected down to a half-hypermultiplet in the bifundamental (this is possible because the two neighbouring gauge factors correspond to $SO\times USp$ or viceversa, so the bifundamental is a pseudoreal representation). Overall, for the case of a linear quiver (i.e. with the direction 6 taken to be non-compact), with $k$ NS5-branes and $N_i$ D4-branes (including orientifold images) on the $i^{th}$ interval, and $N_0$, $N_k$ semi-infinite D4-branes at the extremes, we have the 4d $\NN=2$ symmetry group and hypermultiplet content given by:
\beqa
&[G_0]\times G_1\times\ldots G_{k-1}\times [G_k]&\nonumber \\
& \frac 12 (\fund_0,\fund_1)+\frac 12(\fund_1,\fund_2)+\frac 12 (\fund_2,\fund_3)+\ldots +\frac 12(\fund_{k-1},\fund_k)&
\label{O4-linear}
\eeqa
where the symmetry groups $G_i$ alternate between $SO(N_i)$ and $USp(N_{i+1})$, and again the squared brackets indicate a global flavor symmetry. Clearly, the cancellation of global gauge anomalies for the $USp$ factors require that all the $N_i$ must have the same parity. 

For this class of theories, the conformality condition corresponds to the absence of brane bending, which requires the total D4-brane charge (including O4-planes) to be the same for all intervals, so the group has the structure $\cdots \times SO(N+2) \times USp(N) \times SO(N+2)\times \cdots$.

In the case of even $k=2p$, the two endpoing gauge factors are of the same kind (equivalently, the two semi-infinite O4-planes are of the same kind), so the model can be turned into an elliptic one, by taking the direction 6 to be compact (i.e. gauging the global symmetry groups). The theory is of the kind (\ref{O4-linear}), with the endpoint gauge factors identified. Namely, restricting already to the conformal case
\beqa
&SO(N+2)_0\times USp(N)_1 \times\ldots \times SO(N+2)_{2p-2}\times USp(N)_{2p-1}&\nonumber\\
& \frac 12 (\fund_0,\fund_1)+\frac 12(\fund_1,\fund_2)+\ldots +\frac 12(\fund_{2p-2},\fund_{2p-1})+\frac 12 (\fund_{2p-1},\fund_0)&
\label{O4-elliptic}
\eeqa
The special case of $k=0$, namely no NS5-branes, corresponds to D4-branes on $\IS^1$ on top of an O4$^\pm$-plane, and reproduces (by T-duality) the 4d $\NN=4$ $SO(N)$ or $USp(N)$ SYM theories, which correspond to the theories I.2 and I.3 in Table \ref{table:SCFTs}, see section \ref{sec:classI-123}.

These elliptic models can be T-dualized along the compact direction 6 and obtain a realization of the gauge theory in terms of D3-branes on top of an orientifold of $\IC^2/\IZ_k$, as described in \cite{Uranga:1999mb} (in a T-dual D5-brane setup). The generator of the orientifold group is $\Omega R' (-1)^{F_L}$, with $R':(z_1,z_2)\to(z_2,-z_1)$ on $\IC^2$, which squares to the identity in the quotient $\IC^2/\IZ_k$ when $k$ is even.

\subsubsection{Adding O6-planes}
\label{sec:o6-planes}

Let us now consider the inclusion of O6-planes, which again are of two kinds\footnote{It is possible to have a variant of the O6$^-$-plane with one stuck D6-brane, but this requires the presence of odd IIA Romans mass parameter \cite{Hyakutake:2000mr}, so we will not consider this possibility.}, O6$^\pm$-planes, with RR charge $\pm 4$ in D6-brane charge units in the double cover. In addition, there are two posible ways in which one can align these O6-planes to preserve at least 4 supersymmetries. We may locate the O6-planes along 0123789 (i.e. parallel to D6-branes introduced in section \ref{sec:suspended-d4s}), preserving 4d $\NN=2$, and denoted as O6$^\pm$-planes. The other possibility is to place them along 0123457, preserving only 4d $\NN=1$, and denoted by O6'${}^\pm$-planes. More precisely, we orientifold by an action $\Omega R (-1)^{F_l}$, where $\Omega$ is worldsheet parity, $F_L$ is left-moving fermion number, and $R$ is a $\IZ_2$ action flipping the sign of coordinates transverse to the orientifold plane, namely 456 for O6-planes, and 689 for O6'-planes. 

The fact that the orientifold flips the coordinate 6 implies an important difference between linear and elliptic models. In linear models there is only one orientifold plane, located at $x^6=0$, while in elliptic models there are two fixed points of $R$ in the $\IS^1$ parametrized by the coordinate $x^6$, so there are two orientifold planes. These must necessarily span the same directions (namely, we must have two O6-planes, or two O6'-planes, but not mixed configurations). For simplicity we will focus on linear models, and eventually build elliptic ones by gluing linear ones. We now turn to the discussion of different quiver gauge theories which are obtained, first by using O6-planes, and then by using O6'-planes.

\medskip

\noindent {\bf Linear quiver models with O6-planes}

\noindent Let us consider configurations of $k$ NS5-branes along 012345, and located at different positions in a non-compact direction 6, with stacks of $N_i$ D4-branes along 0123 and along 6, in which they are suspended in the $i^{th}$ interval between neighbouring NS5-branes, and  let us introduce an O6-plane along 0123789 at $x^6=0$. In addition, we allow for the possible presence of $w_i$ D6-branes in the $i^{th}$ interval. These configurations were originally considered in \cite{Landsteiner:1997ei}. In order for the parent configuration to be invariant under the $\IZ_2$ symmetry, the NS5-branes must be located symmetrically with respect to the origin in 6, and the D4-brane stacks (and the D6-branes if present) must also be symmetric both in their positions and multiplicities $N_i$. There are now two different situations, depending on the parity of the number $k$ of NS5-branes.

\begin{figure}[htb]
\begin{center}
\includegraphics[scale=.4]{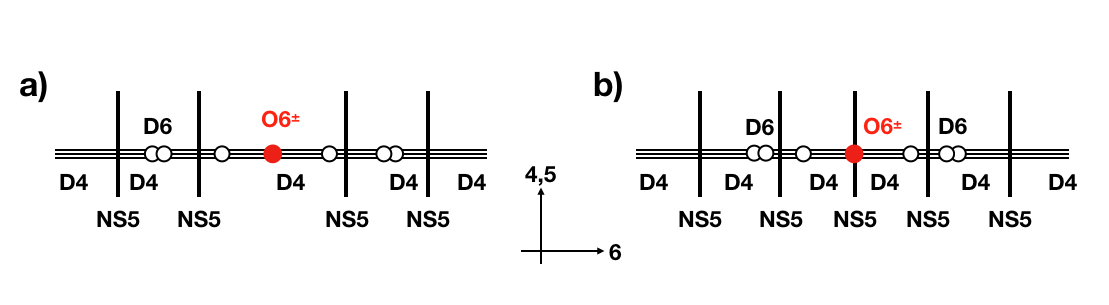}
\caption{\small a) The linear quiver brane configurations with an O6-plane, for the even $k=2p$ case. b) Linear quiver brane configuration with O6-plane for odd $k$.}
\label{fig:o6-linear}
\end{center}
\end{figure}

If the number of NS5-branes is even, $k=2p$, they are located in pairs away from the O6-plane, see Figure \ref{fig:o6-linear}a. This implies that, in the parent gauge theory (\ref{linear-quiver}), the two semi-infinite regions are mapped to each other $SU(N_0)\leftrightarrow SU(N_k)$ (hence we need $N_0=N_k$), resulting in a single global symmetry factor; similarly, the intervals are exchanged pairwise, $SU(N_i)\to SU(N_{k-i})$ (hence we need $N_i=N_{k-i}$), leading to only one copy of each in the quotient; finally, the interval corresponding to $SU(N_p)$ is mapped to itself, so the gauge group is projected down\footnote{The O6-plane projection on D4-brane gauge groups is opposite to that on D6-branes on top of the O6-plane, due to the 4 relative DN+ND boundary conditions.\label{foot:orientifold-flip}} to $SO(N_p)$ for O6$^+$, or to $USp(N_p)$ for O6$^-$. Regarding the hypermultiplet matter content, all bifundamentals are exchanged pairwise, so only one copy per each pair survives in the orientifold quotient. The flavours in the fundamental from the D6-branes are also exchanged pairwise, for D6-branes away from the  O6-plane, resulting in only one copy per pair, and mapped to themselves for D6-branes on top of the O6-plane, resulting in half-hypermultiplets.\footnote{These are possible because the O6-plane projection on the D4- and D6-branes is opposite, leading to $SO\times USp$ patterns for the global and gauge symmetry (or viceversa).}
Overall, the structure of the 4d $\NN=2$ gauge group and matter content is
\beqa
& [SU(N_0)]\times SU(N_1)\times \ldots \times SU(N_{p-1})\times G_p &\nonumber\\
& (\fund_0,\antifund_1)+\ldots+(\fund_{p-2},\antifund_{p-1})+(\fund_{p-1},\fund_p) &\nonumber\\
& w_1\fund_1+\ldots w_{p-1}\fund_{p-1}+\frac 12 w_p\fund_p 
\label{one-O6-ns5-away}
\eeqa
where $G_p=SO(N_p)$ for the O6$^+$-plane and $G_p=USp(N_p)$ for the O6$^-$-plane.

If the number of NS5-branes is odd, $k=2p+1$, there must be one NS5-brane stuck on the O6-plane, see Figure \ref{fig:o6-linear}b, while the $2k$ remaining ones are distributed in symmetric pairs. As before, the two semi-infinite regions are exchanged pairwise, $SU(N_0)\leftrightarrow SU(N_k)$, but now all the remaining gauge factors are exchanged pairwise, $SU(N_i)\to SU(N_{k-i})$, and there is no gauge factor mapped to itself. On the other hand, the bifundamental hypermultiplets are exchanged pairwise, except for that on top of the O6-plane which is mapped to itself. The latter bifundamental is projected down to a 2-index tensor representation, which is the symmetric for the O6$^+$-plane case and the antisymmetric for the O6$^-$-plane.\footnote{This can be checked because Higgsing the $SU(N_p)$ gauge group with a hypermultiplet in the 2-index symmetric (resp. antisymmetric) representation breaks the symmetry down to $SO(N_p)$ (resp. $USp(N_p)$). This is the field theory counterpart to removing the stuck NS5-brane in the directions 789, producing a configuration with effectively an even number of NS5-branes with the O6$^\pm$-plane projection described above.} Regarding the fundamental flavors from the D6-branes, they are all exchanged pairwise in this case, so the final spectrum contains only copy per such pair.

Overall, the structure of the 4d $\NN=2$ gauge group and matter content is
\beqa
& [SU(N_0)]\times SU(N_1)\times \ldots \times SU(N_{p}) &\nonumber \\
& (\fund_0,\antifund_1)+\ldots+(\fund_{p-1},\fund_p)+ T_p &\nonumber \\
&w_1\fund_1+\ldots w_p\fund_p &
\label{one-O6-ns5-ontop}
\eeqa
where $T_p=\Ysymm_p$ for the O6$^+$-plane and $T_p=\Yasymm_p$ for the O6$^-$-plane.

For these classes of 4d $\NN=2$ theories (for general $k$), the conformality condition is the absence of brane bending, or equivalently that all the NS5-branes have the same linking number, including the contribution from D6-brane charges. This can be equivalently expressed as requiring that all the intervals have the same number of D4-branes, after the net D6-brane charge (either from explicit D6-branes, or from the O6-plane charge) has been formally moved across the NS5-branes to infinity, creating D4-branes via Hanany-Witten transitions. The motion of D6-brane charge must be carried out in a $\IZ_2$ symmetric way, so as to preserve the orientifold symmetry (i.e. half the units of charge are moved on each side of the O6-plane).
Explicit examples of this class appear in sections \ref{sec:classI-456}. \ref{sec:classII}, \ref{sec:classIII} and \ref{sec:multiple-orientifolds}.

\medskip

\noindent {\bf Elliptic models with O6-planes}

\noindent The previous linear quiver models can be combined into elliptic models, corresponding to brane configurations with a compact direction 6 with two O6-planes at opposite fixed points, which can be of equal or opposite sign. The model building rules are as above, so the construction of new examples is straightforward, see \cite{Uranga:1998uj,Park:1998zh} for the original references. We skip the detailed discussion of examples, and mention the key ingredients to be used in the main text. Also, note that in compact models the conformality requirement is related to the cancellation of D6-brane charge; therefore, conformal models are possible when both orientifold planes are O6$^-$-planes (if 8 additional D6-branes are included), or when one is an O6$^-$-plane and the other is an O6$^+$-plane (and no extra D6-branes are added). On the other hand, it is not possible to obtain conformal models if both are O6$^+$-planes.

To provide a simple example, we consider an elliptic model with 2 NS5-branes located on top of 2 O6$^-$-planes, and $N$ D4-branes suspended between them in each interval (which are orientifold images of each other). We add 8 D6-branes (4 and their orientifold images) to cancel the RR charge and get a conformal model. The theory can be obtained by simply gluing two linear theories of the kind described in (\ref{one-O6-ns5-ontop}). The resulting 4d $\NN=2$ gauge group and matter content is
\beqa
&SU(N)&\nonumber \\
& 2\Yasymm + 4\fund &
\label{two-o6-ns5-ontop}
\eeqa
This corresponds to the theory I.5 in Table \ref{table:SCFTs}. Similar examples are easily constructed, and will arise in the main text in sections \ref{sec:classI-456}, \ref{sec:classII-123} and \ref{sec:classIII-12}

Elliptic models can be T-dualized to systems of D3-branes at a $\IC^2/\IZ_k$ singularity, quotiented by a suitable orientifold action, whose details encode the different possible models for even or odd $k$. The one-to-one construction of type IIB orientifolds of $\IC^2/\IZ_k$ corresponding to the different type IIA brane configurations was carried out in \cite{Park:1998zh}, to which we refer the reader for details. 

\medskip

\noindent {\bf Linear quiver models with O6'-planes}

\noindent Let us now consider introducing O6'-planes instead of O6-planes in the configurations of NS5- and D4-branes of section \ref{sec:suspended-d4s}. Namely, we consider $k$ NS5-branes along 012345 and located at different positions in 6, and stacks of $N_i$ D4-branes along 0123 and suspended in the $i^{th}$ interval in 6 between a pair of NS5-branes. We first consider the direction 6 to be non-compact, and orientifold the configurations so that there is an O6'-plane along 0123457 and located at $x^6=0$. This requires that the NS5- and D4-branes (and D6-branes if present) are arranged in a $\IZ_2$ symmetric way. The orientifold breaks half of the supersymmetries, leading to 4d $\NN=1$ theories, which we discuss next. We again have two large classes of models, depending on the parity of the number $k$ of NS5-branes.

\begin{figure}[htb]
\begin{center}
\includegraphics[scale=.30]{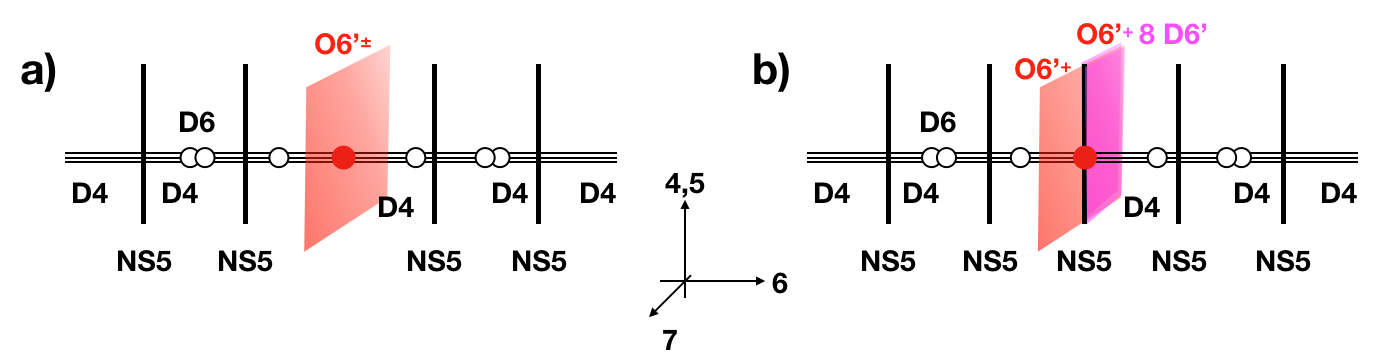}
\caption{\small a) The linear quiver brane configurations with an O6'-plane, for the even $k=2p$ case. b) Linear quiver brane configuration with O6'-plane for odd $k$ The orientifold plane is split in two halves by the stuck NS5-brane, with one half corresponding to an O6'$^+$, and the other to an O6'$^-$ with 8 half D6'-branes on top, to ensure conservation of RR charge.}
\label{fig:o6prime-linear}
\end{center}
\end{figure}

We start by discussing the case of even $k=2p$, so that the NS5-branes are arranged in orientifold image pairs away from the O6'-plane, see Figure \ref{fig:o6prime-linear}a. As in the case of O6-planes, the two endpoint global symmetry factors and $k-2$ gauge factors are exchanged in pairs, while there is one gauge factor (with label $p=k/2$) which is mapped to itself. Also, all the bifundamental hypermultiplets are exchanged in pairs by the orientifold action. The ingredientes exchanged by the orientifold action are insensitive to the presence of the O6'-plane, so their spectrum fills out multiplets of a local 4d $\NN=2$ supersymmetry. On the other hand, the 4d $\NN=2$ vector multiplet of the $SU(N_p)$ gauge factor of the parent theory is mapped to itself, so it feels the breaking to 4d $\NN=1$ by the O6'-plane, as follows. Let us decompose the 4d $\NN=2$ vector multiplet with respect to 4d $\NN=1$; then, the O6'-plane projects the $\NN=1$ vector multiplet in the same way as an O6-plane (namely, an O6'$^+$-plane projects the gauge group down to $SO(N_p)$, while an O6'$^-$-plane projects it down to $USp(N_p)$); on the other hand, the O6'-plane acts on the 4d $\NN=1$ adjoint chiral multiplet with an extra sign as compared with an O6-plane projection (namely, an O6'$^+$-plane projects it down to a 2-index symmetric representation, and an O6'$^-$-plane projects it down to a 2-index antisymmetric).

Overall, the structure of the resulting 4d $\NN=1$ symmetry group and matter content is
\beqa
& [SU(N_0)]\times SU(N_1)\times \ldots \times SU(N_{p-1})\times G_p &\nonumber\\
& {\rm Adj}_0+\ldots +{\rm Adj}_{p-1}+ {\rm T}_p
&\nonumber \\
& (\fund_0,\antifund_1)+\ldots+(\fund_{p-2},\antifund_{p-1})+(\fund_{p-1},\antifund_p) +& \nonumber \\
& (\antifund_0,\fund_1)+\ldots+(\antifund_{p-2},\fund_{p-1})+(\antifund_{p-1},\fund_p) & 
\label{one-o6p-ns5-away}
\eeqa
where for an  O6'$^+$-plane we have $G_p=SO(N_p)$ and ${\rm T}_p=\Ysymm_p$, and for an O6'$^-$-plane we have $G_p=USp(N_p)$ and ${\rm T}_p=\Yasymm_p$. As emphasized above, these fields fail to fill out 4d $\NN=2$ vector multiplets because this sector feels directly the breaking to 4d $\NN=1$ by the O6'-plane.

Consider now the case of odd $k=2p+1$. Then there must necessarily be one NS5-brane on top of the O6'-plane, see Figure \ref{fig:o6prime-linear}b. In fact, it cuts it in two halves which are semi-infinite in the direction 7, producing a non-trivial configuration, studied in \cite{Landsteiner:1998gh,Brunner:1998jr,Elitzur:1998ju}. For starters, the O6'-plane must flip its RR charge sign when crossing the NS5-brane (in analogy with the O4-planes in section \ref{sec:o4-planes}), so the two halves have opposite RR charge, i.e. a half O6'${}^+$- and a half O6'${}^-$-plane. Then, consistency of the configuration requires continuity of the RR charge across the NS5-brane, so we must add 8 half D6'-branes on top of the half O6'$^-$-plane.

The effect of the O6'-plane projection is as follows. Just like in the O6-plane case, there is no symmetry group which is mapped to itself, as all of them (both the endpoint global factors and the gauge factors) are exchanged pairwise, so only one copy within each pair is kept. Since they are insensitive to the O6'-plane projection, they preserve a local 4d $\NN=2$ supersymmetry, so they lead to 4d $\NN=1$ vector multiplets and adjoint chiral multiplets. Regarding the 4d $\NN=2$ matter fields in the parent theory, $k-1$ of the bifundamental hypermultiplets are exchanged pairwise, so one copy within each pair is kept. Decomposing the $p=(k-1)/2$ bifundamental 4d $\NN=2$ hypermultiplets under 4d $\NN=1$ supersymmetry, we obtain chiral multiplets in vector-like pairs $(\fund_i,\antifund_{i+1})+(\antifund_i,\fund_{i+1})$ for $i=0,\ldots, p-1$. Finally, there is one 4d $\NN=2$ bifundamental hypermultiplet in the parent theory which is mapped to itself under the O6'-plane action. The orientifold projection by the two half O6'$^\pm$-plane leads to chiral multiplets in the $\Ysymm+\bYasymm$ of $SU(N_p)$, and the additional 8 half D6'-brane lead to 8 chiral multiplets in the $\antifund_P$. The resulting configuration is chiral, but free of anomalies. The opposite chirality configuration would be obtained by exchanging the two half O6'$^\pm$-planes, i.e. a parity operation in the direction 7, hence there is only one class of theories of this kind.

Overall, the structure of the resulting 4d $\NN=1$ gauge group and matter content is 
\beqa
& [SU(N_0)]\times SU(N_1)\times \ldots \times SU(N_p) &\nonumber\\
& {\rm Adj}_0+\ldots +{\rm Adj}_{p-1}+ {\rm Adj}_p &\nonumber \\
& (\fund_0,\antifund_1)+\ldots +(\fund_{p-1},\antifund_p) + \bYasymm_p+8\antifund_p& \nonumber \\
& (\antifund_0,\fund_1)+\ldots+(\antifund_{p-1},\fund_p) + \Ysymm_p \, & 
\label{one-o6p-ns5-ontop}
\eeqa
For these classes of gauge theories (for general $k$), the conformality condition is not necessarily related to brane bending, so it will be determined by direct requirement of cancellation of the beta functions. Some explicit examples of this class of constructions appears in sections \ref{sec:classII-45}, \ref{sec:classIII-3}.

\medskip

\noindent {\bf Elliptic models with O6'-planes}

\noindent In analogy with the O6-planes, it is possible to make the direction 6 compact and construct elliptic models of brane configurations with two O6'-planes. The resulting gauge theories are obtained by gluing together two linear quiver models in the previous discussion. The model building rules are as above, so the construction of new examples is straightforward, see \cite{Park:1999eb} for the original references. Explicit examples of this class of constructions appear in sections \ref{sec:classII-45}, \ref{sec:classIII-3}. We note that in this $\NN=1$ case, the O6'-plane/D6'-brane RR charge cancellation does not ensure exact conformality (which thus must be checked with direct QFT tools), but only at 1-loop or in the large $N$ limit. 

Elliptic models can be T-dualized to systems of D3-branes at a $\IC^2/\IZ_k$ singularity, quotiented by a suitable orientifold action which preserves only 4d $\NN=1$, whose details encode the different possible models for even or odd $k$. The construction of such orientifolds and their one-to-one map to type IIA brane configurations was carried out in \cite{Park:1999eb}, to which we refer the reader for details.

\subsubsection{Adding O8-planes}
\label{sec:o8-planes}

There is yet another possible orientifold that can be performed for NS5/D4-brane configurations, which is introducing O8-planes along 012345689. Namely we quotient by $\Omega R (-1)^{F_L}$, where now $R:x^7\to -x^7$. These configurations were considered in \cite{Feng:2001rh}, to which we refer the reader for details. The O8-plane can have positive or negative RR charge, denoted as O8$^\pm$-planes, and it stretches along the direction 6, so there is only one O-plane either in linear or elliptic models. In order to preserve the orientifold symmetry, the D4-branes must be located symmetrically, but the NS5-branes can be located at generic positions in 6. In these respect, models with O8-planes are analogous to the models with O4-planes of section \ref{sec:o4-planes}, with the difference that the O8-plane does not flip charge when it crosses the NS5-branes in 6, basically because its worldvolume spans the dimensions 89 which allow to surround the NS5-brane. Another difference is that, while O4-planes preserve the same supersymmetries as the D4-branes, and hence the 4d $\NN=2$ of generic NS5/D4-brane systems, the O8-plane breaks the symmetry down to 4d $\NN=1$. This implies that the orientifold projection on the 4d $\NN=2$ vector and hypermultiplets will reflect this breaking, by acting differently on their 4d $\NN=1$ components.

\medskip

\noindent {\bf Linear quiver models with O8-planes}

\noindent Let us consider a configuration of $k$ NS5-branes with stacks of $N_i$ D4-branes suspended between pairs of them (and semi-infinite ones at the ends), and an O8-plane. Let us focus on the case of non-compact direction 6, and elliptic models can be easily constructed in a subsequent stage. The orientifold action on the spectrum of the 4d $\NN=2$ gauge theory (\ref{linear-quiver}) is as follows (we do not include D6-branes for simplicity). Regarding the 4d $\NN=2$ vector multiplets, each gauge factor is mapped to itself, so the 4d $\NN=1$ $SU(N)$ vector multiplets of the parent theory end up projected down to $SO(N)$ for the O8$^-$-plane case, or to $USp(N)$ for the O8$^+$ case. Because the  O8-plane preserves only 4d $\NN=1$ supersymmetry, there is an extra sign in the orientifold action on the adjoint chiral multiplet in the 4d $\NN=2$ parent vector multiplet. Hence, for the O8$^-$-plane we get the 2-index symmetric representantion, and for the O8$^+$-plane we get the 2-index antisymmetric one. Regarding the 4d $\NN=2$ matter, each bifundamental hypermultiplet of the parent theory is mapped to itself, and is projected down to a 4d $\NN=1$ chiral multiplet in the corresponding bifundamental. Overall, the 4d $\NN=1$ gauge group and chiral matter content is:
\beqa
& [G_0]\times G_1\times\ldots\times G_{k-1}\times [G_k] &\nonumber\\
& T_1+\ldots + T_{k-1} & \nonumber \\
& (\fund_0,\fund_1)+(\fund_1,\fund_2)+\ldots+(\fund_{k-1},\fund_k)&
\eeqa
where, for the O8$^-$-plane case we have $G_i=SO(N_i)$ and $T_i=\Ysymm_i$, and for the O8$^+$-plane case we have $G_i=USp(N_i)$ and $T_i=\Yasymm_i$. As usual, square brackets indicate global symmetry groups. Models with O8-planes will appear (together with other ingredients) in section \ref{sec:classIII-45678}.

\medskip

\noindent {\bf Elliptic models with O8-planes}

\noindent Elliptic models are easily obtained for general $k$ by simply gluing together the ends of the above linear quiver theories, i.e. by taking $N_0=N_k$ and gauging the diagonal combination of the global symmetry groups. The T-dual of elliptic models correspond to 4d $\NN=1$ orientifolds of systems of D3-branes at $\IC^2/\IZ_k$ singularities, as described explicitly in  \cite{Feng:2001rh}.

\subsection{Adding orbifolds}
\label{app:orbifolds}

As already mentioned, the elliptic models of the above brane configurations admit T-dual description in terms of D3-branes in flat space (or orientifolds thereof) or $\IC^2/\IZ_k$ singularities (or orientifolds thereof). There is a natural generalization, which provides a large class of 4d $\NN=1$ theories, obtained by stacks of D3-branes at orbifold singularities $\IC^3/\Gamma$, with $\Gamma$ a discrete subgroup of $SU(3)$ \cite{Kachru:1998ys,Lawrence:1998ja,Hanany:1998it}. These can be constructed using worldsheet techniques, and in the case of Abelian orbifolds (i.e. $\Gamma=\IZ_N$ or $\Gamma=\IZ_N\times \IZ_M$) they are particular examples of the larger class of D3-branes at toric CY threefold singularities, which admit a very efficient description in terms of dimer diagrams \cite{Hanany:2005ve} (see \cite{Kennaway:2007tq} for a review). Some of these models admit a T-duality back to a type IIA NS5/D4-brane configuration, possibly with their own orbifold quotients \cite{Lykken:1997gy}. In this section we simply review these setups for the particular example of $\IC^3/(\IZ_2\times\IZ_2)$, used in the main text for the construction of theories I.7 and II.5.

Consider a stack of ${\tilde N}$ D3-branes at an orbifold $\IC^3/(\IZ_2\times\IZ_2)$ singularity, where the generators of the orbifold group act on $\IC^3$ as 
\beqa
\theta:(z_1,z_2,z_3)\to (-z_1,-z_2,z_3)\quad ,\quad \omega:(z_1,z_2,z_3)\to (-z_1,z_2,-z_3)
\label{z2z2-generators}
\eeqa
There are actually two versions of this orbifold, depending on the choice of discrete torsion \cite{Vafa:1986wx} (see also \cite{Font:1988mk,Vafa:1994rv}). We study them in turns.

\subsubsection{Orbifold $\IC^3/(\IZ_2\times\IZ_2)$ without discrete torsion}
\label{sec:orbifold-nodt}

For gauge theories on D3-branes at Abelian orbifolds, the most studied setup is the case `without discrete torsion', which geometrically corresponds to D3-branes at a singular geometry with collapsed holomorphic cycles. It admits an open string description, as well as a simple description using toric geometry, and T-dual Hanany-Witten pictures. We study them in turns.

\subsubsection*{The open string description}

When locating ${\tilde N}$ D3-branes at orbifolds without discrete torsion, the D3-brane Chan-Paton matrices must obey the orbifold  group law, e.g. $\gamma_{\theta,3}\gamma_{\omega,3} =\gamma_{\omega,3}\gamma_{\theta,3}$, for an abelian group, namely the Chan-Paton indices form a representation of the orbifold group. For the $\IZ_2\times\IZ_2$ orbifold, upon diagonalization, we can split the stack of D3-branes into four sets, of multiplicities $N_{ij}$, $i,j=1,2$, characterized by the $\pm 1$ eigenvalue of the Chan-Paton label of each D3-brane under $\theta$ and $\omega$, so the matrices have the structure
\beqa
\gamma_{\theta,3}={\rm diag}\, (\id_{N_{11}},\id_{N_{12}},-\id_{N_{21}},-\id_{N_{22}}) \quad ,\quad 
\gamma_{\omega,3}={\rm diag}\, (\id_{N_{11}},-\id_{N_{12}},\id_{N_{21}},-\id_{N_{22}})\quad 
\label{cp-matrices-nodt}
\eeqa
The configuration preserves 4d $\NN=1$ supersymmetry. The theory before the orbifold is 4d $\NN=4$ SYM, namely a 4d $\NN=1$ vector multiplet and three adjoint chiral mulitplets, each associated to motion of the D3-branes in the three complex coordinates of $\IC^3$. The orbifold projection on the gauge group requires gauge transformations to commute with (\ref{cp-matrices-nodt}), producing a breaking $SU({\tilde N})\to SU(N_{11})\times SU(N_{12})\times SU(N_{21})\times SU(N_{22})$, with $\sum_{ij}N_{ij}={\tilde N}$. The orbifold projection on the three chiral multiplets requires the invariance under the combined action on the geometry (\ref{z2z2-generators}) (because these fiels behave as coordinates in $\IC^3$) and on the Chan-Paton indices (\ref{cp-matrices-nodt}). The resulting 4d $\NN=1$ vector and chiral multiplets (see e.g. \cite{Morrison:1998cs,Park:1999ep}) are
\beqa
&SU(N_{11})\times SU(N_{12})\times SU(N_{21})\times SU(N_{22})&\nonumber\\
X_{ij}:\quad & (\fund_{11},\antifund_{21})+(\fund_{21},\antifund_{11}) + (\fund_{12},\antifund_{22})+(\fund_{22},\antifund_{12})&\nonumber \\
Y_{ij}:\quad& (\fund_{11},\antifund_{12})+(\fund_{12},\antifund_{11}) + (\fund_{21},\antifund_{22})+(\fund_{22},\antifund_{21})&\nonumber \\
Z_{ij}:\quad & (\fund_{11},\antifund_{22})+(\fund_{22},\antifund_{11}) + (\fund_{12},\antifund_{21})+(\fund_{21},\antifund_{12})&
\label{z2z2}
\eeqa
where we have introduced the notation $X, Y, Z$ for later convenience. There is a cubic superpotential, easy to write down explicitly, but which we skip. The conformal case corresponds to taking all gauge factors with equal rank, i.e. $N_{ij}=N$ \cite{Kachru:1998ys,Hanany:1998ru,Lawrence:1998ja,Hanany:1998it}.

\subsubsection*{The brane tiling \& toric geometry description}

As mentioned above, the geometry is toric, and the gauge theory admits a description in terms of brane tilings (aka dimer diagrams) \cite{Hanany:2005ve,Kennaway:2007tq} (see \cite{Garcia-Etxebarria:2006ngz} for a quick review)\footnote{See also \cite{Hanany:1997tb,Hanany:1998ru,Hanany:1998it} for a further closely related brane box realization with D5-branes suspended in a rectangular array of NS5-branes}. A brane tiling is a graph tiling $\IT^2$ or equivalently an infinite periodic graph tiling $\IR^2$ (namely a set of faces separated by edges which meet at vertices), which is bipartite, i.e. vertices are colored black or white in such a way that all edges join a black and a white note. Each bipartite graph corresponds to a 4d $\NN=1$ gauge theory via the following dictionary: 
\begin{itemize}
\item Each face $F_i$ corresponds to an $SU(n_i)$ gauge factor
\item Each edge $E_{ij}$ separating two faces $F_i$, $F_j$ is a chiral bifundamental $\Phi_{ij}$ in the $(\fund_i,\antifund_j)$ (the ordering of $i$ and $j$ is determined by the orientation of the edge, e.g. we declare that the bifundamental goes from $i$ to $j$ if that corresponds to moving counterclockwise around black nodes, or equivalently clockwise around white nodes)
\item Each vertex $V_a$ at which $p$ edges $E_{i_1i_2}$, $E_{i_2i_3}, \ldots E_{i_p,i_1}$ join corresponds to a superpotential term $W=\pm \tr( \Phi_{i_1i_2}\Phi_{i_2i_3} \ldots \Phi_{i_p,i_1})$, with positive or negative sign for black or white nodes, respectively.
\end{itemize}
The resulting 4d $\NN=1$ gauge theories corresponds to a stack of D3-branes at a toric CY3 singularity (constructible from tiling data, see below), with the conformal case corresponding to the equal rank case, $n_i=N$. 

\begin{figure}[htb]
\begin{center}
\includegraphics[scale=.35]{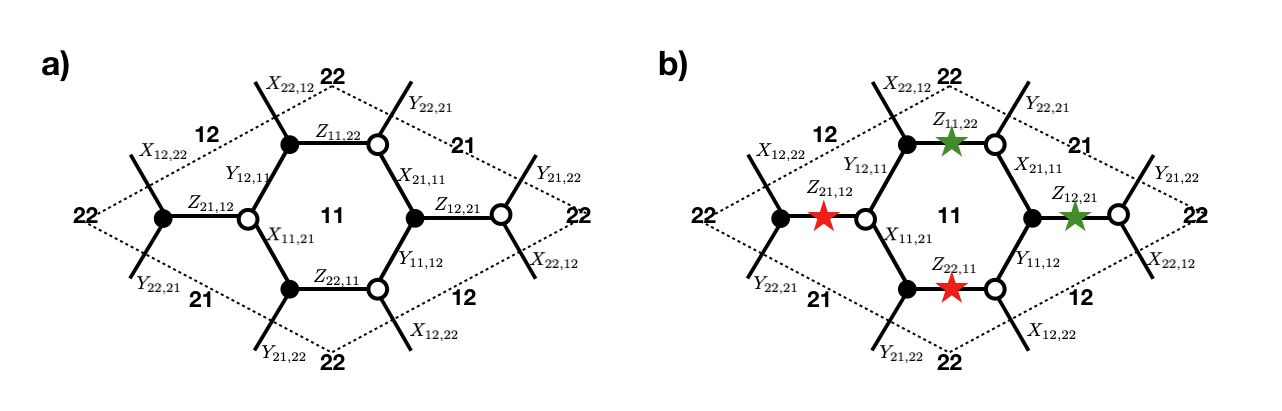}
\caption{\small a) The brane tiling for the  $\IC^3/(\IZ_2\times\IZ_2)$ singularity. The dashed rhombus indicates the unit cell of the periodic array. b)  Orientifold leading to the interpolating model for the theory II.5 in the main text. The red and green stars are fixed points with signs $-$ and $+$, respectively}
\label{fig:tiling-z2z2}
\end{center}
\end{figure}

Let us focus on the case of interest, the orbifold $\IC^3/(\IZ_2\times\IZ_2)$, whose dimer diagram is shown in Figure \ref{fig:tiling-z2z2}a. Note that the chiral multiplets $X_{ij},Y_{ij},Z_{ij}$ corresponding to the three lines in (\ref{z2z2}) arise from edges crossed when one moves in southwest diagonals, southeast diagonals, and the nortbound vertical directions, respectively. It is possible to recover the CY3 geometry as the moduli space of the gauge theory in the abelian theory (i.e. formally taking all gauge factors to be $U(1)$'s). Namely gauge invariant mesonic operators give coordinates for the space transverse to the D3-brane, i.e. the CY3. These mesons are given by closed loops in the brane tiling, modulo homology relations (due to F-term equivalences). In our case, we have the mesons associated to loops in the southwest, southeast or north directions 
\beqa
& x\simeq X_{11,21}X_{21,11}\simeq X_{22,12}X_{12,22}& ,\nonumber \\
& y\simeq Y_{11,12}Y_{12,11}\simeq Y_{22,21}Y_{21}Y_{22}& , \nonumber \\
&z\simeq Z_{11,22}Z_{22,11}\simeq Z_{12,21}Z_{21,12}& .
\label{mesons1}
\eeqa
There is also a meson corresponding to the  loop around any node, e.g.
\beqa
t=X_{11,21}Y_{21,22}Z_{22,11} \, .
\label{meson2}
\eeqa
These coordinates are not independent, as they satisfy $xyz=t^2$. This is precisely the hypersurface equation defining $\IC^3/(\IZ_2\times\IZ_2)$. This is made clear by defining combinations of the coordinates of $\IC^3$ invariant under (\ref{z2z2-generators}), namely 
\beqa
x=z_1^2\quad , \quad y=z_2^2\quad, \quad z=z_3^2\quad, \quad t=z_1z_2z_3\, ,
\label{inv-mono}
\eeqa
which indeed satisfy $zyw=t^2$. This gives a precise map between fields in the gauge theory and coordinates in the CY3 geometry.

\subsubsection*{The type IIA NS5- and D4-brane description}

For systems of D3-branes at $\IC^3/(\IZ_n\times\IZ_m)$ orbifolds there is a T-dual type IIA NS5/D4-brane brane configuration, of the kind introduced in \cite{Lykken:1997gy}. To discuss it, it is convenient to start from the $\IZ_n\times\IZ_m$ generalization of (\ref{z2z2-generators})
\beqa
\theta:(z_1,z_2,z_3)\to (e^{2\pi i/n} z_1,e^{-2\pi i/n}z_2,z_3)\;\;,\;\;\omega:(z_1,z_2,z_3)\to (e^{2\pi i/m} z_1,z_2,e^{-2\pi i/m}z_3) \quad\quad
\label{orbifold-generators}
\eeqa
To perform the T-duality, we pick the $\IS^1$ orbit of e.g. the $\IZ_n$ orbifold, namely $(e^{i\varphi} z_1,e^{-i\varphi}z_2,z_3)$. Upon this T-duality, the $\IZ_n$ orbifold becomes a set of $n$ NS5-branes equally spaced in the T-dual $\IS^1$ \cite{Ooguri:1995wj}, and the D3-branes become D4-branes suspended between them, so we get a 4d $\NN=2$ elliptic model with group $\prod_{i=1}^n SU(N_i)$ and bifundamental hypermultiplets for consecutive factors. The coordinate $z_3$ becomes the 45 coordinates in the NS5/D4- brane configuration, and $z_1$ becomes 89\footnote{More precisely, the part of $z_1,z_2$ after removing the $\IS^1$ orbit of $\theta$, maps the coordinates 789.}. Hence, we now have to perform the $\IZ_m$ quotient generated by $\omega$ in (\ref{orbifold-generators}), which stays as orbifold quotient in the Hanany-Witten T-dual, acting on those coordinates. Each NS5-brane is mapped to itself, while the D4-branes are mapped to themselves up to the orbifold action on their Chan-Paton indices. The latter splits, each $SU(N_i)$ factor into $m$ factors, producing a gauge group $\prod_{ij} SU(N_{ij})$ with $i=1,\ldots, n$, $j=1,\ldots, m$. One can similarly study the orbifold projection on the adjoint 4d $\NN=1$ multiplet in the 4d $\NN=2$ vector multiplet, and on the hypermultiplets. The final theory is
\beqa
&\prod_{ij} SU(N_{ij}) &\nonumber \\
&\sum_{ij} [(\fund_{ij},\antifund_{i+1,j})+(\fund_{ij},\antifund_{i,j+1})+(\fund_{ij},\antifund_{i-1,j-1})]&
\label{orbifold-theory}
\eeqa
For $n=m=2$ we see that we recover the spectrum (\ref{z2z2}) of the $\IC^3/(\IZ_2\times\IZ_2)$ theory. For this theory, in the conformal case $N_{ij}=N$, the type IIA picture is an elliptic model with 2 NS5-branes along 012345, with $N$ D4-branes along 0123 and suspended in 6 between the NS5-branes in the two intervals they define, and with quotient by the $\IZ_2$ generated by $\omega$ acting non-trivially on the complex planes 45 and 89. We will exploit this construction for theory II.5 in section \ref{sec:classII-45}.

\subsubsection*{Orientifolding brane tilings}

It is possible to introduce orientifold quotients in the above setups. The most efficient way to do this is to consider orientifolds of the brane tilings, which were considered in \cite{Park:1999ep}, to which we refer the reader for details. For our present purposes we focus on quotients by $\IZ_2$ symmetries of the tiling, flipping the two coordinates, i.e. leaving 4 fixed points. As shown in \cite{Park:1999ep}, such orientifolds must satisfy certain rules:
\begin{enumerate}
\item The orientifold must exchange black and white nodes. This implies that each orientifold point lies either in the middle of a face or in the middle of an edge in the tiling.
\item Orientifold points carry a $\pm$ sign, and the total number of points with a given sign is even(odd) in tilings for which the number of black (or white) nodes $N_{\rm b/w}$ is even(odd). 
\end{enumerate}

The sign $+$ or $-$ choice determines the orientifold projection on gauge factors, when the orientifold point is in the middle of a face (the $+$ leads to $SO$ and the $-$ leads to $USp$), or on chiral multiplets, when the orientifold point is in the middle of an edge (the $+$ leads to a $\Ysymm$ and the $-$ leads to a $\Yasymm$).

In our $\IC^3/(\IZ_2\times\IZ_2)$ case, rule 1 implies that the 4 fixed points must fall either all in the middle of faces, or all in the middle of edges. For concreteness we focus on the latter case, see Figure \ref{fig:tiling-z2z2}b, where we see that the gauge factor 11 is exchanged with 22, and 12 with 21. 

Using rule 2, and since there are 4 black (or white) nodes, the number of orientifold points with either sign must be even. We can choose the four fixed points to have all $+$ signs, all $-$ signs, or two $+$ and two $-$ signs. For concreteness, let us consider the latter case, already displayed in Figure \ref{fig:tiling-z2z2}b. The resulting theory is
\beqa
& SU(N_{11})\times SU(N_{21})&\nonumber \\
&(\fund,\antifund)+(\antifund,\fund) + (\fund,\fund)+(\antifund,\antifund)+(\Ysymm,1) + (\bYasymm,1) + (1,\Yasymm)+(1,\bYsymm)& \nonumber \\
& 8(\antifund,1)+8(1,\fund)&
\label{interp-ii5}
\eeqa
where we have added some flavors in the $\fund$ or $\antifund$ to cancel anomalies (by introducing suitable additional flavor branes).
Other distributions of the two $O^+$ and $O^-$ fixed points lead to equivalent theories, up to overall complex conjugations.
As discussed in section \ref{sec:classII-45}, this model is closely related to the brane construction for theory II.5 in Table \ref{table:SCFTs}.

It is possible to express the explicit orientifold action on the variables $z_i$ of $\IC^3/(\IZ_2\times\IZ_2)$, which will facilitate translating it to the type IIA picture. Recalling the realization of coordinates as mesons (\ref{mesons1}), the result in \cite{Franco:2007ii} is that under an orientifold action, a meson passing through two orientifold points picks up a sign equal to the product of the signs of these orientifold points. We obtain 
\beqa
x\to -x \quad,\quad  y\to y\quad , \quad z\to -z\quad ,\quad t\to -t
\eeqa
where the last transformation is a general rule for a mesonic operator appearing as a superpotential term \cite{Franco:2007ii}. Recalling (\ref{inv-mono}), the orientifold must act as 
 \beqa
 z_1\to e^{\pi i/2}z_1\quad , \quad z_2\to z_2\quad , \quad z_3\to e^{\pi i/2} z_3
 \eeqa
Equivalently, the orientifold action is $\Omega (-1)^{F_L} R\alpha $, where $R:z_i\to -z_i$, $i=1,2,3$ and  $\alpha^2=\omega$. This orientifold can be shown to contain an O7-plane along the 4-plane $z_1=0$, which requires the introduction of 8 D7-branes, at $z_1=0$ as well, providing the extra fundamental flavors as mentioned above.

This orientifold can be T-dualized to a type IIA description, as follows. Let us T-dualize along the orbit of $\omega$, and so the orbifold fixed plane $z_2$ becomes the coordinates 45 along which the 2 T-dual NS5 branes stretch. We thus get an elliptic model with 2 NS5-branes (along 0123 and 45 (i.e. $y$) and D4-branes along 0123 and suspended between them in 6 in the two possible intervals. In this T-dual, the action $\theta$ remains a $\IZ_2$ orbifold action on $z_1$ and $z_2$, hence  $z_1$ corrresponds to the coordinates 89. As explained above, the 4d $\NN=1$ gauge theory before the orientifold is (\ref{z2z2}). In the type IIA picture, the D4-branes in the two intervals lead to the gauge symmetries $SU(N_{11})\times SU(N_{22})$ and $SU(N_{12})\times SU(N_{21})$, respectively. 

The T-duality along the orbit of $\omega$ turns the O7-plane (and D7-branes) into an O8-plane (and D8-branes). Each interval is mapped to itself, and the orientifold relates the gauge factors $11\leftrightarrow 22$, $12\leftrightarrow 21$. The bifundamentals $Y_{ij}$ arise from open strings between D4-branes across the NS5-brane, and they are exchanged under the orientifold action, leading to bifundamentals in the quotient. Finally, each of the bifundamentals $X_{ij}$ and $Z_{ij}$ is mapped to itself and projects down to a 2-index tensor, in the combination $\Ysymm+\bYasymm+8\antifund$ (once additional flavor D8-branes are introduced), or its complex conjugate. The rules of the detailed orientifold projetions are not easily derived in the type IIA picture, but they simply follow from the T-duality with the previous type IIB orientifold.

\subsubsection{Orbifold $\IC^3/(\IZ_2\times\IZ_2)$ with discrete torsion}
\label{sec:orbifold-dt}

\subsubsection*{The open string description}

For Abelian orbifolds $\IC^3/(\IZ_n\times\IZ_m)$ it is possible to introduce additional discrete phases in the partition function of the orbifold worldsheet CFT which modify the theory and its D-branes. Morally, the extra phases, known as discrete torsion \cite{Vafa:1986wx} (see also \cite{Font:1988mk,Vafa:1994rv}) modify the action of the generator $\omega$ on the twisted sectors of the generator $\theta$ (and viceversa). At the closed string sector, the geometry is modified and admits complex structure moduli, corresponding to collapsed 3-cycles at the singular point. In the level of D-branes, the Chan-Paton matrices form a projective representation of the orbifold group \cite{Douglas:1998xa,Douglas:1999hq}, with the extra phases being determined by those of the discrete torsion.

For the $\IC^3/(\IZ_2\times\IZ_2)$ orbifold with discrete torsion, the action of $\omega$ on the $\theta$-twisted sector states has an extra sign (and viceversa). The singular geometry is now to be smoothed out by complex deformations, rather than by blowing up holomorphic cycles, and hence there is no description in toric geometry. Its D3-brane Chan-Paton matrices obey the orbifold group law up to a sign, i.e. $\gamma_{\theta,3}\gamma_{\omega,3} =-\gamma_{\omega,3}\gamma_{\theta,3}$. The basic structure of these matrices and the resulting 4d $\NN=1$ gauge theories where studied in \cite{Douglas:1998xa,Douglas:1999hq}, as we now review. We consider ${\tilde N}=2N$ D3-branes with Chan-Paton matrices
\beqa
\gamma_{\theta,3}=\id_N\otimes \sigma_1\quad \gamma_{\omega,3}=\id_N\otimes\sigma_2
\eeqa
The symmetry is broken down as $SU(2N)\to SU(N)$, as follows. The orbifold by $\theta$ splits $SU(2N)\to SU(N)\times SU(N)$, just like in the case without discrete torsion, but then the orbifold by $\omega$ acts by swapping the two factors, so that only one combination survives. Regarding the 4d $\NN=1$ adjoint chiral multiplets in the $SU(N)\times SU(N)$ vector multiplet, they are exchanged (with an extra sign) by $\omega$, so again we get one combination, in the adjoint of the final $SU(N)$. Finally, a the 4d $\NN=2$ bifundamental hypermultiplets in the $(\fund,\antifund)+(\antifund,\fund)$ of $SU(N)\times SU(N)$ after the $\IZ_2$ quotient by $\theta$, are exchanged by $\omega$. Hence, we keep only one copy of each, which lead to 2 additional 4d $\NN=1$  adjoint chiral multiplets of the final diagonal $SU(N)$. The complete field content is hence a 4d $\NN=1$ $SU(N)$ vector multiplet and 3 adjoint chiral multiplets, which coincides with that of 4d $\NN=4$ SYM. However, the theory has only 4d $\NN=1$ supersymmetry, because there is a superpotential $W=XYZ+XZY$, which has an extra sign (dictated by the discrete torsion) compared with familiar 4d $\NN=4$ SYM commutator superpotential.

The addition of orientifold planes in theories with discrete torsion has been considered (sometimes implicitly) in certain particular cases in compact examples in e.g. \cite{Berkooz:1996dw,Aldazabal:1998mr,Klein:2000tf,Klein:2000qw}, so a systematic discussion is beyond the scope of this work. We will discuss one concrete example in section \ref{sec:classI-78}, to reproduce the theory I.7 of Table \ref{table:SCFTs}.

Similarly, the application of T-duality to D3-branes at orbifold singularities with discrete torsion to reach type IIA NS5- and D4-brane configurations has not been carried out in the literature. We will consider the basic ingredients on one such construction for the particular case of the $\IC^3/(\IZ_2\times\IZ_2)$ in the discussion of the theory I.7 in section \ref{sec:classI-78}.

\subsection{Brane rotations}
\label{sec:rotated-branes}

The configurations of NS5- and D4-branes (and possibly D6-branes) in section \ref{app:basic}, and their orientifolds in section \ref{app:oplanes} admit a class of deformations which preserves 4d $\NN=1$ supersymmetry. This corresponds to an $SU(2)$ rotation between the complex planes corresponding to 45 and 89. That is, rotating the NS5-branes (and possibly the D6-branes) by angles $\theta$ and $-\theta$ in the 2-planes 48 and 59, respectively. We refer to them as NS5$_\theta$-branes. Rotated NS5-branes along the directions 012389, dubbed NS5'-branes (and similarly rotated D6'-branes\footnote{Indeed, the O6'-planes and D6'-branes in section \ref{sec:o6-planes} were rotated objects in the sense we are now discussing.}) appeared already in the early literature on gauge theories from brane configurations, see e.g. \cite{Elitzur:1997fh,Barbon:1997zu}, and \cite{Giveon:1998sr} for a review. 

The effect of relative rotation by a small angle between two NS5-branes with D4-branes suspended between them was shown in \cite{Barbon:1997zu} to correspond to adding a superpotential mass term for the 4d $\NN=1$ adjoint chiral multiplet in the $\NN=2$ vector multiplet. For small angle, the relation was shown to be $m\sim \tan \theta$, but one expects deviations from these relation for large angles. Upon integrating out the massive adjoint, the resulting theory contains quartic couplings between the flavors of the corresponding gauge factor (or bifundamentals with respect to other gauge factors, or 2-index tensor representations descending from them in orientifold models). For small angles, the quartic superpotential couplings have the structure $\frac 1m(Q{\tilde Q}Q{\tilde Q})$. However, we emphasize that such quartic couplings survive even for $\theta=\pi/2$, even if the small angle relation would suggest this corresponds to $m\to\infty$ limit \cite{Aharony:1997ju,Klebanov:1998hh}.

In the presence of orientifold planes, it is still possible to perform the rotations of NS5-branes, as long as they remain compatible with the orientifold symmetry. For instance, in the presence of O4-planes, c.f. section \ref{sec:o4-planes}, the orientifold acts in the same way in the directions 4, 5, 8, 9, hence the rotation of each NS5-brane by an arbitrary angle $\theta$ is compatible with the symmetry. A similar comment holds for models with O8-planes in section \ref{sec:o8-planes}. Finally, O6-planes (or O6'-planes) c.f. section \ref{sec:o6-planes} act differently in the directions 45 and 89; hence, an NS5-brane located at a point away from the O6-plane in the direction 6 and rotated by an angle $\theta$ must map to an NS5-brane at the orientifold image point in 6 and rotated by the opposite angle $-\theta$. This also implies that NS5-branes on top of O6-planes (hence mapped to themselves) cannot be rotated.

\subsection{Kutasov models and `glued' NS5-branes}
\label{sec:kutasov}

In this section we review certain brane configurations which are useful to describe 4d $\NN=1$ theories with extra adjoints (or other fields descending from them). The construction is based on the use of brane configurations with coincident NS5-branes. We emphasize that these are not supposed to be separated by some distance and then made coincident in some limit, rather, they are supposed to be taken coincident to start with. In other words, they do not lead to intervals with D4-branes suspended between them. Rather, these configurations with coincident NS5-branes were introduced in \cite{Elitzur:1997fh} to reproduce 4d $\NN=1$ $SU(N)$ gauge theories with extra chiral multiplets in the adjoint representation and their $\NN=1$ dualities \cite{Kutasov:1995ve} (see also \cite{Kutasov:1995np,Kutasov:1995ss}). 

These stacks of $k$ coincident NS5-branes can be used in brane constructions as follows \cite{Elitzur:1997fh}. Consider one set of $k$ coincident NS5-branes along 012345 and one additional NS5'-brane along 012389, separated in the non-compact coordinate 6, and we suspend $N$ D4-branes between them. We can introduce $N_f$ D6-branes along 0123789, to produce extra flavours, or we can introduce them via semi-infinite D4-branes stretching out from the NS5'-brane (both possibilities are related by a Hanany-Witten brane creation process). The resulting theory on the suspended D4-branes is a 4d $\NN=1$ $SU(N)$ gauge theory 
with $N_f$ flavors, i.e. chiral multiplets in the $\fund+\antifund$ , and one adjoint $X$ with superpotential
\beqa
W=\tr X^{k+1} \, .
\label{mono-supo}
\eeqa
The usual case of SQCD is obtained for $k=1$ because the superpotential is a mass term, so the adjoint chiral multiplet may be integrated out in the IR. The above brane configuration were shown in \cite{Elitzur:1997fh} to reproduce the Seiberg-like dualities of these gauge theories via motions of the NS5-branes.

\begin{figure}[htb]
\begin{center}
\includegraphics[scale=.35]{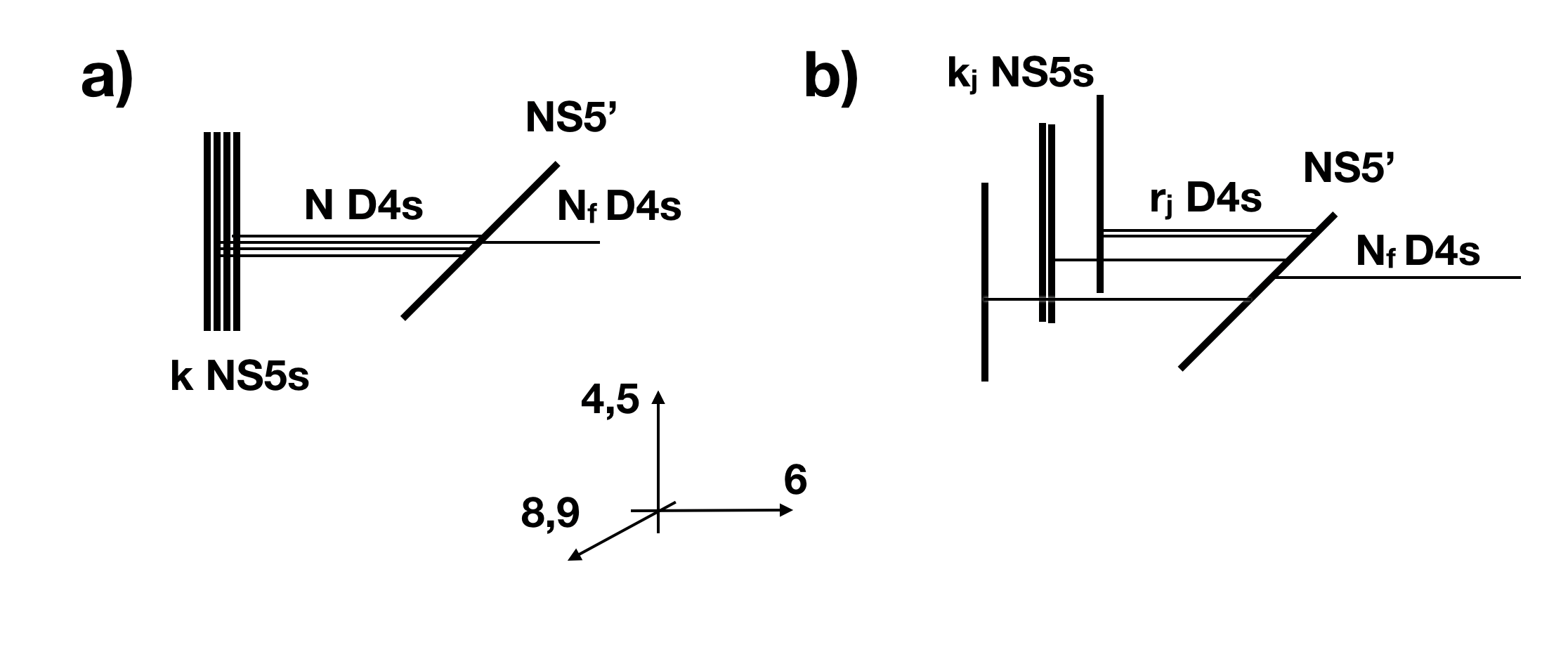}
\caption{\small a) Brane configuration for Kutasov theories with $k$ coincident NS5-branes and one NS5'-brane, corresponding to the theory with monomial superpotential (\ref{mono-supo}). b) Configuration for a more generic degree $(k+1)$ superpotential, with NS5-brane stacks of multiplicity $k_j$ for the $j^{th}$ critical point (i.e. different positions in 89),  with the D4-branes distributed in subsets of multiplicity $r_j$.}
\label{fig:kutasov}
\end{center}
\end{figure}

In order to get further intuition about the presence of the adjoint chiral multiplet, it is interesting to consider deformations of the above superpotential and their corresponding brane realization. Consider deforming the superpotential (\ref{mono-supo}) to a more general polynomial, with $k$ critical points  $\partial_X W=0$, which we initially take to be all different. In the gauge theory there is a potential, whose minima corresponds to distributing $r_j$ eigenvalues of the $N\times N$ matrix $X$ in the $j^{th}$ critical point. The gauge symmetry is broken as
\beqa
SU(N)\to SU(r_1)\times\ldots \times SU(r_k)
\eeqa
Around each such critical point, the superpotential for the adjoint is locally quadratic, so it can be integrated out, and we are left with a bunch of SQCD theories. In the brane picture, shown in Figure \ref{fig:kutasov}b, the deformation of the superpotential corresponds to separating the stack of $k$ NS5-branes in the directions 89, with their locations corresponding to the critical points $\partial_X W=0$. We can now distribute the $N$ D4-branes in sets of $r_j$ suspended between the $j^{th}$ NS5-brane and the NS5'-brane. We get $k$ different sectors of $r_j$ D4-branes which locally are suspended between one NS5- and one NS5'-brane, and hence lead to SQCD theories, i.e. without adjoint. 

It is now easy to consider the possibility that several critical points of $W$ coincide, e.g. we have $p$ different critical points with multiplicity $k_j$, with $\sum_{j=1}^p k_j=k$, and we again distribute the $N$ eigenvalues among the $p$ critical points. In the brane configuration, we move the NS5-branes in 89 but allow some of them to coincide, so we have $p$ stacks, with multiplicities $k_j$, adding up to $k$. We distribute the $N$ D4-branes and suspend $r_j$ of them between the $j^{th}$ stack of $k_j$ NS5-branes and the NS5'-brane. The resulting theory, either from the field theory or the brane configuration, is a set of $p$ sectors, labeled by $j$, each corresponding to a $SU(r_j)$ Kutasov theory with an adjoint and monomial superpotential with exponent $k_j+1$.

The above arguments motivated in \cite{Elitzur:1997fh} the intuition that the adjoint morally parametrizes motions of D4-branes in the directions 89. This will be useful in the construction of SCFTs in section \ref{sec:classIII-45678}.

\begin{figure}[htb]
\begin{center}
\includegraphics[scale=.35]{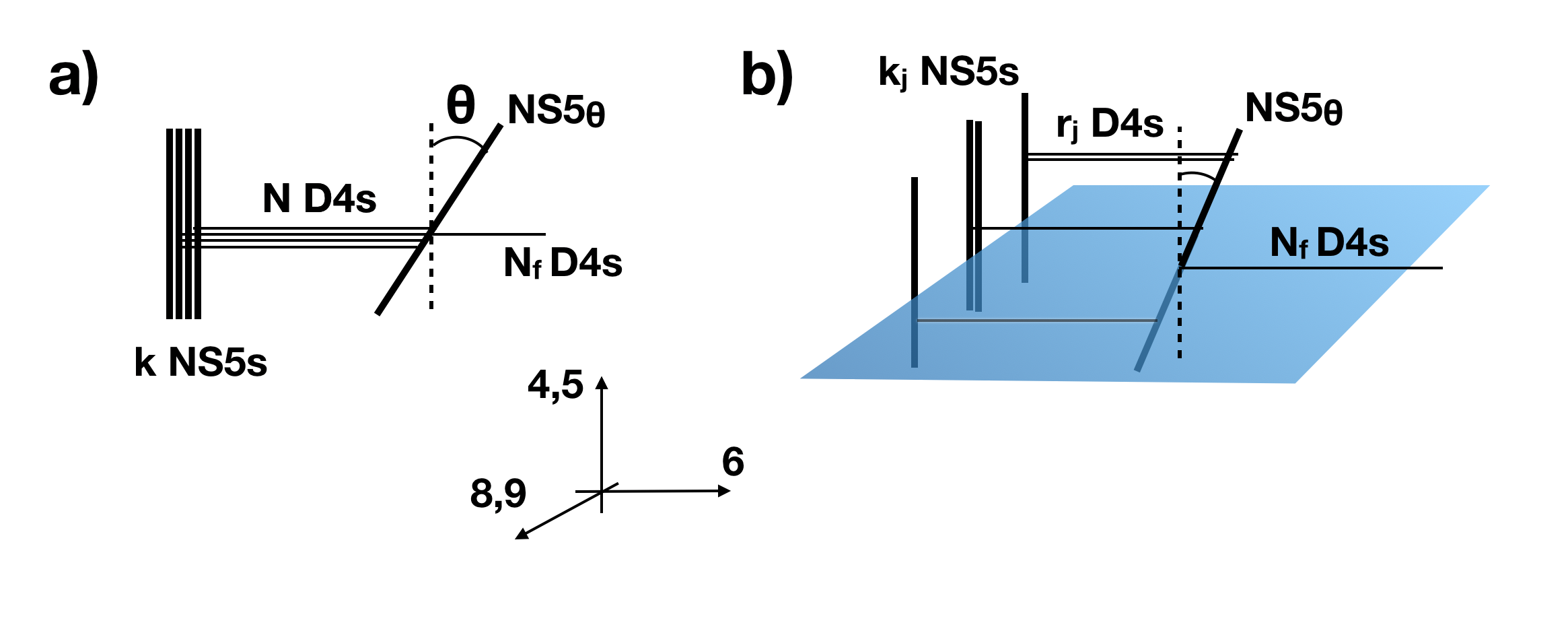}
\caption{\small a) Brane configuration for Kutasov theories with $k$ coincident NS5-branes and 1 rotated NS5$_\theta$-brane, corresponding to the theory with monomial superpotential (\ref{mono-supo}). b) Configuration for a more generic degree $(k+1)$ superpotential, with NS5-brane stacks of multiplicity $k_j$ for the $j^{th}$ critical point (i.e. different positions in 89),  with the D4-branes distributed in subsets of multiplicity $r_j$ in different positions in both 89 and 45. We have introduced a blue plane at the origin in 45 for reference.}
\label{fig:kutasov-rotated}
\end{center}
\end{figure}

Finally, let us note that we can consider a simple generalization of the above brane configuration by allowing the NS5'-brane to have a more general rotation between the 45 and 89 directions, i.e. we take it to be an NS5$_\theta$-brane as in section \ref{sec:rotated-branes}, see Figure \ref{fig:kutasov-rotated}a. It is easy to see that for $k$ NS5-branes and one NS5$_\theta$-brane for generic $\theta$ we still have an adjoint with monomial superpotential of the kind (\ref{mono-supo}). For instance, we can deform the monomial superpotential to a generic degree-$(k+1)$ polynomial, and show that the D4-branes can still distribute among different vacua, by simply adjusting their positions in the 89 and 45 coordinates simultaneously, see Figure \ref{fig:kutasov-rotated}b. Hence, the generalizations involving NS5-brane rotations also exist in brane configurations with coincident NS5-branes and leading to extra adjoint chiral multiplets. 

\bibliographystyle{JHEP}
\bibliography{references}
\end{document}